\documentclass[12pt, a4paper]{article}
\usepackage{hyperref}
\usepackage{graphicx}
\usepackage{mathrsfs}
\usepackage{amssymb}
\usepackage{color}
\usepackage{amsmath}
\usepackage{ascmac}
\usepackage{amsthm}
\usepackage{booktabs}
\usepackage{pdflscape}
\usepackage{dcolumn}
\usepackage{longtable}
\usepackage{dcolumn}
\usepackage{arydshln}
\usepackage{bm}
\usepackage{caption}
\usepackage{subfig}
\usepackage{setspace}
\usepackage{cases}
\usepackage{comment}
\usepackage{multirow}
\usepackage{booktabs}
\usepackage[square, sort,comma,numbers]{natbib}% Citation support using natbib.sty
\bibpunct[, ]{(}{)}{;}{a}{}{,}% Citation support using natbib.sty
\usepackage{url}

\newcommand{\eps}{\epsilon}

\begin{document}

\title{
\LARGE{Coal Mining, Women's Work, and Local Demographic Change: Evidence from Interwar Japan}
}

\author{Kota Ogasawara\thanks{Department of Industrial Engineering and Economics, School of Engineering, Institute of Science Tokyo, 2-12-1, Okayama, Meguro-ku, Tokyo 152-8552, Japan (E-mail: ogasawara.k.ab@m.titech.ac.jp).\newline
I wish to thank Daniel Gallardo Albarr\'{a}n, Walker Hanlon, Daisuke Haraguchi, Bernard Harris, Kate Ho, Janet Hunter, Yuzuru Kumon, Kazushige Matsuda, Melanie Meng Xue, Stephen Morgan, Tirthankar Roy, Tom Learmouth, Eric Schneider, Christopher Sims, Masayuki Tanimoto, and seminar participants at Groningen, LSE, Northwestern, PSE, Tokyo Tech, and Warwick for their helpful comments.
I would like to thank the Manuscript Library at Kyushu University and the Fukuoka Communal Archives for providing access to the archives on the Chikuh\=o coalfield.
This project was supported by the Japan Society for the Promotion of Science Grant 24K21409.
There are no conflicts of interest to declare.
All errors are my own.
A previous version of this paper was circulated under the title ``Technology, institution, and regional growth: Evidence from mineral mining industry in industrializing Japan.''
}
}
\date{
Institute of Science Tokyo\\[2ex]%
\today}
\maketitle

\begin{abstract}
\begin{spacing}{0.85}
This paper examines how coal mining shaped local demographic change in interwar Japan, focusing on the role of women's work.
Coal mines initially employed married couples as underground teams, offering women wages well above those in textile mills.
The mechanization of haulage and the subsequent revision of labor regulations, enforced in 1933, removed most women from the pits.
Linking mine locations with municipal-level census and vital statistics, I find that coal mining substantially increased local populations, mainly through the inflow of workers, whereas marriage and fertility rates in mining areas were lower than in surrounding areas.
After women were removed from the pits, these gaps narrowed.
Early-life mortality was higher in mining areas and increased with the number of female miners, consistent with occupational hazards to pregnant women, whereas there is little evidence of pollution channels.
\end{spacing}
\bigskip


\bigskip

\end{abstract}

\newpage
%-------------------------------------------------------------------------------
\section{Introduction}

Natural resource extraction has profound effects on economies.
Resource abundance can foster industrialization and regional development by providing energy and raw materials, but it can also hinder growth by crowding out other sectors.\footnote{See \citet{Sachs:1999ul, Sachs:2001vc} for the resource curse hypothesis, \citet{Corden:1982ub, Corden:1984tt} for theories of the effects of resource booms on other sectors, and \citet{Caselli:2009to, Lederman2007} for broader reviews of the economic consequences of natural resources.}
These effects are particularly pronounced in the regions where resources are extracted, because mining attracts workers and reshapes local labor markets.
Population growth in resource-abundant regions has typically been attributed to the inflow of workers attracted by higher wages.
However, local population also depends on marriage, fertility, and mortality, and much less is known about how resource extraction shaped these demographic outcomes.

Coal mining created earning opportunities not only for men but also for women.
In many coalfields, married couples worked underground as teams, and women earned substantially more in the pits than in other industries.
Economic theories of fertility predict that such opportunities raise the opportunity cost of childrearing and reduce fertility \citep{Becker1960-af, Galor1996-gw}, a relationship documented in historical settings where women's earning opportunities expanded \citep{Schultz1985-ch}.
Women's work in mining could also affect the health of their children, because heavy physical labor and poor working conditions during pregnancy may increase the risk of fetal and infant death.
Mining regions therefore offer a setting in which to examine how women's work shaped the demographic consequences of resource extraction.

This paper studies coal mining in interwar Japan.
In prewar Japanese coal mines, a skilled male miner extracted coal at the face, while a female miner, often his wife, hauled it to the gangway \citep{sumiya1968}.
Underground work offered women wages one and a half to two times higher than those in the silk and spinning industries \citep{Nishinarita1985}.
From the late 1920s, the transition to the longwall method and the introduction of conveyors replaced the hauling work of female miners, and the labor regulations enforced in 1933 prohibited underground work by women \citep{Tanaka:1977}.
The number of female miners fell from 65,147 in 1924 to 15,695 in 1933, and the female-to-male wage ratio in coal mining declined, whereas that in agriculture remained stable.
Contemporary observers noted that women who left the pits took on new roles in the home \citep{Iwaya1997}.

To examine these changes, I construct a municipal-level dataset by digitizing the 1925, 1930, and 1935 Population Censuses and Vital Statistics and linking them with the locations of coal mines documented in official directories of mines.
I also digitize municipal-level statistics on fetal and infant deaths in 1933 and 1938 from official reports on births and early-life deaths.
I first compare coal-mining municipalities with surrounding municipalities in each census year, using the distribution of coal-bearing geological strata as an instrumental variable for mine locations.
I then examine how outcomes in mining municipalities changed before and after the regulation enforcement.

The analysis yields three main findings.
First, coal mining substantially increased local populations, mainly through the inflow of workers.
Mining municipalities had more male-biased populations and smaller households, whereas their marriage and fertility rates were lower than those of surrounding areas, so that natural increase contributed little to population growth before the regulation.
Second, women's work shaped fertility in mining areas.
Before the regulation, female labor force participation in mining areas was lower than in surrounding areas, a substantial share of employed women worked in mining, and marital fertility was lower than in surrounding areas.
After women were removed from the pits, the fertility and marriage gaps between mining and surrounding areas narrowed.
Third, early-life mortality was higher in mining areas, and fetal and infant mortality increased with the number of female miners in mining municipalities, consistent with occupational hazards to pregnant women.
By contrast, I find little evidence that air or water pollution from coal mining increased early-life mortality.

This paper contributes to the literature on the local effects of resource extraction.
Studies of resource booms have documented their effects on employment, earnings, and population \citep{Black:2005uh, Michaels:2010ks}, and economic historians have debated the role of coal in regional development \citep{Mokyr1977, Fernihough2020}.
I show that the population growth associated with coal mining resulted mainly from migration, whereas the low marriage and fertility rates of mining areas limited natural increase.
The demographic consequences of resource extraction thus depended on how mining employment was organized, including the extent to which it relied on women.
This study also adds evidence from an industrializing Asian economy to a literature that has focused mainly on Europe, the United States, Latin America, and Africa.\footnote{See, for example, \citet{Fernihough2020} for Europe, \citet{Black:2005uh, Michaels:2010ks} for the United States, \citet{Aragon:2013ib} for Latin America, and \citet{Aragon:2015db, BenshaulTolonen:2019kx} for Africa. Recent studies also show that mining affects women's employment in local economies \citep{Kotsadam:2016du, Aragon2018-rs}.}

This paper also provides historical evidence on the relationship between women's work and fertility.
Mining areas, where women had access to high-wage mining jobs, had lower marriage and fertility rates than surrounding areas, and these gaps narrowed after women were removed from the pits.
These patterns are consistent with the prediction that women's earning opportunities raise the opportunity cost of children \citep{Becker1960-af, Galor1996-gw} and complement evidence from periods of rising female wages \citep{Schultz1985-ch}.
They also shed light on the consequences of protective labor legislation for women, which has been studied mainly in terms of its political economy and employment effects \citep{Humphries1981-pl, Goldin1988-mh}.
%However, because the areas that relied on female miners also tended to have larger mines, I cannot fully distinguish this channel from other changes in mining communities, such as firms' efforts to encourage miners to form families \citep{ogino1993}.

Finally, this paper contributes to the literature on the health costs of resource extraction and industrialization.
Studies have focused mainly on pollution, showing that mining and coal use increased infant mortality and reduced health in local populations \citep{Aragon:2015db, vonderGoltz:2019br, Beach2017kja, Heblich2017wp}.
I find that higher early-life mortality in mining areas was associated with the number of female miners rather than with proxies for pollution, suggesting that the occupational hazards faced by pregnant women were an important channel through which mining affected the health of the next generation.
This finding highlights a cost of women's work in hazardous industries that has received little attention in the literature on resource extraction, and it relates to studies showing that conditions in utero have long-lasting effects \citep{Almond2018-da}.

%--- (9) Roadmap
The remainder of this paper is organized as follows.
Section~\ref{sec:sec2} provides the historical background, and Section~\ref{sec:sec_cf} presents the conceptual framework.
Section~\ref{sec:sec_data} describes the data.
Section~\ref{sec:sec_ea} examines the effects of coal mining on local population, marriage, and fertility, and Section~\ref{sec:sec_pe} examines how these outcomes changed after women were removed from the pits.
Section~\ref{sec:sec_aa} analyzes early-life mortality.
Section~\ref{sec:sec_rob} discusses robustness, and Section~\ref{sec:sec_conclusion} concludes.

%--------------------------------------------------
\section{Historical Background} \label{sec:sec2}

\subsection{Coal Mining and Local Labor Markets} \label{sec:sec21}

%----------
%Figure 1
\begin{figure}[]
\centering
\captionsetup{justification=centering,margin=1.5cm}
\includegraphics[width=0.65\textwidth]{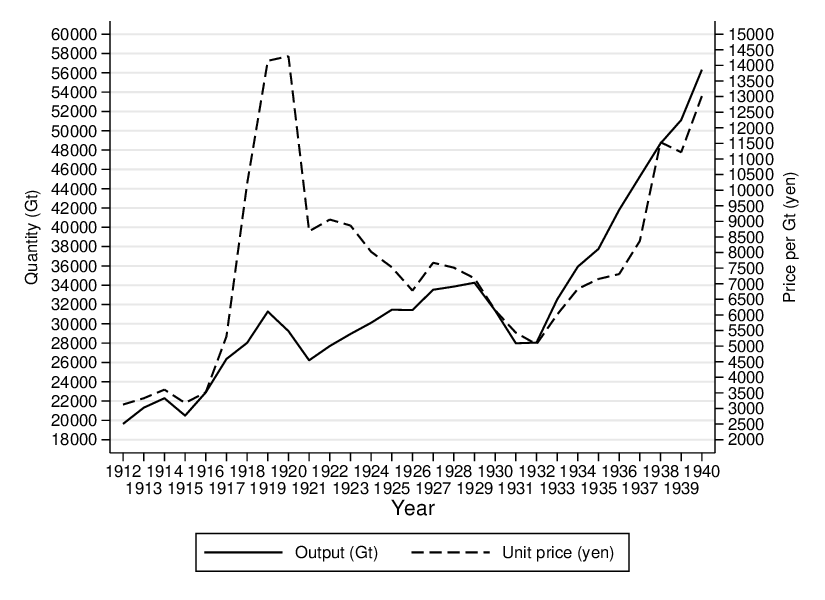}
\caption{Coal Production in Japan from 1912 to 1940}
\label{fig:ts_coal}
\scriptsize{\begin{minipage}{400pt}
\setstretch{0.85}
Notes:
The solid line shows the coal output in gigatons (Gt).
The dashed line indicates the coal price in yen per gigaton.
The unit price is defined as the total price of coal output divided by the total coal output in each year.
Source: Created by the author. Data on coal output are from \citet{miti1954}.
\end{minipage}}
\end{figure}
%----------

The Mining Code of 1892 allowed private companies to have mining concessions, accelerating the installation of modern technologies in mines and the development of Japan's mining sector.
Winding engines were installed in coal and copper mines in the 1890s, and the flame-smelting process was adopted in gold and silver mines in the 1900s \citep[pp.~220--221]{Ishii1991}.
Before the First World War (WWI), however, the Japanese coal industry had developed as a labor-intensive export industry.
Then, the outbreak of WWI increased miners' average wage rates, reducing the coal industry's comparative advantage.
Figure~\ref{fig:ts_coal} illustrates the development of the coal industry and shows the negative shock in the early 1920s.
This motivated coal mining firms to reduce the number of miners and adopt new extraction machines to improve productivity.\footnote{See \citet[pp.~387--393]{sumiya1968} for details on the machinery adoption process.}
Mining firms reallocated resources to relatively productive mines to improve their average productivity.
Consequently, average labor productivity in the coal mining sector increased from the 1920s, particularly in the 1930s \citep{Okazaki:2021}.\footnote{Generally, during the interwar period, the mining and manufacturing industries experienced steady capital investment, particularly in production facilities and machines, and public investments \citep[pp.~138--143]{Nakamura1971}. Consequently, the relative contribution of manufacturing and mining, which provided energy and resources to manufacturing, was the largest at 42-51\% among all economic sectors during this period \citep[p.90]{Minami1994}.}
Figure~\ref{fig:ts_coal} shows that total coal production increased during the interwar period.\footnote{As the moderate decline in the early 1930s confirmed, the effect of the depression was limited and temporary compared to that in other countries. For details, see Online Appendix~\ref{sec:seca_gd}.}

In the early 1930s, coal accounted for approximately $55$\% of total mineral output in Japan, meaning that coal extraction was an influential economic activity at that time \citep[p.44]{mb1934}.
Higher wage rates for miners motivated peasants to move into the mining industry, shifting the local industrial structure from agriculture to the production sector.
For example, the average daily wage of female miners working in the pits was one and a half to two times higher than that of female workers in the silk and spinning industries \citep[pp.~84--85]{Nishinarita1985}.
As a result, the populations of mining areas consisted not only of original inhabitants but also of a substantial number of migrant workers.
The inflow of workers and the employment of women in the pits may have changed marriage, fertility, and early-life mortality in mining areas, and thus how local populations grew.
However, little quantitative evidence exists on how coal mining shaped these demographic outcomes in local economies.\footnote{Most research on mines in prewar Japan comprises business history studies that investigated the features of the management and industrial organization in coal mining companies \citep[pp.~220-233]{Ishii1991}. \citet{tanaka1984} provided a comprehensive review of the operations of the Japanese coal mining industry.}

\subsection{Mining Technology and Labor Patterns}\label{sec:sec22}

By the end of the 19th century, most mines had mechanized the hauling processes in the main shaft.
This was done by installing equipment such as drainage pumps and hoisting machines.
However, this did not mean that the miners' workload related to hauling operations was reduced.
Instead, such mechanization in the main shafts increased the workload of miners extracting and transporting coal around the coalface \citep[pp.~310--312]{sumiya1968}.
Notably, while miners included single male workers, many of them were married couples.
This was because miners usually worked in teams of two to three \citep[pp.~300--308]{sumiya1968}.
The male skilled miner (\textit{saki yama}) extracted coal at the face; meanwhile, the female miner (\textit{ato yama}) carried that coal to the coal wagon at the gangway.
She traveled through steep pits while carrying wooden baskets \citep[p.~319]{sumiya1968}.
In addition, the environment in the pit was generally poor.
Accidents caused by rockfall occurred frequently.\footnote{Available statistics indicate that the average number of accidents per year in the pits between $1917$ and $1926$, causing miners' injuries and deaths, was $142,595$, of which $61,112$ were from rockfall.}
The air was polluted by dust, oxygen levels were low, and there was a risk of carbon monoxide poisoning.
There was also a fire risk from dynamite blasting and spontaneous combustion of gas and coal.\footnote{Accidents caused by explosions, poisoning, and suffocation were fewer than those caused by rockfall, but they still occurred $311$ times per year on average between $1917$ and $1926$. These accident statistics are from \citet[p.~table.6]{jma1928}.}

An institutional change that transformed the harsh working conditions of female miners took place during the interwar period.
Following the International Labor Conference after WWI, the Social Affairs Bureau held a consultation with mining companies in the 1920s to revise the Miners' Labor Assistance Regulations of 1916.
Initially, mining companies, which relied heavily on female miners for production, opposed the Social Affairs Bureau's insistence on prohibiting women from working underground and late at night.
However, the renewal of the mining method changed the attitudes of mining companies in the late 1920s.
In fact, the transition from the traditional room-and-pillar mining method to the longwall method was accompanied by the adoption of technologies such as coal cutters and conveyors.
Compared to the room-and-pillar method, the coal extraction space of the longwall method is larger, allowing for the introduction of those machines \citep[p.~91]{Nishinarita1985}.
Specifically, the conveyors moved extracted coal from pits to the gangway more efficiently than female miners, strongly incentivizing mining companies to prohibit female miners from underground work \citep[p.~79]{Tanaka:1977}.\footnote{Online Appendix~\ref{sec:seca0} provides finer details of this mechanization process. Online Appendix~\ref{sec:seca_tech} provides cross-sectional evidence on the replacement of female miners by machines.}
Finally, the Revised Miners' Labor Assistance Regulations of September 1928 prohibited some underground and late-night work by female miners.
As the grace period for enforcement of the regulations lasted until 1933, the number of female miners gradually declined from around 1930.
This change, induced by mechanization, was commonly observed in Japanese coal mines \citep[p.~392]{tanaka1984}.\footnote{A detailed chronology of technology adoptions in each coal mine in Chikuh\=o coalfield confirms this movement \citep{cccmi1973}.}

Panel A of Table~\ref{tab:nminers} shows that the number of female miners working in the pits declined sharply between 1930 and 1933, when the revised regulation was fully enforced.
The full enforcement of the revised labor regulations did not eliminate female miners altogether, as females were still allowed to work in the thin-layer pits.
They were primarily responsible for coal selection (\textit{sentan}) and other out-of-pit labor.
In the late 1930s, more than $15,000$ women still worked in the mines, representing approximately $10$\% of all the miners measured.
From this perspective, female miners played an important role in the coal mines throughout the interwar period.
The enforcement of regulations, however, impacted the market values of female workers' skills in the mining area.
Column 1 of Panel B in Table~\ref{tab:nminers} presents evidence that female miners' relative wage rate dropped after the enforcement.
While there was a decreasing trend in the relative wage rate during the mechanization process in the 1920s, the difference between 1930 and 1933 shows the largest decline ($0.66$ v.~$0.48$).
This reduction was not a macroeconomic trend around the mining region, given that the relative wage rate in the agricultural sector was stable during the same period (Column 2 of Panel B).
Moreover, the rate for in-pit workers decreased from $0.78$ to $0.62$, whereas that for out-pit workers remained relatively stable, from $0.49$ to $0.45$.
The mean difference before and after the enforcement year is statistically significant only in the in-pit workers' category (bottom of Panel B).
This indicates that regulating in-pit female miners did not improve the relative wage rate of the out-pit female miners.
Therefore, although the prohibition of underground work may have reduced occupational hazards for female miners, it also lowered the wages of female miners relative to those of male miners.
These figures are national averages, and part of the decline in the overall ratio reflects the shift of female miners from in-pit to out-of-pit work.

Although systematic statistics are unavailable, unemployed female miners were viewed as likely to become housewives.
This is plausible because most female miners were married: more than 70\% of female miners were married in both the 1920 and 1930 Population Censuses \citep[pp.~160--161]{census1930v2}.\footnote{The age distribution of female miners was also similar in the two censuses, indicating that the composition of female miners changed little during this period (Online Appendix~\ref{sec:seca_age_dist}).}
\citet[p.~11]{Iwaya1997} describes this shift, stating:
\begin{quote}
``...the mining community gave women a new role, not in labor, but in supporting labor in the home.''
\end{quote}
The official report of the 1930 census shows that the female labor force participation rate in the mining area was significantly lower than in the surrounding agricultural municipalities (Online Appendix~\ref{sec:test_lfpr}).
Women who left the pits had few alternative jobs in mining areas, and those available, such as domestic work, paid less than mining.

%Table 1
\begin{table}[]
\def\arraystretch{1.0}
\begin{center}
\captionsetup{justification=centering}
\caption{Number of Coal Miners and Relative Wage Rates:\\ Labor Statistics Field Survey Reports}
\label{tab:nminers}
\footnotesize
\scalebox{0.95}[1]{
{\setlength\doublerulesep{2pt}
\begin{tabular}{lccccc}
\toprule[1pt]\midrule[0.3pt]
\multicolumn{6}{l}{\textbf{Panel A: Number of miners}}\\
&\multicolumn{2}{c}{Female}&\multicolumn{2}{c}{Male}&\multicolumn{1}{c}{\% share of}\\
\cmidrule(rr){2-3}\cmidrule(rr){4-5}
Year
&\multicolumn{1}{c}{Miners}
&\multicolumn{1}{c}{Work in the pits (\%)}
&\multicolumn{1}{c}{Miners}
&\multicolumn{1}{c}{Work in the pits (\%)}
&\multicolumn{1}{c}{female miners}\\\hline
1924    &65147    &$-$    &179986    &$-$    &27    \\
1927    &56956    &71    &176524    &76    &24    \\
1930    &32680    &60    &153876    &76    &18    \\
1933    &15695    &37    &133460    &77    &11    \\
1936    &18694    &25    &178816    &77    &9        \\
&&&&&\\
\multicolumn{6}{l}{\textbf{Panel B: Relative wage rate (female/male)}}\\
&\multicolumn{3}{c}{(1) Mining Sector}&\multicolumn{2}{c}{(2) Agricultural Sector}\\
\cmidrule(rrr){2-4}\cmidrule(rr){5-6}
Year
&\multicolumn{1}{c}{Overall}
&\multicolumn{1}{c}{In-pit worker}
&\multicolumn{1}{c}{Out-pit worker}
&\multicolumn{1}{c}{Daily labor}
&\multicolumn{1}{c}{Annual labor}        \\\hline
1924        &0.81    &$-$        &$-$        &$-$        &$-$        \\
1927        &0.74    &0.83    &0.52    &0.77    &0.72    \\
1930        &0.66    &0.78    &0.49    &0.77    &0.72    \\\hdashline[0.5pt/3pt]
1933        &0.48    &0.62    &0.45    &0.75    &0.73    \\
1936        &0.45    &0.67    &0.46    &0.77    &0.70    \\\hline
Mean diff.        &0.27    &0.16    &0.05    &0.01    &0.01    \\
($p$-value)    &(0.016)    &(0.046)    &(0.124)    &(0.423)    &(0.769)    \\\midrule[0.3pt]\bottomrule[1pt]
\end{tabular}
}
}
{\scriptsize
\begin{minipage}{430pt}
\setstretch{0.85}
Notes:
The statistics on miners in this table are drawn from the Labor Statistics Field Survey Reports, which surveyed mines employing 50 or more miners.
\% share of female miners in Panel A shows the percentage share of females relative to the total number of miners.
The relative wage rate in Panel B is the female average daily wage divided by the male average daily wage.
The overall ratios for the miners between 1927 and 1936 (column 1) are recalculated using the number of miners working inside and outside of the pits and their daily wages.
The mean difference in the wage ratio around 1933 and $t$-statistic $p$-values for the mean difference are listed at the bottom of Panel B.\\
Sources:
Statistics are from \citet[pp.~5; 32]{lsfsr1924}, \citet[pp.~6--7; 26; 30]{lsfsr1927_4}, \citet[pp.~6--7; 26; 30; 38]{lsfsr1930}, \citet[pp.~41; 64; 68; 76]{lsfsr1933}, and \citet[pp.~45; 78; 79]{lsfsr1936}.
The average daily wages for agricultural laborers, both daily and annual, are obtained from \citet[p.~107]{ltes1966}.
%Statistics Bureau of the Cabinet (1926, p.~5; 1930, pp.~6--7; 1932b, pp.~6--7; 1936, p.~41; 1937, p.~45).
The survey subject description is from \citet[p.~1]{lsfsr1927_1} and \citet[p.~1]{lsfsr1936}.
% the Statistics Bureau of the Cabinet (1932a, p.~1; 1937, p.~1).
\end{minipage}
}
\end{center}
\end{table}

\subsection{Occupational Hazards and Pollution}\label{sec:sec23}

%--- (1) Occupational hazards in the pits
Working conditions in the pits were harsh.
Accidents caused by rockfall occurred frequently.\footnote{Available statistics indicate that the average number of accidents per year in the pits between $1917$ and $1926$, causing miners' injuries and deaths, was $142,595$, of which $61,112$ were from rockfall.}
The air contained dust, oxygen levels were low, and there was a risk of carbon monoxide poisoning.
There was also a fire risk from dynamite blasting and the spontaneous combustion of gas and coal.\footnote{Accidents caused by explosions, poisoning, and suffocation were fewer than those caused by rockfall, but they still occurred $311$ times per year on average between $1917$ and $1926$. These accident statistics are from \citet[p.~table.6]{jma1928}.}
Dust in the pits damaged miners' lungs and increased the risk of respiratory diseases such as tuberculosis \citep[pp.~410--411]{sumiya1968}.
A health survey of miners found that coal miners had the highest prevalence of respiratory disease among all types of miners at that time \citep[p.~3]{jma1928}.

%--- (2) Hazards for pregnant women
These conditions posed particular risks to pregnant women.
Under the traditional extraction system, female miners carried heavy baskets of coal through steep pits to the wagons at the gangway (Section~\ref{sec:sec22}).
Heavy physical work during pregnancy can increase the risk of adverse pregnancy outcomes \citep{Ahmed2007-wp}.
Medical studies have also shown that exposure to particulates from coal increases the risk of stillbirths and premature deaths through maternal respiratory diseases \citep{Landrigan:2017fq}.
These risks did not disappear with the full enforcement of the regulation in 1933.
Women were still allowed to work in thin-layer pits and engaged in coal selection (\textit{sentan}) and other out-of-pit work, and more than $15,000$ women worked in coal mines in the 1930s (Table~\ref{tab:nminers}).
Therefore, early-life mortality in mining areas may have depended on the number of female miners even after the enforcement.

%--- (3) Pollution
Pollution from coal mining may also have increased early-life mortality in mining areas.
Industrial pollution was not generally recognized as a social problem in prewar Japan, and there is little documentation of pollution caused by coal mining.\footnote{An exception is the Ashio Copper Mine Incident in the late 19th century \citep[e.g.,][]{Notehelfer:1975vg}. In the field of development economics, \citet{vonderGoltz:2019br} provides evidence that mining is associated with child stunting in developing nations.}
However, anecdotal evidence suggests two potential pathways.

The first is air pollution from soot and smoke generated by burning coal in the boilers of steam-powered winding engines.
These engines were introduced early in the mechanization of the hauling process (Section~\ref{sec:sec21}), but they were gradually replaced by electric winders during the interwar period \citep{jma1930, jma1931}.
The second, and presumably more relevant, is water pollution from wastewater generated during mining.
As coal mining became increasingly mechanized, some mines introduced water-washing machines for coal sorting, and coal sludge from the sorting process began to contaminate rivers.
Although little historical documentation systematically describes this pollution, the Onga River in the Chikuh\=o coalfield was called ``\textit{zenzai}'' (sweet red-bean soup) because its water was contaminated by liquid waste \citep{chiba1964}.
Coal sludge contains toxic metals, including mercury (Hg) and cadmium (Cd), which are associated with miscarriages and stillbirths \citep{Amadi:2017fl, Sergeant2022-qm}.
If this pathway was important, early-life mortality should have been higher in mining areas with greater access to rivers.
From the 1920s, however, some mines installed settling ponds to purify wastewater and recirculate it for coal cleaning \citep[pp.~313--316]{acm3}, which may have mitigated this risk.
Section~\ref{sec:sec_aa} examines these occupational hazard and pollution channels using municipal-level data on fetal and infant deaths in 1933 and 1938.

%----------------------------------------
\section{Conceptual Framework}\label{sec:sec_cf}

%--- (1) Motivation
Economic theories of fertility emphasize that women's earning opportunities affect fertility through the opportunity cost of childrearing \citep{Becker1960-af, Galor1996-gw}.
In mining areas, underground work offered women higher wages than other available jobs (Section~\ref{sec:sec22}), and the revised regulation reduced these opportunities.
The effect of such a change on fertility is theoretically ambiguous.
A decline in women's wages lowers the opportunity cost of childrearing, which increases fertility (substitution effect), but it also reduces household income, which decreases fertility (income effect).
This section presents a simple model to clarify which effect is likely to dominate, given the historical conditions of coal mining areas in prewar Japan.
I consider the collective model of \citet{Siegel2017-sn}, in which a married couple chooses fertility to maximize household utility.\footnote{For brevity, I consider a simplified model that abstracts from the utility from leisure and from some home production parameters. This model offers an interpretation of the optimal fertility choice materially similar to that of the original generalized model. See \citet{Browning2014} for a comprehensive review of cooperative and non-cooperative models of couples.}

%--- (2) Model
Both spouses derive utility from consumption ($c$) and children ($b$) under the additively separable utility function $u_{g} (c_{g}, b) = \log c_{g} + \tau \log b$, where $g \in \{m, f\}$ indicates the husband and the wife, respectively.
Raising children requires home production $y(b) = \pi b$ with $\pi > 0$, which is produced using labor-saving services purchased in the market ($s$) and the parents' total home labor input ($\Omega$) as $y(b) = s^{\gamma} \Omega^{1-\gamma}$.
The parameter $\gamma$ ($0 \leq \gamma < 1$) represents the availability of childcare services in the economy.
The parents' total home labor input is given by the constant elasticity of substitution function $\Omega = \left( \kappa_{m} h_{m}^{1-\rho} + \kappa_{f} h_{f}^{1-\rho} \right)^{\frac{1}{1-\rho}}$, where $\kappa_{g}$ is the productivity of housework including childcare, $h_{g} \geq 0$ is the home labor input, and $\rho > 0$ is the inverse of the elasticity of substitution.
Labor income is the product of the wage $w_{g}$ and the labor input $n_{g} \geq 0$, and it is used for consumption and for purchasing labor-saving services.

The household maximizes the weighted sum of utilities $\theta u_{m} (c_{m}, b) + (1 - \theta) u_{f} (c_{f}, b)$ subject to the budget constraint $c_{m} + c_{f} + s = w_{m} n_{m} + w_{f} n_{f}$ and the time constraints $n_{g} + h_{g} = 1$ for $g \in \{m, f\}$.
The optimal fertility ($b^{*}$) is then given in closed form.
The derivative of optimal fertility with respect to the wife's wage ($\partial b^{*}/\partial w_{f}$) is negative under the following condition:
\begin{eqnarray}\label{sign}
\footnotesize{
\begin{split}
\gamma \kappa_{f}^\frac{1}{\rho} w_{f}^\frac{\rho-1}{\rho}
+ w_{m} \left[\left(\frac{\kappa_{m}}{w_{m}} \right)^{\frac{1}{\rho}} -  (1-\gamma)\left(\frac{\kappa_{f}}{w_{f}} \right)^{\frac{1}{\rho}} \right] < 0.
\end{split}
}
\end{eqnarray}
Online~\ref{sec:secb} provides the derivation.

%--- (3) Applying the condition to the historical setting
Under the paternalistic social norm stipulated in the Old Civil Code, the husband's productivity in housework, particularly childrearing, was negligible ($\kappa_{m} = 0$).\footnote{The same predictions can be derived under the more plausible assumption that $\kappa_{m} \simeq 0$ (Online~\ref{sec:secb}).}
Condition~(\ref{sign}) then reduces to $w_{f}/w_{m} < (1-\gamma)/\gamma$.
The right-hand side is at least one when $\gamma \leq 0.5$, whereas the female-to-male wage ratio in coal mining was below one throughout the period (Table~\ref{tab:nminers}).
Because childcare services were scarcely available at that time, the condition is likely to hold, so that the substitution effect dominates and fertility decreases with women's wages.\footnote{As mechanization progressed, miners were required to accumulate knowledge and experience in machinery operation. Mining firms therefore had an incentive to encourage miners to form families and continue their employment over the long term \citep{ogino1993}, and many coal mines established nurseries and other welfare facilities from the 1920s \citep{bm1926}. However, these nurseries charged fees and met only a small part of childcare needs, suggesting that $\gamma$ remained low (Online~\ref{sec:secb}). These family policies could also have raised fertility in mining areas independently of women's wages, a possibility I discuss when interpreting the results.}
This result yields the following two predictions:

\begin{enumerate}
\item[]
Prediction 1: Before the regulation, fertility was lower in mining areas than in surrounding areas because women in mining areas had access to high-wage underground work.
\item[]
Prediction 2: After the regulation, the fertility gap between mining and surrounding areas narrowed because women's wages in mining declined relative to men's.
\end{enumerate}

Because the model describes the fertility choice of married couples, these predictions apply most directly to marital fertility.
The model does not describe marriage decisions, and I interpret differences in marriage rates together with the composition of the population, which was shaped by the inflow of workers.

%--- (4) Local population change
These predictions also have implications for local population change.
Over a given period, the change in population is the sum of natural change and net migration:
\begin{equation*}
\Psi_{t+1} - \Psi_{t} = (\Upsilon_{t} - \Theta_{t}) + \Pi_{t},
\end{equation*}
where $\Psi$, $\Upsilon$, $\Theta$, and $\Pi$ indicate the number of people, live births, deaths, and net migrants, respectively.
First, Prediction 1 implies that births contributed little to population growth in mining areas before the regulation, whereas Prediction 2 implies that their contribution increased after the regulation.
Second, deaths are determined outside the household's fertility choice in the short run.
Heavy work and poor conditions in the pits could increase fetal and infant deaths among female miners (Section~\ref{sec:sec23}).
Because women continued to work in and around the mines after the full enforcement of the regulation, early-life mortality in mining areas may have increased with the number of female miners even after 1933.
Third, net migration was positive in mining areas because higher wages attracted workers from surrounding agricultural areas.
The regulation reduced the attractiveness of mining work for women, but it is unlikely to have offset the benefits of moving for couples, because the regulation did not reduce male miners' wages.\footnote{Because female miners usually migrated with their husbands, the migration of couples was less affected by the regulation than that of single women.}
In summary, mining areas experienced an inflow of workers throughout the period, whereas their fertility and early-life mortality may have been shaped by women's work in the mines and its regulation.

%----------------------------------------
\section{Data}\label{sec:sec_data}

This study created a unique dataset of the demographics across mining areas in the Japanese archipelago.
As mining is a localized economic activity, a smaller lattice dataset is preferred to identify the impact of mines.
A set of official census-based municipal-level statistics published in the interwar period was collected and digitized.

\subsection{Dependent Variables}\label{sec:sec_data_dv}

The primary dependent variable, population, was obtained by digitizing municipal-level statistics from the 1925, 1930, and 1935 Population Censuses \citep{census1925v3, census1930v1, census1935v1}.
To investigate the natural change mechanism underlying local population growth due to mining, I also digitized the 1925, 1930, and 1935 Municipal Vital Statistics \citep{mvs1925, mvs1930, mvs1935}.
They document the total numbers of marriages, live births, and deaths across all municipalities.
I consider four dependent variables: the crude marriage rate, the crude birth rate, the marital fertility rate, and sex-specific death rates.\footnote{The crude marriage and birth rates are the numbers of marriages and live births per $1,000$ people, respectively. The marital fertility rate is the number of live births per $1,000$ married women. Because the number of married women by age is unavailable at the municipal level, this measure does not account for differences in the age composition of married women. This limitation is unlikely to explain the lower marital fertility in mining areas: because mining areas attracted young couples, married women there were likely younger on average, which would bias marital fertility upward rather than downward. Moreover, out-of-wedlock births accounted for a small share of births in interwar Japan \citep{Ogasawara2022-fx}, and the results are similar when the crude birth rate is used. The sex-specific death rates are the numbers of male and female deaths per $1,000$ males and females, respectively.}
These are used to test miners' fertility responses and changes in mortality associated with mining activities (Section~\ref{sec:sec_ea}) and female labor regulations (Section~\ref{sec:sec_pe}).

The infant mortality rate and fetal death rate are considered in Section~\ref{sec:sec_aa} to analyze the occupational hazard channel.\footnote{The infant mortality rate is the number of infant deaths per $1,000$ live births. The number of censored observations in infant mortality is negligible, with 5 or fewer. The fetal death rate is defined as the number of fetal deaths per $1,000$ births. The censored observations in the fetal death rates are $9.3$\% and $14$\% in 1933 and 1938, respectively. I confirmed that the estimates from the Tobit estimator \citep{Tobin:1958uq} are materially similar to those from the OLSE (not reported). This confirms the evidence that censoring is not material in the empirical setting.}
These rates allow testing the impact of occupational hazards on female miners' mortality selection before birth within the biological framework.
The crude death rate among children aged 1--4 is also considered to assess the potential impacts of air pollution.\footnote{The child mortality rate is defined as the number of deaths of children aged 1--4 in 1933 (or 1938) per $1,000$ children aged 1--5 in 1930, because the number of children by age is documented only in the 1930 Population Census. The denominator therefore does not correspond to the population at risk in the year of death, particularly for 1938, and it should be interpreted as a measure of the size of the child population in each municipality.}
To obtain both variables, I digitized the 1933 and 1938 reports documenting municipal-level statistics on births and infant deaths, published by \citet{mvs1933aiiku} and \citet{mvs1938aiiku}.

Comprehensive statistics on migration are unavailable for the prewar period.
I therefore infer the role of migration indirectly.
Because population change is the sum of natural change and net migration (Section~\ref{sec:sec_cf}), population growth in mining areas that exceeds their natural increase indicates net in-migration.
To examine the characteristics of migrants, I also consider the sex ratio and household size.
An inflow of single male workers raises the sex ratio and reduces the average household size, whereas an inflow of married couples changes the sex ratio little.
These two variables thus indicate both the presence of in-migration and whether it consisted mainly of single men or of couples.
The data on the sex ratio and household size were collected from the official reports of the 1925, 1930, and 1935 Population Censuses.\footnote{The sex ratio is the number of males divided by the number of females. The average household size is the number of people in households divided by the number of households.}

These variables are standard measures of local demographic outcomes in the literature \citep[e.g.,][]{Angrist:2010tz, Abramitzky:2011bu}.
The census documents the total number of people in all municipalities in the census years.
Similarly, the vital statistics are compiled through a national registration system that covers all births and deaths in a given year.
This comprehensiveness helps avoid sample selection issues in statistical inference.\footnote{The potential imprecision of birth data in prewar Japan was mostly eliminated by the 1920s \citep{Drixler2016}.}
Online Appendix Table~\ref{tab:sum_outcome} presents the summary statistics on these variables.

To construct the base municipal dataset, I used the 1938 municipal classifications, the final year of the statistics used to construct the dependent variables.
Consequently, data for municipalities that had been consolidated by 1938 were re-tabulated according to the post-consolidation municipal classifications.
Since municipal divisions are rare, I excluded them from the analysis.
Small islands are excluded from the base municipalities because they had no mines and were far from the main island.
The total number of municipalities observed in these census years is $11,142$, based on the municipal boundaries in the final sample year, 1938.
To construct the analytical sample, I retain observations surrounding coal mines.
This means that the generation of analytical samples depends on the locations of mines.
I introduce the data on mine deposits in the next subsection.

\subsection{Locations of Coal Mines}\label{sec:sec_data_mine}

I use official reports named the National Directory of Factories and Mines (\textit{Zenkoku K\=ojy\=o K\=ozan Meibo}) published by \citet[][]{cs1924, cs1932, cs1937} (hereafter, the NDFM), which document coal mine locations measured in December 1922, October 1931, and October 1936.
The NDFM listed all mines employing $50$ or more miners.
The main target of this study is therefore the average impact of small- to large-scale coal mines, excluding very small collieries employing fewer than $50$ miners.\footnote{Coal mines were geographically clustered: for example, $90$\% of the municipalities with coal mines had one to three mines in 1930. My main findings are not sensitive to the exclusion of the small number of municipalities with exceptionally large numbers of coal miners (Online Appendix~\ref{sec:r_inf}).}
Because the NDFM includes small and medium-sized mines with $50$ or more miners, it covers a wider range of mines than studies that focus on large mines owned by major firms.

The treatment group is defined as the municipalities located within $0$--$5$ km of the centroid of a municipality with coal mines.\footnote{The study used an official shapefile provided by the Ministry of Land, Infrastructure, Transport and Tourism for geocoding. See Online Appendix~\ref{sec:secc1} for the details.}
A $5$ km threshold is plausible given that the mean distance to the nearest neighboring municipality is roughly $3.7$ km, so that the treatment group includes the municipalities with mines and their immediate neighbors.
Consistent with this, estimates by distance band show that the effects of coal mines are concentrated within $5$ km and are not statistically significant beyond this range (Online Appendix~\ref{sec:test_threshold}).
For the control group, the threshold is set as $5$--$30$ km from the centroid.\footnote{The definition of both groups is conservative compared with that in the literature, which uses 10--20 km and 100--200 km as the thresholds of the treatment and control groups, respectively \citep{Wilson:2012ic, Aragon:2015db, Kotsadam:2016du, BenshaulTolonen:2019kx}. This is because Japan has a smaller, thinner archipelago than those of countries studied in the literature, such as South Africa. Online Appendix~\ref{sec:test_threshold} provides evidence that 5--30 km is a plausible distance for the control group given that the impacts of the coal mines are concentrated within 5 km of the mines.}
The dominant heavy metal mines were gold, silver, and copper mines (Online Appendix~\ref{sec:seca2}).
Generally, coal and heavy metal mines have different mining points and did not coexist.
To be conservative, however, I excluded several duplications with these gold, silver, and/or copper mines from the coal mine sample.
For similar reasons, I do not consider the oil mines, oil refinery, and smelting places in this study.
The generation step of the analytical sample will be presented in Sections~\ref{sec:sec_ea}, \ref {sec:sec_pe}, and~\ref{sec:sec_aa}.

%----------
%Figure 2
\begin{figure}[]
\centering
\captionsetup{justification=centering}
\subfloat[December 1922]{\label{fig:map_coal_1922}\includegraphics[width=0.35\textwidth]{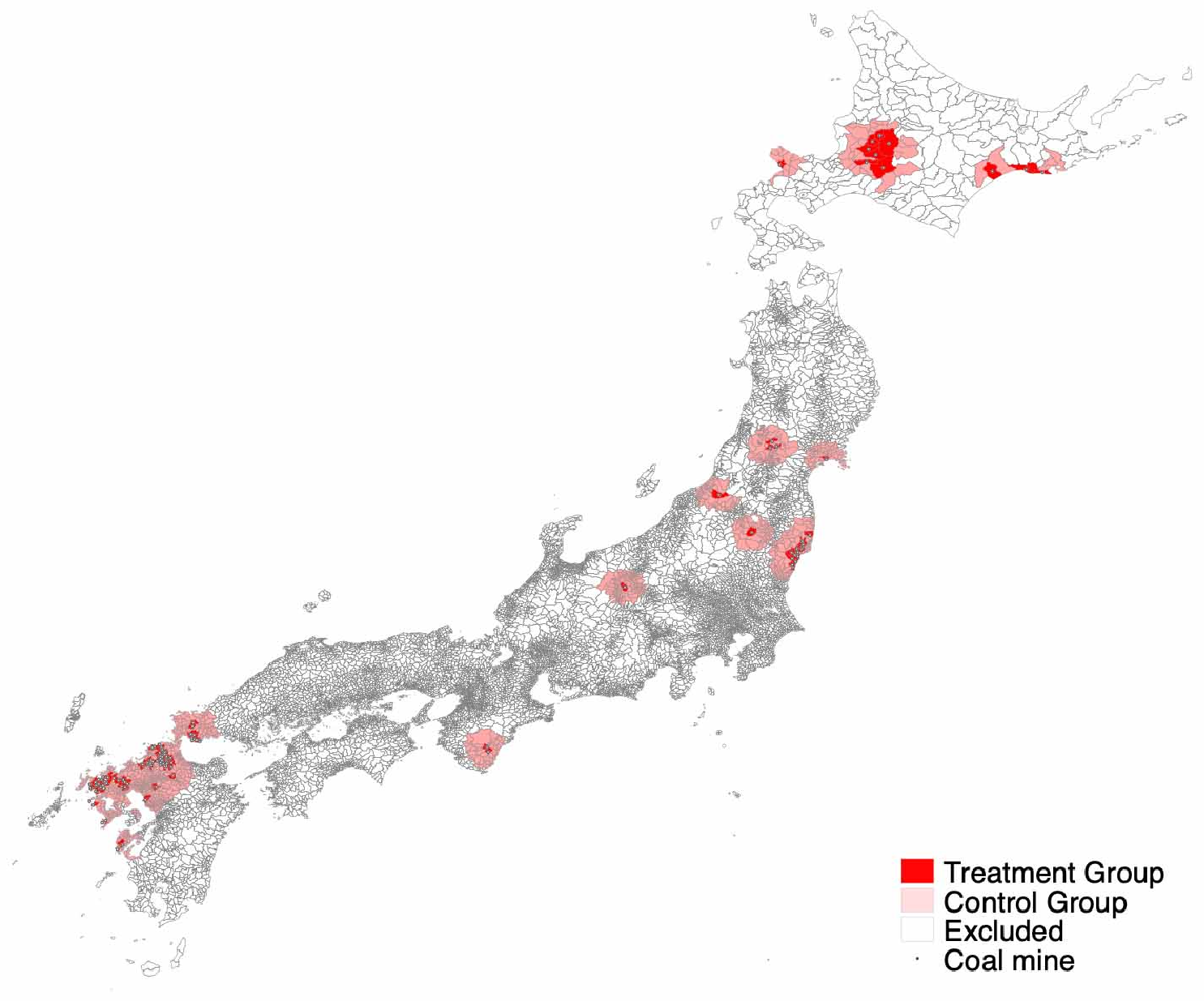}}
\subfloat[October 1931]{\label{fig:map_coal_1931}\includegraphics[width=0.35\textwidth]{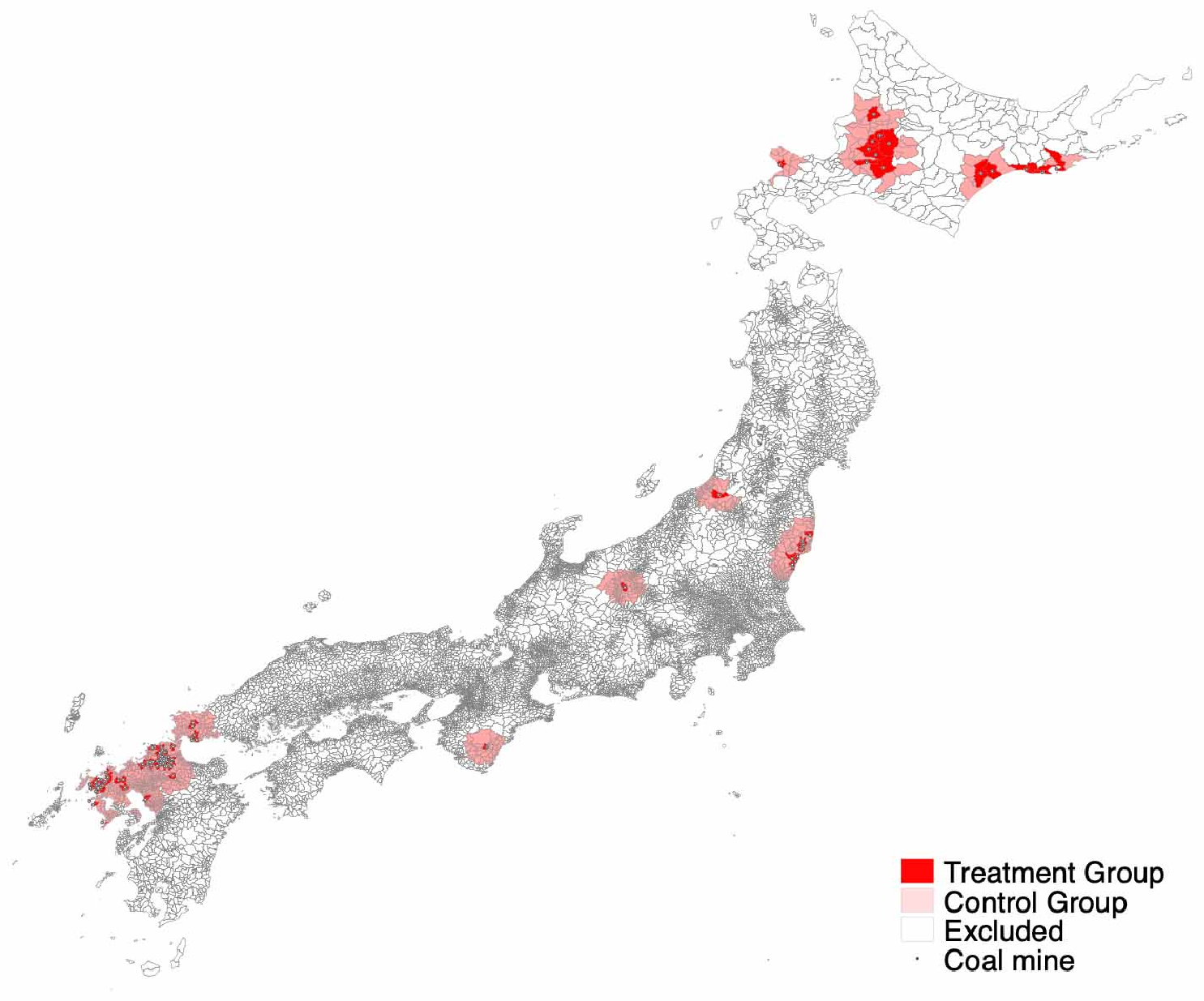}}
\subfloat[October 1936]{\label{fig:map_coal_1936}\includegraphics[width=0.35\textwidth]{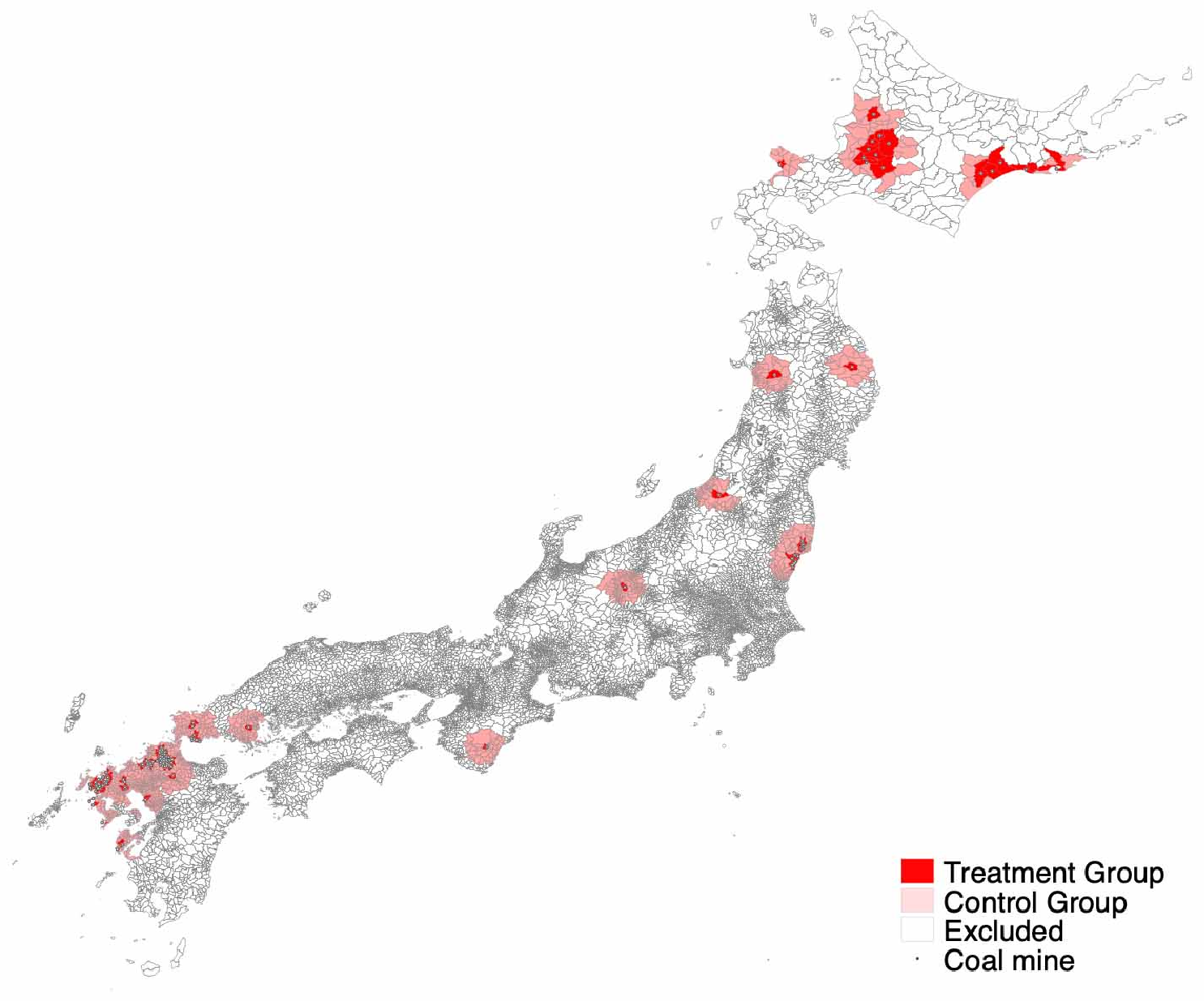}}
\caption{Spatial Distribution of Treated and Control Municipalities}
\label{fig:map_coal_mines}
\scriptsize{\begin{minipage}{450pt}
\setstretch{0.9}
Notes:
The white circles mark the locations of municipalities with coal mines, as measured in December 1922, October 1931, and October 1936.
The treatment group highlighted in red includes the municipalities within 5 kilometers of a mine.
The control group, highlighted in pink, includes municipalities located between 5 and 30 kilometers from a mine.
The excluded municipalities are shown as empty lattices in the figures.
Sources: Created by the author using the NDFM \citep[][]{cs1924, cs1932, cs1937}.
\end{minipage}}
\end{figure}
%----------

Figure~\ref{fig:map_coal_mines} indicates the coal mine locations from the NDFM and the corresponding treatment and control groups by year.
Most coal mines are located in three specific areas in the southernmost, central, and northeastern regions.
These are representative coalfields: Chikuh\=o coalfield in Ky\=ush\=u (the southernmost island), Jy\=oban coalfield in T\=ohoku (the northeastern district), and Ishikari-Kushiro coalfield in Hokkaid\=o (the northernmost island).
Notably, a comparison of these three figures shows that these agglomerations were relatively stable over time.
I will take a closer look at this point when I generate the analytical sample in each empirical analysis section.

\subsection{Geological Stratum}\label{sec:sec_data_gs}

The spatial distribution of Paleogene sedimentary rock, which contains the coal-bearing formations of Japan's major coalfields, is used as an instrumental variable (IV) for the location of coal mines.
Data on the geological strata are obtained from the official database of the Ministry of Land, Infrastructure, Transport and Tourism, which includes roughly nine thousand stratum points.
Each municipality in the analytical sample was matched to the nearest stratum point to identify the relevant stratum.
Online Appendix~\ref{sec:secc2} provides finer details of the definition and an example of a geological columnar section of a representative coal mine to show the relevance of the IV.
Online Appendix Table~\ref{tab:sum_key} presents the summary statistics of the indicator variables for the stratum.

\subsection{Control Variables}\label{sec:sec_data_cv}

As coal is bulky, railways were the primary means of transporting it.
Firms may prefer to set extraction points closer to railway stations to reduce transportation costs \citep[pp.~440--441]{sumiya1968}.
Further, the local population size could be larger as one moves closer to stations.
Therefore, controlling for the distance between each municipality and its nearest neighboring station is preferable to address potential endogeneity.
The location information for all stations, obtained from the official dataset provided by the Ministry of Land, Infrastructure, Transport and Tourism, was used to compute the distance to the nearest-neighbor station.

Although railways were primarily used to transport coal because coal mines were usually inland, marine transportation was also used for secondary logistics.
Hence, as in railway transportation, the distance to the nearest seaport is included as a variable to control for accessibility to marine transportation.
The locations of seaports are obtained from the official dataset provided by the Ministry of Land, Infrastructure, Transport and Tourism and supplemented with the ports listed in the Harbor Statistics of the Great Empire of Japan.

Accessibility to rivers is also considered.
Not all mining areas have large rivers suitable for water transportation, meaning that rivers did not dominate the locations of coal mines.
However, several mines used river transport as their primary mode of logistics until the development of railways (before the late 1920s) \citep{chiba1964}.
This implies that the river's spatial distribution could have influenced the locations of mines in such mining areas.
To control for this possibility, the distance to the nearest neighboring river is included as a control variable.
The locations of rivers managed by the national government or prefectures are obtained from the official dataset provided by the Ministry of Land, Infrastructure, Transport and Tourism.

Finally, the potential influence of the elevation is taken into account.
The mine's location might have been correlated with municipal elevations because terrain may relate to locational fundamentals such as climatic conditions and fixed transportation costs.
The grid cell elevation data provided by the Ministry of Land, Infrastructure, Transport, and Tourism were used to systematically calculate the average elevation in each municipality.

As shown below, including these control variables has a more moderate influence on the estimates once indicators for cities and towns are included.\footnote{The city and town indicators are based on municipal status in 1938, the year used to define the municipal boundaries (Section~\ref{sec:sec_data_dv}). Because coal mining may have contributed to the promotion of villages to towns or of towns to cities by 1938, these indicators may partly reflect the effects of mining. In this case, conditioning on them absorbs part of the effect of coal mining, and the estimates that include them should be interpreted as conservative. In the panel analysis, the indicators are time-invariant and absorbed by the municipal fixed effects, so this issue does not arise.}
Nevertheless, I include all the controls in the main cross-sectional specifications, because they make the exclusion restriction of the instrument more plausible.
The river accessibility and average elevation variables are time-constant and not included in the panel data analysis.
Online Appendices~\ref{sec:secc41}--\ref{sec:secc44} summarize the details of the data and maps for railways, seaports, rivers, and elevation.
Online Appendix Table~\ref{tab:sum_control} presents the summary statistics.\footnote{All the controls are log-transformed to better model the skewness of distances and elevations in nature, although this transformation does not affect the results.}

%----------------------------------------
\section{Estimating the Impact of Coal Mining on Regional Population Growth}\label{sec:sec_ea}

This section investigates the impact of coal mining on local population growth.
The following subsections first describe the construction of the analytical sample in light of the spatiotemporal distribution of coal mines and then present the empirical specification.
The subsequent section reports the estimation results.

\subsection{Analytical Sample}\label{sec:sec_ea_as}

\begin{figure}[]
\centering
\captionsetup{justification=centering}
\subfloat[December 1922]{\label{fig:map_coal_1922_Kyushu}\includegraphics[width=0.35\textwidth]{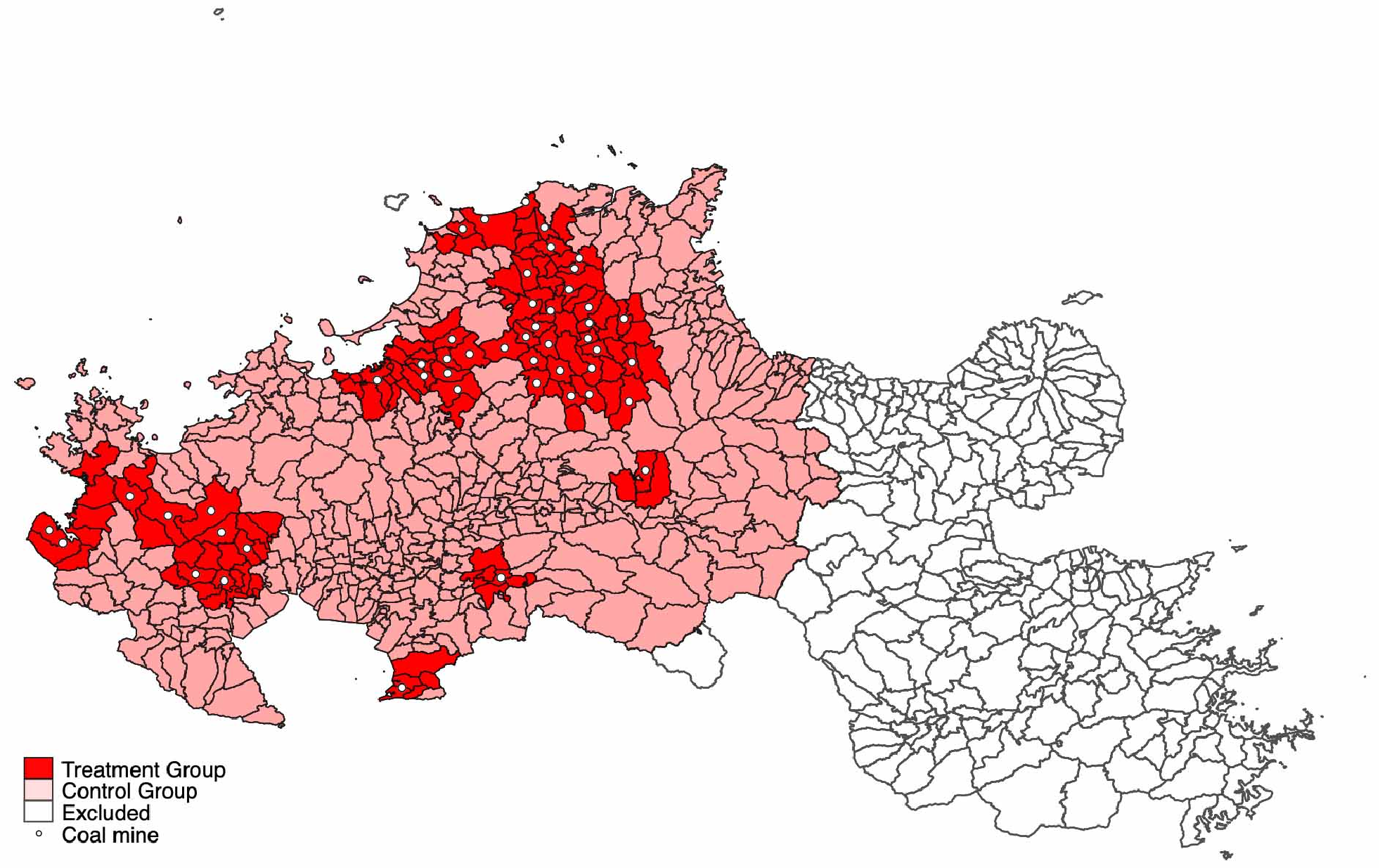}}
\subfloat[October 1931]{\label{fig:map_coal_1931_Kyushu}\includegraphics[width=0.35\textwidth]{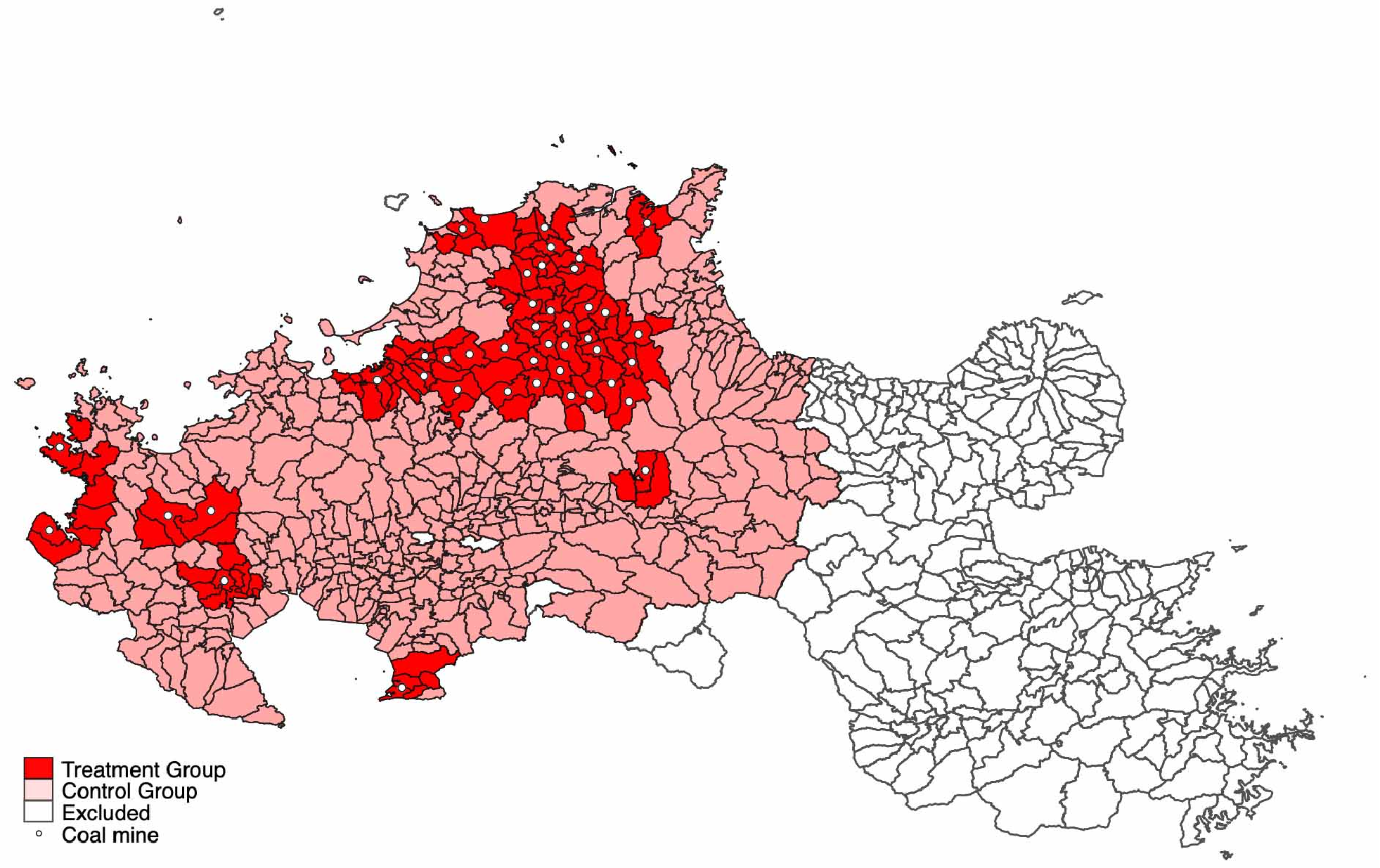}}
\subfloat[October 1936]{\label{fig:map_coal_1936_Kyushu}\includegraphics[width=0.35\textwidth]{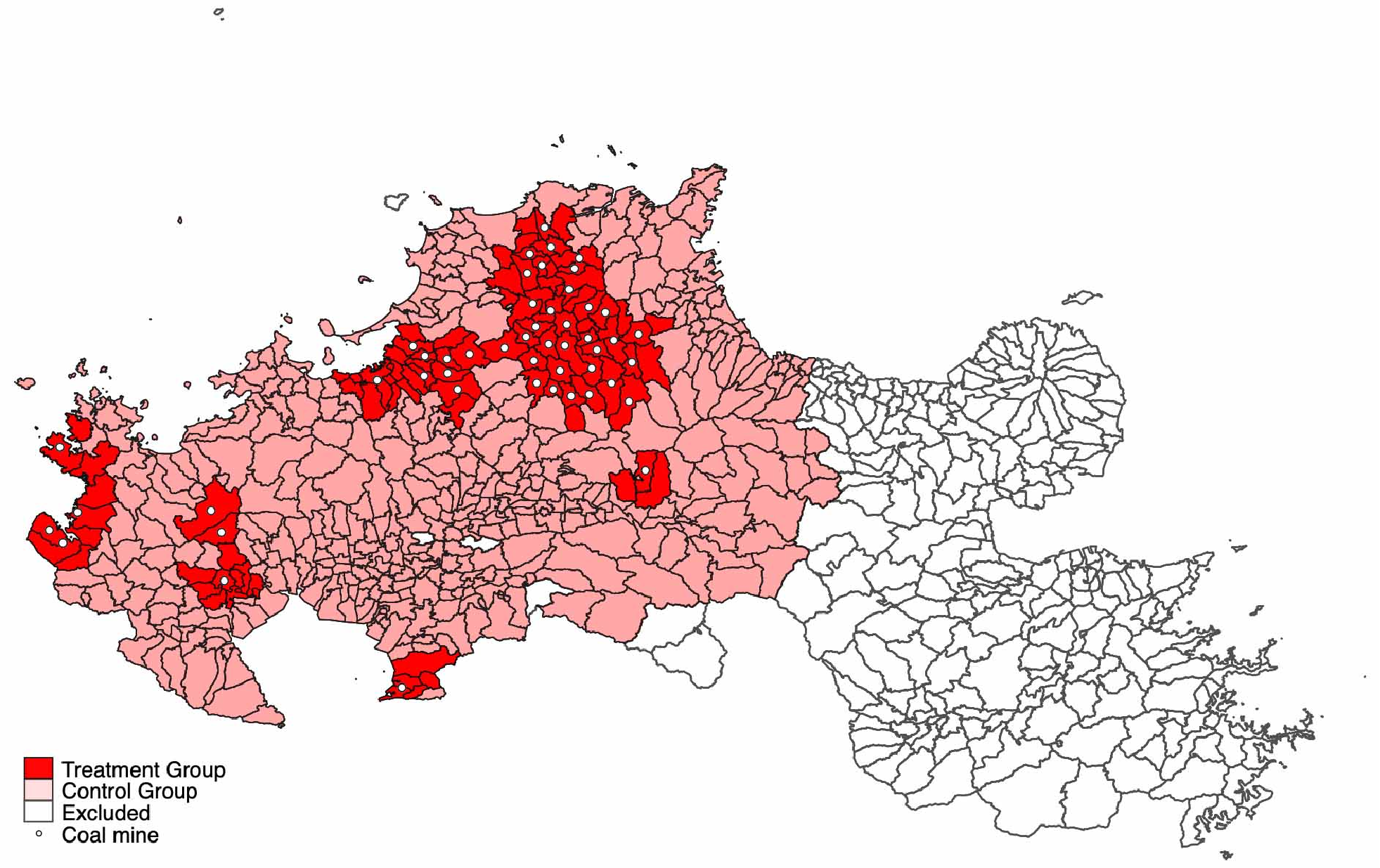}}
\caption{Location of Coal Mines surrounding Chikuh\=o Coalfield in Ky\=ush\=u}
\label{fig:map_coal_mines_Kyushu}
\scriptsize{\begin{minipage}{450pt}
\setstretch{0.9}
Notes:
White circles in Figure~\ref{fig:map_coal_mines_Kyushu} mark the coal mines in the Chikuh\=o, Kasuya, and Miike coalfields in Fukuoka prefecture, and in the Karatsu coalfield in Saga prefecture.
\=Oita prefecture has no coal mines but appears in the figures to indicate the sample boundary.
All these prefectures are in the Ky\=ush\=u region, the southernmost of Japan's four main islands (Figure~\ref{fig:map_coal_mines}).
The treatment (control) group, highlighted in red (pink), includes municipalities within 5 km (5-30 km) of a mine.
Excluded municipalities are shown as empty lattices in the figures.
Sources: Created by the author using the NDFM \citep[][]{cs1924, cs1932, cs1937}.
\end{minipage}}
\end{figure}

The distribution of coal mines is relatively stable over time, as examined in Figure~\ref{fig:map_coal_mines}.
To see a specific example, Figure~\ref{fig:map_coal_mines_Kyushu} provides a map for the most representative coalfield surrounding Chikuh\=o Coalfield in Ky\=ush\=u, Japan's southernmost island.
One can observe that the distribution of mines has changed little, with only a few mines opening or closing between December 1922 and October 1936.
This feature makes it infeasible to use within-variation in the spatiotemporal distribution of the mines for identification under the difference-in-differences setting.\footnote{Online Appendix~\ref{sec:app_did} discusses this point in detail. It also presents results from a difference-in-differences setting for a small subsample that includes the Iwate and Miyoshimuen coal mines in the T\=ohoku region. The results are consistent with the main findings presented in this section.}
I therefore rely on cross-sectional variation in mine locations.
Because mine locations may be correlated with unobserved local characteristics, I use the spatial distribution of Paleogene coal-bearing strata as an instrumental variable for the locations of coal mines (Section~\ref{sec:sec_ea_es}).

%----------
%Figure 3
%----------

The generation of the analytical samples is as follows.
From the $11,142$ municipalities, those within a $30$ km radius of the centroid of municipalities with coal mines are first retained.
The intersection of the municipalities within $5$ km from a gold, silver, and copper (GSC) mine is then excluded.
Overlap among the different types of mines was rare, and the number of treated municipalities excluded from this trimming step was small (fewer than 30).
Online Appendix~\ref{sec:secc1} provides finer details of the trimming and overlaps.
The number of municipalities in the analytical sample based on the 1922, 1931, and 1936 NDFM is $1,402$, $1,142$, and $1,316$, respectively.
Meanwhile, the proportions of treated municipalities are stable at $14$\%, $15$\%, and $13$\%, respectively, across the samples.\footnote{The slight reduction in the share of the treatment group in 1935 is due to the increase in the number of enclaves (Figure~\ref{fig:map_coal_mines_Kyushu}). A mine opening in the enclave area increases the number of controlled (i.e., surrounding) municipalities, thereby decreasing the relative share of the treatment group. Examples are the Iwate and Miyoshimuen coal mines in T\=ohoku region (Online Appendix~\ref{sec:app_did}).}

The NDFM (December 1922, October 1931, and October 1936) is matched to the Population Censuses of 1925, 1930, and 1935, respectively.
The gaps between the two sources may cause some municipalities to be misclassified if mines opened or closed in the interim.
However, such misclassification is likely to be limited, because the distribution of mining municipalities was stable over time, particularly in the major coalfields (Figure~\ref{fig:map_coal_mines_Kyushu}).

\subsection{Estimation Strategy}\label{sec:sec_ea_es}

I consider a linear regression model as follows:
\begin{eqnarray}\label{eq1}
\footnotesize{
\begin{split}
y_{i} = \alpha + \beta \text{\textit{MineDeposit}}_{i} + \mathbf{x}_{i}' \boldsymbol{\varphi} + e_{i},
\end{split}
}
\end{eqnarray}
where $i$ indexes municipalities, $y$ is the outcome variable, $\text{\textit{MineDeposit}}$ is an indicator variable that equals one for municipalities within 5 km from a mine, $\mathbf{x}$ is a vector of control variables, and $e$ is a random error term.
Although the locations of coal deposits are determined by geology, whether and where a deposit was mined depended on the decisions of mining firms.
The least-squares estimate of $\beta$ may therefore be biased if mine locations are correlated with unobserved local characteristics that also affect the outcomes.

The direction of this bias is ambiguous.
On the one hand, among municipalities near a deposit, mining firms may have chosen those with better infrastructure as their main sites.
If a deposit is found between a village and a city, for example, the firm may choose the city because it is likely to have better infrastructure.
In this case, $\text{\textit{MineDeposit}}$ is positively correlated with the error term in the population equation, leading to an upward bias.\footnote{As an example, consider a simplified projection of $y$ on $\textit{MineDeposit}$ and $\textit{City}$, $y = \alpha + \beta \text{\textit{MineDeposit}} + \varphi \textit{City} + e$. When the municipality type ($\text{\textit{City}}$) is omitted, the linear projection coefficient can be written as $\beta^{*}=\beta + \Xi \varphi$, where $\Xi = (E[\text{\textit{MineDeposit}}^{2}]^{-1})(E[\text{\textit{MineDeposit}}\cdot\text{\textit{City}}])$. If $\text{\textit{MineDeposit}}$ and $\text{\textit{City}}$ are positively correlated ($\Xi > 0$) and cities have more people than towns and villages ($\varphi > 0$), then $\beta^{*}=\beta + \Xi \varphi > \beta$.\label{foot:obv}}
On the other hand, coal deposits were often located in hilly areas with limited farmland and poor access to markets, which would have had smaller populations in the absence of mining.
In this case, $\text{\textit{MineDeposit}}$ is negatively correlated with the error term, leading to a downward bias.

I address these concerns in two steps.
First, to address the choice of mining sites with better infrastructure, I include two indicator variables, each equal to one for cities and towns, respectively.
Second, to address the location of deposits in areas with poor geographic conditions, I include the average elevation, which captures hilly terrain, and the distances to the nearest railway station, seaport, and river, which capture access to markets (Section~\ref{sec:sec_data_cv}).\footnote{The differences between the estimates with and without these variables indicate the extent to which observable local characteristics are correlated with mine locations. As shown below, the estimates are materially similar across specifications once the city and town indicators are included. To be conservative, the specification including all control variables is preferred.}
However, these variables may not fully capture local characteristics correlated with mine locations, such as the quality of farmland.
I therefore estimate equation~(\ref{eq1}) by two-stage least squares, using the distribution of geological strata as an instrumental variable for $\text{\textit{MineDeposit}}$.
The first-stage equation is as follows:
\begin{eqnarray}\label{eq2}
\footnotesize{
\begin{split}
\textit{MineDeposit}_{i} = \iota + \zeta \textit{Stratum}_{i} + \mathbf{x}'_{i}\boldsymbol{\xi} + \eps_{i},
\end{split}
}
\end{eqnarray}
where $\textit{Stratum}$ is a binary variable that equals one for municipalities with rock formed during the Paleogene period of the Cenozoic era, and $\eps$ is a random error term.

The relevance condition holds because, unlike the Carboniferous coal measures of Europe \citep{Fernihough2020}, most of Japan's major coalfields, including Chikuh\=o, Miike, Ishikari, and J\=oban, lie in coal-bearing formations of the Paleogene period \citep{Nagao1975, tanaka1984}.
The exclusion restriction requires that, conditional on the control variables, the geological stratum affects the outcomes only through the presence of coal mines.
This is plausible because the stratum was formed tens of millions of years ago and is unlikely to be related to the local characteristics that determined population and fertility in the interwar period, other than through coal mining.
A potential concern is that the stratum may be correlated with terrain, which could affect agricultural productivity and population directly.
The controls for elevation and accessibility to rivers and transportation are included to address this concern.\footnote{Online Appendix~\ref{sec:secc2} summarizes the geological strata variable in finer detail and shows that the location of coal mines is determined by specific strata created in this era. Online Appendix~\ref{sec:app_as} discusses the other identification assumptions.}

The IV estimate should be interpreted as a local average treatment effect \citep{Imbens1994, Angrist1996}.
It represents the effect of coal mining in municipalities where mines were located because of the Paleogene coal-bearing formations, which include the major coalfields of Japan.
The heteroskedasticity-consistent covariance matrix estimator is used as a baseline estimator.\footnote{The results are materially similar between the Eicker-White type covariance matrix estimator suggested by \citet{Hinkley:1977} and the HC2 estimator proposed by \citet{Horn:1975tu}.}
To assess the potential influence of local-scale spatial correlations, I also report standard errors clustered at the county level using the cluster-robust covariance matrix estimator \citep{Arellano1987} in Online Appendix~\ref{sec:app_crve}.
The results are materially similar across the two variance estimators.

\subsection{Results}\label{sec:sec_ea_r}

%------
%Table 2
%\begin{landscape}
\begin{table}[htbp]
\def\arraystretch{1.0}
\begin{center}
\captionsetup{justification=centering}
\caption{Coal Mining on Regional Development: \\Local Population Growth in the Mining Area}
\label{tab:r_pop}
\scriptsize
\scalebox{1.0}[1]{
{\setlength\doublerulesep{2pt}
\begin{tabular}{lD{.}{.}{-2}D{.}{.}{-2}D{.}{.}{-2}D{.}{.}{-2}}
\toprule[1pt]\midrule[0.3pt]
\multicolumn{5}{l}{\textbf{Panel A: 1925 Census}}\\
&\multicolumn{4}{c}{Dependent Variable: ln(Population)}\\
\cmidrule(rrrr){2-5}
&\multicolumn{1}{c}{(1)}&\multicolumn{1}{c}{(2)}&\multicolumn{1}{c}{(3)}&\multicolumn{1}{c}{(4)}\\\hline
\textit{MineDeposit}
&1.335$***$	&0.775$***$	&0.583$***$	&0.342$***$    \\
&(0.297)		&(0.198)		&(0.202)		&(0.051)        \\
City and town indicators
&\multicolumn{1}{c}{No}&\multicolumn{1}{c}{Yes}&\multicolumn{1}{c}{Yes}&\multicolumn{1}{c}{Yes}\\
Railway accessibility
&\multicolumn{1}{c}{No}&\multicolumn{1}{c}{No}&\multicolumn{1}{c}{Yes}&\multicolumn{1}{c}{Yes}\\
Port accessibility
&\multicolumn{1}{c}{No}&\multicolumn{1}{c}{No}&\multicolumn{1}{c}{Yes}&\multicolumn{1}{c}{Yes}\\
River accessibility
&\multicolumn{1}{c}{No}&\multicolumn{1}{c}{No}&\multicolumn{1}{c}{Yes}&\multicolumn{1}{c}{Yes}\\
Elevation
&\multicolumn{1}{c}{No}&\multicolumn{1}{c}{No}&\multicolumn{1}{c}{Yes}&\multicolumn{1}{c}{Yes}\\
Observations
&\multicolumn{1}{c}{1,402}&\multicolumn{1}{c}{1,402}&\multicolumn{1}{c}{1,402}&\multicolumn{1}{c}{1,402}\\
Estimator
&\multicolumn{1}{c}{IV/Wald}&\multicolumn{1}{c}{IV}&\multicolumn{1}{c}{IV}&\multicolumn{1}{c}{OLS}\\
First-stage $F$-statistic	&38.658&35.301&33.128&$--$\\
Mean of the DV			&8.230&8.230&8.230&8.230\\
Std. Dev of the DV		&0.705&0.705&0.705&0.705\\\hdashline[0.5pt/3pt]
Magnitude	 in original-scale&&&\multicolumn{1}{c}{79\%}&\\
\hspace{10pt}Geometric mean for the TG/CG	&&&\multicolumn{1}{c}{$7,044/3,932$}&\\\hline
&&&&\\
\multicolumn{5}{l}{\textbf{Panel B: 1930 Census}}\\
&\multicolumn{4}{c}{Dependent Variable: ln(Population)}\\
\cmidrule(rrrr){2-5}
&\multicolumn{1}{c}{(1)}&\multicolumn{1}{c}{(2)}&\multicolumn{1}{c}{(3)}&\multicolumn{1}{c}{(4)}\\\hline
\textit{MineDeposit}
&1.384$***$    &0.824$***$    &0.673$***$    &0.460$***$    \\
&(0.290)        &(0.196)        &(0.195)        &(0.058)        \\
City and town indicators
&\multicolumn{1}{c}{No}&\multicolumn{1}{c}{Yes}&\multicolumn{1}{c}{Yes}&\multicolumn{1}{c}{Yes}\\
Railway accessibility
&\multicolumn{1}{c}{No}&\multicolumn{1}{c}{No}&\multicolumn{1}{c}{Yes}&\multicolumn{1}{c}{Yes}\\
Port accessibility
&\multicolumn{1}{c}{No}&\multicolumn{1}{c}{No}&\multicolumn{1}{c}{Yes}&\multicolumn{1}{c}{Yes}\\
River accessibility
&\multicolumn{1}{c}{No}&\multicolumn{1}{c}{No}&\multicolumn{1}{c}{Yes}&\multicolumn{1}{c}{Yes}\\
Elevation
&\multicolumn{1}{c}{No}&\multicolumn{1}{c}{No}&\multicolumn{1}{c}{Yes}&\multicolumn{1}{c}{Yes}\\
Observations
&\multicolumn{1}{c}{1,142}&\multicolumn{1}{c}{1,142}&\multicolumn{1}{c}{1,142}&\multicolumn{1}{c}{1,142}\\
Estimator
&\multicolumn{1}{c}{IV/Wald}&\multicolumn{1}{c}{IV}&\multicolumn{1}{c}{IV}&\multicolumn{1}{c}{OLS}\\
First-stage $F$-statistic    &43.733&38.099&36.692&$--$\\
Mean of the DV			&8.288&8.288&8.288&8.288\\
Std. Dev of the DV		&0.740&0.740&0.740&0.740\\\hdashline[0.5pt/3pt]
Magnitude	 in original-scale&&&\multicolumn{1}{c}{96\%}&\\
\hspace{10pt}Geometric mean for the TG/CG	&&&\multicolumn{1}{c}{$7,180/3,663$}&\\\hline
&&&&\\
\multicolumn{5}{l}{\textbf{Panel C: 1935 Census}}\\
&\multicolumn{4}{c}{Dependent Variable: ln(Population)}\\
\cmidrule(rrrr){2-5}
&\multicolumn{1}{c}{(1)}&\multicolumn{1}{c}{(2)}&\multicolumn{1}{c}{(3)}&\multicolumn{1}{c}{(4)}\\\hline
\textit{MineDeposit}
&1.484$***$    &0.923$***$    &0.758$***$    &0.495$***$    \\
&(0.295)        &(0.202)        &(0.195)        &(0.057)        \\\hline
City and town indicators
&\multicolumn{1}{c}{No}&\multicolumn{1}{c}{Yes}&\multicolumn{1}{c}{Yes}&\multicolumn{1}{c}{Yes}\\
Railway accessibility
&\multicolumn{1}{c}{No}&\multicolumn{1}{c}{No}&\multicolumn{1}{c}{Yes}&\multicolumn{1}{c}{Yes}\\
Port accessibility&\multicolumn{1}{c}{No}&\multicolumn{1}{c}{No}&\multicolumn{1}{c}{Yes}&\multicolumn{1}{c}{Yes}\\
River accessibility&\multicolumn{1}{c}{No}&\multicolumn{1}{c}{No}&\multicolumn{1}{c}{Yes}&\multicolumn{1}{c}{Yes}\\
Elevation
&\multicolumn{1}{c}{No}&\multicolumn{1}{c}{No}&\multicolumn{1}{c}{Yes}&\multicolumn{1}{c}{Yes}\\
Observations
&\multicolumn{1}{c}{1,316}&\multicolumn{1}{c}{1,316}&\multicolumn{1}{c}{1,316}&\multicolumn{1}{c}{1,316}\\
Estimator
&\multicolumn{1}{c}{IV/Wald}&\multicolumn{1}{c}{IV}&\multicolumn{1}{c}{IV}&\multicolumn{1}{c}{OLS}\\
First-stage $F$-statistic		&44.573	&39.980	&37.999	&$--$\\
Mean of the DV		&8.266	&8.266	&8.266	&8.266\\
Std. Dev of the DV			&0.739	&0.739	&0.739	&0.739\\\hdashline[0.5pt/3pt]
Magnitude	 in original-scale	&		&		&\multicolumn{1}{c}{113\%}&\\
\hspace{10pt}Geometric mean for the TG/CG	&&&\multicolumn{1}{c}{$7,998/3,748$}&\\\midrule[0.3pt]\bottomrule[1pt]
\end{tabular}
}
}
{\scriptsize
\begin{minipage}{405pt}
\setstretch{0.85}
***, **, and * represent statistical significance at the 1\%, 5\%, and 10\% levels, respectively.
Standard errors based on the heteroskedasticity-robust covariance matrix estimator are reported in parentheses.\\
Notes:
Panels A, B, and C present the results for the 1925, 1930, and 1935 cross-sectional census samples, respectively.
Columns (1)--(3) present the estimates from the IV estimator for the system of Equations~\ref{eq1} and~\ref{eq2}.
Column (4) shows the estimates from the OLS estimator for Equation~\ref{eq1}.
Geometric mean in the TG and CG show the $\text{exp}(\hat{\alpha} + \hat{\beta})$ and $\text{exp}(\hat{\alpha})$ from the estimated system, respectively.
Panel A of Online Appendix Table~\ref{tab:sum_outcome} shows the summary statistics for the population (before log-transformation) by treatment and control groups.
\end{minipage}
}
\end{center}
\end{table}
%\end{landscape}
%------

Panel A of Table~\ref{tab:r_pop} presents the main results for the population in 1925.
The estimate from a simple IV estimation is shown in Column (1). \footnote{The first-stage $F$-statistics reported in the same panel exceed the rule-of-thumb threshold of 10 proposed by \citet{Staiger1997} in all regressions, indicating that the instrument is not weak.}
Then I consider the city and town indicators (Column (2)), and both the indicators and control variables (Column (3)).
The estimated coefficient decreases from $1.335$ to $0.775$ when the city and town indicators are included.
This suggests that municipalities on the Paleogene strata differed from other municipalities in municipal type, and that conditioning on municipal type is important for the exclusion restriction.
Including the additional controls for transportation accessibility and terrain has a more moderate influence on the estimate ($0.583$), suggesting that, conditional on municipal type, the instrument is not strongly correlated with these observable characteristics.
Panels B and C confirm similar results for the 1930 and 1935 populations, using the same column layouts.
The estimated coefficient from the specification including the full set of controls in Column (3) is $0.673$ for the 1930 sample and $0.758$ for the 1935 sample.

The estimated magnitude is economically meaningful.
The estimates from the preferred specification of Column (3) indicate that coal mines increased the local population by $79$\% in 1925 (Panel A), $96$\% in 1930 (Panel B), and $113$\%\footnote{Statistically, it is more precise to state that the geometric mean of population in the treatment group with mines was approximately 113\% (i.e., $\text{exp}(0.758) = \text{exp}(\hat{\alpha}+0.758)/\text{exp}(\hat{\alpha}) \approx 2.13$) greater than that of the control group. However, simple interpretations are used throughout this study to avoid redundancy.} in 1935 (Panel C), respectively.
The estimated magnitude is similar across the three years, although the average number of miners per coal mine declined during the sample period.\footnote{Table~\ref{tab:nminers} shows that the total number of miners had decreased in the 1920s and 1930s. In the coal mines in my analytical sample, the average number of miners per coal mine was $1,091$ in December 1922, $865$ in October 1931, and $824$ in October 1936. The average number of mines per mining municipality is stable, roughly two in all three samples.}
The following subsections examine the channels through which coal mining increased local populations.

Column (4) in each panel presents the OLS estimates from the specification with the full set of controls.
The OLS estimates are smaller than the IV estimates in Column (3).
Two explanations are consistent with this pattern.
First, as discussed in Section~\ref{sec:sec_ea_es}, mines may have been located in areas that would otherwise have had smaller populations, such as hilly areas with limited farmland, which biases the OLS estimates downward.
Second, the IV estimates reflect the effects in municipalities where mining was induced by the Paleogene formations, that is, the major coalfields with large mines, whereas the OLS estimates also include smaller mines in other formations.
Note that the IV estimates also decline when the city and town indicators are included (Columns 1--2), suggesting that the stratum is correlated with municipal type; the exclusion restriction is therefore more plausible conditional on these indicators, as in the preferred specification.

\subsubsection*{The Natural Growth Channel}

Table~\ref{tab:r_mech} summarizes the estimates from the IV estimation for the demographic variables to assess potential channels behind the local population growth.
All regressions include city and town indicators and the control variables as in my preferred specification of Column (3) in Table~\ref{tab:r_pop}.

I first examine the marriage and fertility rates.
The estimates for 1925 and 1930 in Panels A and B of Table~\ref{tab:r_mech} are similar.
The coefficient on the crude marriage rate is negative and statistically significant (Column 1), consistent with the large number of single male miners in mining areas.
The coefficient on the crude birth rate is also negative and statistically significant (Column 2).
Because a lower crude birth rate could simply reflect a smaller share of married women in the population, Column (3) uses the marital fertility rate, which accounts for the number of married women.
The estimate remains negative and statistically significant, and its magnitude ($-36.567$ in 1930) exceeds one standard deviation of marital fertility ($29.669$).
Thus, fertility in mining areas was lower than in surrounding areas even among married couples, consistent with the first prediction from the model in Section~\ref{sec:sec_cf}.

The results for 1935 in Panel C show a different pattern.
The coefficient on the crude marriage rate is smaller in magnitude than in 1930 ($-1.656$ versus $-2.709$), and the coefficients on the fertility rates are smaller and no longer statistically significant; for example, the coefficient on marital fertility is $-9.745$, compared with $-36.567$ in 1930.
These results suggest that the marriage and fertility gaps between mining and surrounding areas narrowed after the full enforcement of the regulation, consistent with the second prediction from the model.

I next examine the crude death rates.
In Panels A and B, the coefficients for both males and females are statistically insignificant (Columns 4 and 5), indicating that overall mortality in mining areas was similar to that in surrounding areas in 1925 and 1930.\footnote{The results are unchanged if log-transformed rates are used instead of the raw rates (not reported).}
In Panel C, the coefficient on the female death rate is negative and weakly significant ($-2.001$), roughly half a standard deviation.
However, the coefficient in 1925 is of a similar magnitude ($-1.999$), although statistically insignificant, so this result is unlikely to reflect the enforcement of the regulation.
Crude death rates may also fail to capture the effects of working conditions on pregnant women and their children, which I examine in Section~\ref{sec:sec_aa}.

%------------------
%Table 3
\begin{landscape}
\begin{table}[htbp]
\def\arraystretch{1.0}
\begin{center}
\captionsetup{justification=centering}
\caption{Assessing Mechanisms: The Natural and Social Growth Channels}
\label{tab:r_mech}
\scriptsize

\scalebox{1.0}[1]{
{\setlength\doublerulesep{2pt}
\begin{tabular}{lD{.}{.}{-2}D{.}{.}{-2}D{.}{.}{-2}D{.}{.}{-2}D{.}{.}{-2}D{.}{.}{-2}D{.}{.}{-2}}
\toprule[1pt]\midrule[0.3pt]
\textbf{Panel A: 1925 Vital Statistics}&\multicolumn{7}{c}{Dependent Variable}\\
\cmidrule(rrrrrr){2-8}
&\multicolumn{3}{c}{Fertility Channel}
&\multicolumn{2}{c}{Mortality Channel}
&\multicolumn{2}{c}{Migration Channel}\\
\cmidrule(rrr){2-4}\cmidrule(rr){5-6}\cmidrule(rr){7-8}
&\multicolumn{1}{c}{(1) Marriage}
&\multicolumn{1}{c}{(2) Crude}
&\multicolumn{1}{c}{(3) Marital}
&\multicolumn{1}{c}{(4) Male}
&\multicolumn{1}{c}{(5) Female}
&\multicolumn{1}{c}{(6) Sex Ratio}
&\multicolumn{1}{c}{(7) Size}\\
\hline
$\textit{MineDeposit}$
&-2.676$***$	&-7.183$***$    &-41.107$***$	&-0.631    &-1.999   &0.112$***$	&-1.112$***$    \\
&(0.826)		&(2.230)		&(11.835)        	&(1.674)	&(1.470)  &(0.043)		&(0.216)  \\\hline
City and town indicators
&\multicolumn{1}{c}{Yes}&\multicolumn{1}{c}{Yes}&\multicolumn{1}{c}{Yes}&\multicolumn{1}{c}{Yes}&\multicolumn{1}{c}{Yes}&\multicolumn{1}{c}{Yes}&\multicolumn{1}{c}{Yes}\\
Control variables
&\multicolumn{1}{c}{Yes}&\multicolumn{1}{c}{Yes}&\multicolumn{1}{c}{Yes}&\multicolumn{1}{c}{Yes}&\multicolumn{1}{c}{Yes}&\multicolumn{1}{c}{Yes}&\multicolumn{1}{c}{Yes}\\
Observations
&\multicolumn{1}{c}{1,402}&\multicolumn{1}{c}{1,402}&\multicolumn{1}{c}{1,402}&\multicolumn{1}{c}{1,402}&\multicolumn{1}{c}{1,402}&\multicolumn{1}{c}{1,402}&\multicolumn{1}{c}{1,402}\\
Estimator
&\multicolumn{1}{c}{IV}&\multicolumn{1}{c}{IV}&\multicolumn{1}{c}{IV}&\multicolumn{1}{c}{IV}&\multicolumn{1}{c}{IV}&\multicolumn{1}{c}{IV}&\multicolumn{1}{c}{IV}\\
First-stage $F$-statistic
&\multicolumn{1}{c}{33.129}&\multicolumn{1}{c}{33.129}&\multicolumn{1}{c}{33.129}
&\multicolumn{1}{c}{33.129}&\multicolumn{1}{c}{33.129}&\multicolumn{1}{c}{33.129}&\multicolumn{1}{c}{33.129}\\
Sample Mean of the DV
&\multicolumn{1}{c}{9.577}&\multicolumn{1}{c}{37.005}&\multicolumn{1}{c}{183.956}
&\multicolumn{1}{c}{20.682}&\multicolumn{1}{c}{19.235}&\multicolumn{1}{c}{0.987}&\multicolumn{1}{c}{5.307}\\
Std. Dev of the DV
&\multicolumn{1}{c}{2.407}&\multicolumn{1}{c}{6.074}&\multicolumn{1}{c}{32.173}
&\multicolumn{1}{c}{4.492}&\multicolumn{1}{c}{4.356}&\multicolumn{1}{c}{0.082}&\multicolumn{1}{c}{0.611}\\
&&&&&&&\\
\textbf{Panel B: 1930 Vital Statistics}&\multicolumn{7}{c}{Dependent Variable}\\
\cmidrule(rrrrrrr){2-8}
&\multicolumn{3}{c}{Fertility Channel}
&\multicolumn{2}{c}{Mortality Channel}
&\multicolumn{2}{c}{Migration Channel}\\
\cmidrule(rrr){2-4}\cmidrule(rr){5-6}\cmidrule(rr){7-8}
&\multicolumn{1}{c}{(1) Marriage}
&\multicolumn{1}{c}{(2) Crude}
&\multicolumn{1}{c}{(3) Marital}
&\multicolumn{1}{c}{(4) Male}
&\multicolumn{1}{c}{(5) Female}
&\multicolumn{1}{c}{(6) Sex Ratio}
&\multicolumn{1}{c}{(7) Size}\\
\hline
$\textit{MineDeposit}$
&-2.709$***$	&-6.757$***$    &-36.567$***$    &0.701    &-1.540   &0.114$***$	&-0.533$***$\\
&(0.787)		&(1.926)		&(10.250)        	&(1.436)	&(1.377)  &(0.034)		&(0.140)  \\\hline
City and town indicators
&\multicolumn{1}{c}{Yes}&\multicolumn{1}{c}{Yes}&\multicolumn{1}{c}{Yes}&\multicolumn{1}{c}{Yes}&\multicolumn{1}{c}{Yes}&\multicolumn{1}{c}{Yes}&\multicolumn{1}{c}{Yes}\\
Control variables
&\multicolumn{1}{c}{Yes}&\multicolumn{1}{c}{Yes}&\multicolumn{1}{c}{Yes}&\multicolumn{1}{c}{Yes}&\multicolumn{1}{c}{Yes}&\multicolumn{1}{c}{Yes}&\multicolumn{1}{c}{Yes}\\
Observations
&\multicolumn{1}{c}{1,142}&\multicolumn{1}{c}{1,142}&\multicolumn{1}{c}{1,142}&\multicolumn{1}{c}{1,142}&\multicolumn{1}{c}{1,142}&\multicolumn{1}{c}{1,142}&\multicolumn{1}{c}{1,142}\\
Estimator
&\multicolumn{1}{c}{IV}&\multicolumn{1}{c}{IV}&\multicolumn{1}{c}{IV}&\multicolumn{1}{c}{IV}&\multicolumn{1}{c}{IV}&\multicolumn{1}{c}{IV}&\multicolumn{1}{c}{IV}\\
First-stage $F$-statistic
&\multicolumn{1}{c}{36.693}&\multicolumn{1}{c}{36.693}&\multicolumn{1}{c}{36.693}
&\multicolumn{1}{c}{36.693}&\multicolumn{1}{c}{36.693}&\multicolumn{1}{c}{36.693}&\multicolumn{1}{c}{36.693}\\
Sample Mean of the DV
&\multicolumn{1}{c}{8.588}&\multicolumn{1}{c}{34.069}&\multicolumn{1}{c}{174.162}
&\multicolumn{1}{c}{20.061}&\multicolumn{1}{c}{18.596}&\multicolumn{1}{c}{0.987}&\multicolumn{1}{c}{5.340}\\
Std. Dev of the DV
&\multicolumn{1}{c}{2.278}&\multicolumn{1}{c}{5.516}&\multicolumn{1}{c}{29.669}
&\multicolumn{1}{c}{4.476}&\multicolumn{1}{c}{4.324}&\multicolumn{1}{c}{0.077}
&\multicolumn{1}{c}{0.470}\\
&&&&&&\\
\textbf{Panel C: 1935 Vital Statistics}&\multicolumn{7}{c}{Dependent Variable}\\
\cmidrule(rrrrrrr){2-8}
&\multicolumn{3}{c}{Fertility Channel}
&\multicolumn{2}{c}{Mortality Channel}
&\multicolumn{2}{c}{Migration Channel}\\
\cmidrule(rrr){2-4}\cmidrule(rr){5-6}\cmidrule(rr){7-8}
&\multicolumn{1}{c}{(1) Marriage}
&\multicolumn{1}{c}{(2) Crude}
&\multicolumn{1}{c}{(3) Marital}
&\multicolumn{1}{c}{(4) Male}
&\multicolumn{1}{c}{(5) Female}
&\multicolumn{1}{c}{(6) Sex Ratio}
&\multicolumn{1}{c}{(7) Size}\\
\hline
$\textit{MineDeposit}$
&-1.656$**$	&-1.552	&-9.745	&-0.648	&-2.001$*$	&0.068$***$	&-0.527$***$	\\
&(0.693)		&(1.612)	&(8.712)	&(1.254)	&(1.167)		&(0.025)		&(0.148)\\\hline
City and town indicators
&\multicolumn{1}{c}{Yes}&\multicolumn{1}{c}{Yes}&\multicolumn{1}{c}{Yes}&\multicolumn{1}{c}{Yes}&\multicolumn{1}{c}{Yes}&\multicolumn{1}{c}{Yes}&\multicolumn{1}{c}{Yes}\\
Control variables
&\multicolumn{1}{c}{Yes}&\multicolumn{1}{c}{Yes}&\multicolumn{1}{c}{Yes}&\multicolumn{1}{c}{Yes}&\multicolumn{1}{c}{Yes}&\multicolumn{1}{c}{Yes}&\multicolumn{1}{c}{Yes}\\
Observations
&\multicolumn{1}{c}{1,316}&\multicolumn{1}{c}{1,316}&\multicolumn{1}{c}{1,316}&\multicolumn{1}{c}{1,316}&\multicolumn{1}{c}{1,316}&\multicolumn{1}{c}{1,316}&\multicolumn{1}{c}{1,316}\\
Estimator
&\multicolumn{1}{c}{IV}&\multicolumn{1}{c}{IV}&\multicolumn{1}{c}{IV}&\multicolumn{1}{c}{IV}&\multicolumn{1}{c}{IV}&\multicolumn{1}{c}{IV}&\multicolumn{1}{c}{IV}\\
First-stage $F$-statistic
&\multicolumn{1}{c}{40.000}&\multicolumn{1}{c}{40.000}&\multicolumn{1}{c}{40.000}
&\multicolumn{1}{c}{40.000}&\multicolumn{1}{c}{40.000}&\multicolumn{1}{c}{40.000}&\multicolumn{1}{c}{40.000}\\
Sample Mean of the DV
&\multicolumn{1}{c}{9.060}&\multicolumn{1}{c}{34.186}&\multicolumn{1}{c}{177.264}
&\multicolumn{1}{c}{18.390}&\multicolumn{1}{c}{16.958}&\multicolumn{1}{c}{0.985}&\multicolumn{1}{c}{5.330}\\
Std. Dev of the DV
&\multicolumn{1}{c}{2.414}&\multicolumn{1}{c}{5.472}&\multicolumn{1}{c}{29.754}
&\multicolumn{1}{c}{4.133}&\multicolumn{1}{c}{4.210}&\multicolumn{1}{c}{0.073}&\multicolumn{1}{c}{0.566}\\\midrule[0.3pt]\bottomrule[1pt]
\end{tabular}
}
}
{\scriptsize
\begin{minipage}{570pt}
\setstretch{0.85}
***, **, and * represent statistical significance at the 1\%, 5\%, and 10\% levels, respectively.
Standard errors based on the heteroskedasticity-robust covariance matrix estimator are reported in parentheses.\\
Notes:
This table summarizes the results for the demographic outcomes using the IV estimator for the system of equations~\ref{eq1} and~\ref{eq2}.
Panels A, B, and C present the results for the 1925, 1930, and 1935 vital statistics samples, respectively.
Column (1) presents the result for the crude marriage rate.
Columns (2) and (3) show the results for the crude birth rate and marital fertility rate, respectively.
Columns (4) and (5) present the results for the crude death rates among men and women.
Columns (6) and (7) present the results for the sex ratio and average household size.
All regressions include the control variables: railway, port, and river accessibility, and elevation.
\end{minipage}
}
\end{center}
\end{table}
\end{landscape}
%------------------

\subsubsection*{The Migration Channel}

The fertility and mortality results suggest that natural increase cannot explain the larger populations of mining municipalities before the regulation.
Because population change is the sum of natural change and net migration (Section~\ref{sec:sec_cf}), these results imply that in-migration was the main source of population growth in mining areas.
To examine this interpretation, Columns (6) and (7) of Table~\ref{tab:r_mech} present the results for the sex ratio (males per female) and the average household size.
In Panels A and B, the estimate for the sex ratio is positive and that for household size is negative, both statistically significant and economically meaningful.
These results indicate that the inflow of workers, particularly single men, made the population of mining areas male-biased and reduced household size.\footnote{This tendency is in line with the higher sex ratio in urban areas in prewar Japan, which resulted from the internal migration of male workers \citep[pp.~243--247]{Ito1990}.}
In Panel C, the estimate for the sex ratio is smaller ($0.068$ in 1935 versus $0.114$ in 1930), whereas that for household size is similar.
This pattern is consistent with the regulation: as the fertility gap narrowed after the full enforcement, births, for which the sex ratio is balanced, may have reduced the male bias in the population of mining areas.

\subsubsection*{Summary}

The cross-sectional results show that coal mining increased local populations mainly through the inflow of workers, whereas marriage and fertility rates in mining areas were lower than in surrounding areas before the regulation.
In 1935, the fertility gap was smaller and no longer statistically significant.
Because the cross-sectional samples differ across years, Section~\ref{sec:sec_pe} uses a balanced panel of municipalities to examine how these outcomes changed after the enforcement of the regulation.

%----------------------------------------
\section{Evaluating the Effects of Labor Regulation Enforcement on Mining Municipalities}\label{sec:sec_pe}

This section examines how the enforcement of the revised labor regulation changed population, marriage, and fertility in coal-mining municipalities relative to surrounding municipalities.
I first describe the construction of the panel sample and the estimation strategy, and then present the results.

\subsection{Analytical Sample}\label{sec:sec_pe_as}

To evaluate the effects of the regulation, I compare outcomes in coal-mining and surrounding municipalities before and after its enforcement.
I construct a balanced panel of municipalities for 1925, 1930, and 1935 from the cross-sectional datasets described in Section~\ref{sec:sec_ea_as}.
The treatment group consists of the municipalities within $5$ km of the municipalities with coal mines that are listed in all three NDFM (December 1922, October 1931, and October 1936).
The control group consists of the municipalities located $5$--$30$ km from these mines.
I exclude the municipalities within $5$ km of the coal mines that appear only in the 1931 and/or 1936 NDFM, as well as those within $5$ km of the GSC mines, so that the treatment status of each municipality is constant over the sample period.
The final sample consists of $1,039$ municipalities observed in all three years, of which $142$ are in the treatment group, and $897$ are in the control group.
Online Appendix Table~\ref{tab:dup} summarizes the trimming steps.

\subsection{Estimation Strategy}\label{sec:sec_pe_es}

I estimate the following difference-in-differences equation for municipality $i$ in year $t$:
\begin{eqnarray}\label{pe_did1}
\footnotesize{
\begin{split}
y_{i,t} = \varpi \text{\textit{MineDeposit}}_{i} \times \text{\textit{Post}}_{t} + \mathbf{x}_{i,t}' \boldsymbol{\varTheta} + \varsigma_{i} + o_{t} + \backepsilon_{i,t},
\end{split}
}
\end{eqnarray}
where $y_{i,t}$ is the outcome variable, $\text{\textit{MineDeposit}}_{i}$ is the time-invariant indicator for the treatment group, $\text{\textit{Post}}_{t}$ is an indicator for 1935, $\varsigma_{i}$ and $o_{t}$ are municipal and year fixed effects, respectively, and $\backepsilon_{i,t}$ is an error term.
The vector $\mathbf{x}_{i,t}$ includes the distances to the nearest railway station and seaport, which are measured in each NDFM year.
The distance to the nearest river and the average elevation are absorbed by the municipal fixed effects.

The municipal fixed effects remove time-invariant differences between mining and surrounding municipalities, including those arising from the endogenous location of mines discussed in Section~\ref{sec:sec_ea_es}.
The remaining concern is that mining municipalities may have followed different trends from surrounding municipalities for reasons unrelated to the regulation.
To address this concern, I instrument $\text{\textit{MineDeposit}}_{i} \times \text{\textit{Post}}_{t}$ with $\text{\textit{Stratum}}_{i} \times \text{\textit{Post}}_{t}$.
The first-stage equation is as follows:
\begin{eqnarray}\label{pe_did2}
\footnotesize{
\begin{split}
\text{\textit{MineDeposit}}_{i} \times \text{\textit{Post}}_{t} = \varkappa \text{\textit{Stratum}}_{i} \times \text{\textit{Post}}_{t} + \mathbf{x}_{i,t}' \boldsymbol{\varPsi} + \varLambda_{i} + \varPhi_{t} + \eth_{i,t},
\end{split}
}
\end{eqnarray}
where $\text{\textit{Stratum}}_{i}$ is the stratum indicator introduced in Section~\ref{sec:sec_ea_es}, $\varLambda_{i}$ and $\varPhi_{t}$ are municipal and year fixed effects, respectively, and $\eth_{i,t}$ is an error term.
The identifying assumption is that, in the absence of the regulation, outcomes in municipalities on and off the Paleogene strata would have followed parallel trends, conditional on the control variables.
In other words, the stratum affects changes in the outcomes only through the presence of coal mines.

I also estimate an expanded specification that replaces $\text{\textit{MineDeposit}}_{i} \times \text{\textit{Post}}_{t}$ with interactions between $\text{\textit{MineDeposit}}_{i}$ and the 1925 and 1935 dummies, using 1930 as the reference year, to examine whether outcomes evolved differently before the enforcement.
Standard errors are clustered at the municipal level to account for serial correlation and heteroskedasticity \citep{Arellano1987}.
Because the instrument varies with geological formations that extend over wide areas, I also report standard errors clustered at the county level in Online Appendix~\ref{sec:app_crve}; the results are materially similar.

\subsection{Results}\label{sec:sec63}

Panel A of Table~\ref{tab:r_did_pe} summarizes the baseline results.
Column (1) shows that the population of mining municipalities increased by approximately $8.5$\% ($\exp(0.082) - 1$) relative to surrounding municipalities after the enforcement, although this estimate is only marginally significant.

Columns (2)--(4) present the results for marriage and fertility.
The estimated coefficient for the crude birth rate is positive and statistically significant (Column 3).
The result is similar for the marital fertility rate, which accounts for changes in the number of married women (Column 4).
The estimate of $18.357$ corresponds to $59$\% of the standard deviation of marital fertility in the treatment group during the pre-enforcement period.
These results suggest that the fertility gap between mining and surrounding municipalities documented in Section~\ref{sec:sec_ea_r} narrowed after the enforcement.
The estimate for the crude marriage rate is also positive and weakly significant (Column 2), indicating that the gap in marriage rates narrowed as well.
Because the result for marital fertility is similar to that for the crude birth rate, the increase in births is unlikely to be explained only by the increase in marriages.

By contrast, the estimates for the crude death rates of males and females are not statistically significant (Columns 5 and 6).
The estimates for the sex ratio and average household size are also close to zero (Columns 7 and 8).
Thus, there is no evidence that the enforcement changed overall mortality or the composition of the population in mining municipalities.
This is consistent with the argument in Section~\ref{sec:sec_cf} that the regulation was unlikely to reduce the migration of couples, because it did not reduce male miners' wages.

Panel B of Table~\ref{tab:r_did_pe} shows the results from the expanded system for testing the pre-enforcement trend assumption.
In all columns, the estimated coefficient for the interaction term with respect to the 1925 dummy is statistically insignificant.
This confirms that the pre-enforcement trends are practically similar in both the treatment and control groups.

\subsubsection*{Summary}

Fertility and marriage rates in mining municipalities increased relative to surrounding municipalities after the enforcement of the regulation, whereas mortality and the composition of the population did not change significantly.
These results are consistent with the second prediction of the model: as women's earning opportunities in the pits disappeared, the opportunity cost of childrearing declined, and the fertility gap between mining and surrounding municipalities narrowed (Section~\ref{sec:sec_cf}).\footnote{The labor force participation rate is available only in the 1930 Population Census. Online Appendix~\ref{sec:test_lfpr} shows that female labor force participation was lower in the mining area than in the surrounding agricultural area, with a shift in employment from agriculture to mining. Although participation data are unavailable for 1925 and 1935, the time-series statistics confirm that the number of female miners declined after the revision of the regulation (Section~\ref{sec:sec22}).}
However, the estimates capture the combined effects of the regulation and the mechanization that accompanied it, as well as other changes in mining communities.
In particular, I cannot rule out that firms' efforts to encourage miners to form families \citep{ogino1993} contributed to the increase in fertility.

The absence of changes in crude death rates suggests that the reduction in occupational hazards did not substantially change overall mortality in mining municipalities.
However, crude death rates may not capture the effects of working conditions on pregnant women and their children.
Section~\ref{sec:sec_aa} examines these effects using data on fetal and infant deaths.

%------------------
%Table 4
\begin{landscape}
\begin{table}[htbp]
\def\arraystretch{1.0}
\begin{center}
\captionsetup{justification=centering}
\caption{Estimated Effects of Regulatory Enforcement: \\Evidence from the Ever- and Never-Treated Panel Dataset}
\label{tab:r_did_pe}
\scriptsize
\scalebox{1.0}[1]{
{\setlength\doublerulesep{2pt}
\begin{tabular}{lD{.}{.}{-2}D{.}{.}{-2}D{.}{.}{-2}D{.}{.}{-2}D{.}{.}{-2}D{.}{.}{-2}D{.}{.}{-2}D{.}{.}{-2}}
\toprule[1pt]\midrule[0.3pt]
\multicolumn{9}{l}{\textbf{Panel A: The Effects of Regulatory Enforcement}}\\
&\multicolumn{8}{c}{Dependent Variable}\\
\cmidrule(rrrrrrrr){2-9}
&\multicolumn{1}{c}{}
&\multicolumn{3}{c}{Fertility Channel}
&\multicolumn{2}{c}{Mortality Channel}
&\multicolumn{2}{c}{Migration Channel}\\
\cmidrule(rrr){3-5}\cmidrule(rr){6-7}\cmidrule(rr){8-9}
&\multicolumn{1}{c}{(1) Population}
&\multicolumn{1}{c}{(2) Marriage}
&\multicolumn{1}{c}{(3) Crude}
&\multicolumn{1}{c}{(4) Marital}
&\multicolumn{1}{c}{(5) Male}
&\multicolumn{1}{c}{(6) Female}
&\multicolumn{1}{c}{(7) Sex Ratio}
&\multicolumn{1}{c}{(8) Size}\\
\hline
$\textit{MineDeposit} \times \textit{Post}$
&0.082$*$    &1.328$*$	&3.525$**$	&18.357$**$    &0.079    &1.218   &-0.035     &0.020\\
&(0.046)        &(0.753)	&(1.446)		&(7.515)        	&(1.334)	&(1.355)  &(0.028)&(0.074) \\\hline
Municipal and Year FEs
&\multicolumn{1}{c}{Yes}&\multicolumn{1}{c}{Yes}&\multicolumn{1}{c}{Yes}&\multicolumn{1}{c}{Yes}&\multicolumn{1}{c}{Yes}&\multicolumn{1}{c}{Yes}&\multicolumn{1}{c}{Yes}&\multicolumn{1}{c}{Yes}\\
Railway accessibility
&\multicolumn{1}{c}{Yes}&\multicolumn{1}{c}{Yes}&\multicolumn{1}{c}{Yes}&\multicolumn{1}{c}{Yes}&\multicolumn{1}{c}{Yes}&\multicolumn{1}{c}{Yes}&\multicolumn{1}{c}{Yes}&\multicolumn{1}{c}{Yes}\\
Port accessibility
&\multicolumn{1}{c}{Yes}&\multicolumn{1}{c}{Yes}&\multicolumn{1}{c}{Yes}&\multicolumn{1}{c}{Yes}&\multicolumn{1}{c}{Yes}&\multicolumn{1}{c}{Yes}&\multicolumn{1}{c}{Yes}&\multicolumn{1}{c}{Yes}\\
Observations
&\multicolumn{1}{c}{3,117}&\multicolumn{1}{c}{3,117}&\multicolumn{1}{c}{3,117}&\multicolumn{1}{c}{3,117}&\multicolumn{1}{c}{3,117}&\multicolumn{1}{c}{3,117}&\multicolumn{1}{c}{3,117}&\multicolumn{1}{c}{3,117}\\
Clusters
&\multicolumn{1}{c}{1,039}&\multicolumn{1}{c}{1,039}&\multicolumn{1}{c}{1,039}&\multicolumn{1}{c}{1,039}&\multicolumn{1}{c}{1,039}&\multicolumn{1}{c}{1,039}&\multicolumn{1}{c}{1,039}&\multicolumn{1}{c}{1,039}\\
First-stage $F$-statistic
&\multicolumn{1}{c}{39.253}&\multicolumn{1}{c}{39.253}&\multicolumn{1}{c}{39.253}
&\multicolumn{1}{c}{39.253}&\multicolumn{1}{c}{39.253}&\multicolumn{1}{c}{39.253}
&\multicolumn{1}{c}{39.253}&\multicolumn{1}{c}{39.253}\\
Sample Mean of the DV (TG in 1925; 30)
&\multicolumn{1}{c}{8.900}&\multicolumn{1}{c}{7.511}&\multicolumn{1}{c}{32.716}&\multicolumn{1}{c}{163.825}
&\multicolumn{1}{c}{20.265}&\multicolumn{1}{c}{18.130}&\multicolumn{1}{c}{1.030}&\multicolumn{1}{c}{5.044}\\
Std. Dev of the DV (TG in 1925; 30)
&\multicolumn{1}{c}{0.953}&\multicolumn{1}{c}{2.183}&\multicolumn{1}{c}{6.095}&\multicolumn{1}{c}{31.242}
&\multicolumn{1}{c}{4.067}&\multicolumn{1}{c}{3.808}&\multicolumn{1}{c}{0.115}&\multicolumn{1}{c}{0.412}\\
&&&&&&&&\\
\multicolumn{9}{l}{\textbf{Panel B: Assessment of Pretreatment Trends}}\\
&\multicolumn{8}{c}{Dependent Variable}\\
\cmidrule(rrrrrrrr){2-9}
&\multicolumn{1}{c}{}
&\multicolumn{3}{c}{Fertility Channel}
&\multicolumn{2}{c}{Mortality Channel}
&\multicolumn{2}{c}{Migration Channel}\\
\cmidrule(rrr){3-5}\cmidrule(rr){6-7}\cmidrule(rr){8-9}
&\multicolumn{1}{c}{(1) Population}
&\multicolumn{1}{c}{(2) Marriage}
&\multicolumn{1}{c}{(3) Crude}
&\multicolumn{1}{c}{(4) Marital}
&\multicolumn{1}{c}{(5) Male}
&\multicolumn{1}{c}{(6) Female}
&\multicolumn{1}{c}{(7) Sex Ratio}
&\multicolumn{1}{c}{(8) Size}\\
\hline
$\textit{MineDeposit} \times I\text{(Year = 1925)}$
&-0.039		&0.496		&2.584	&14.931	&-2.808	&-2.387	&0.018  &-0.033   \\
&(0.038)		&(0.889)		&(1.833)	&(9.443)	&(1.803)	&(1.702)	&(0.015) &(0.046) \\
$\textit{MineDeposit} \times I\text{(Year = 1935)}$
&0.062		&1.576$*$		&4.819$***$	&25.828$***$	&-1.326	&0.023	&-0.026     &0.003\\
&(0.038)		&(0.917)		&(1.840)		&(9.596)		&(1.583)	&(1.543)	&(0.025)  &(0.076)\\\hline
Municipal and Year FEs
&\multicolumn{1}{c}{Yes}&\multicolumn{1}{c}{Yes}&\multicolumn{1}{c}{Yes}&\multicolumn{1}{c}{Yes}
&\multicolumn{1}{c}{Yes}&\multicolumn{1}{c}{Yes}&\multicolumn{1}{c}{Yes}&\multicolumn{1}{c}{Yes}\\
Railway accessibility
&\multicolumn{1}{c}{Yes}&\multicolumn{1}{c}{Yes}&\multicolumn{1}{c}{Yes}&\multicolumn{1}{c}{Yes}
&\multicolumn{1}{c}{Yes}&\multicolumn{1}{c}{Yes}&\multicolumn{1}{c}{Yes}&\multicolumn{1}{c}{Yes}\\
Port accessibility
&\multicolumn{1}{c}{Yes}&\multicolumn{1}{c}{Yes}&\multicolumn{1}{c}{Yes}&\multicolumn{1}{c}{Yes}
&\multicolumn{1}{c}{Yes}&\multicolumn{1}{c}{Yes}&\multicolumn{1}{c}{Yes}&\multicolumn{1}{c}{Yes}\\
Observations
&\multicolumn{1}{c}{3,117}&\multicolumn{1}{c}{3,117}&\multicolumn{1}{c}{3,117}&\multicolumn{1}{c}{3,117}
&\multicolumn{1}{c}{3,117}&\multicolumn{1}{c}{3,117}&\multicolumn{1}{c}{3,117}&\multicolumn{1}{c}{3,117}\\
Clusters
&\multicolumn{1}{c}{1,039}&\multicolumn{1}{c}{1,039}&\multicolumn{1}{c}{1,039}&\multicolumn{1}{c}{1,039}&\multicolumn{1}{c}{1,039}&\multicolumn{1}{c}{1,039}&\multicolumn{1}{c}{1,039}&\multicolumn{1}{c}{1,039}\\
First-stage $F$-statistic (1925)
&\multicolumn{1}{c}{20.235}&\multicolumn{1}{c}{20.235}&\multicolumn{1}{c}{20.235}&\multicolumn{1}{c}{20.235}
&\multicolumn{1}{c}{20.235}&\multicolumn{1}{c}{20.235}&\multicolumn{1}{c}{20.235}&\multicolumn{1}{c}{20.235}\\
First-stage $F$-statistic (1935)
&\multicolumn{1}{c}{22.162}&\multicolumn{1}{c}{22.162}&\multicolumn{1}{c}{22.162}&\multicolumn{1}{c}{22.162}
&\multicolumn{1}{c}{22.162}&\multicolumn{1}{c}{22.162}&\multicolumn{1}{c}{22.162}&\multicolumn{1}{c}{22.162}\\
Sample Mean of the DV (TG in 1925; 30)
&\multicolumn{1}{c}{8.900}&\multicolumn{1}{c}{7.511}&\multicolumn{1}{c}{32.716}&\multicolumn{1}{c}{163.825}
&\multicolumn{1}{c}{20.265}&\multicolumn{1}{c}{18.130}&\multicolumn{1}{c}{1.030}&\multicolumn{1}{c}{5.044}\\
Std. Dev of the DV (TG in 1925; 30)
&\multicolumn{1}{c}{0.953}&\multicolumn{1}{c}{2.183}&\multicolumn{1}{c}{6.095}&\multicolumn{1}{c}{31.242}
&\multicolumn{1}{c}{4.067}&\multicolumn{1}{c}{3.808}&\multicolumn{1}{c}{0.115}&\multicolumn{1}{c}{0.412}\\\midrule[0.3pt]\bottomrule[1pt]
\end{tabular}
}
}
{\scriptsize
\begin{minipage}{635pt}
\setstretch{0.85}
***, **, and * represent statistical significance at the 1\%, 5\%, and 10\% levels, respectively.
Standard errors based on the cluster-robust covariance matrix estimator are reported in parentheses.\\
Notes:
Panel A presents the estimate ($\hat{\varpi}$) from the second-stage structural equation~\ref{pe_did1} in each outcome variable.
Columns (1)--(8) show the results for the log-transformed population, crude marriage rate, crude birth rate, marital fertility rate, crude death rates for men and women, sex ratio, and average household size, respectively.
Panel B summarizes the results from the expanded regression of equation~\ref{pe_did1}, which includes interaction terms between the ever-treated indicator variable and year dummies (with 1930 as the reference year).
Columns (1)--(8) follow the same column layout as in Panel A.
\end{minipage}
}
\end{center}
\end{table}
\end{landscape}
%------------------

%----------
\section{Early-Life Mortality: Occupational Hazards and Pollution}\label{sec:sec_aa}

The previous sections showed that crude death rates in mining municipalities did not differ substantially from those in surrounding municipalities.
However, crude death rates may not capture the effects of mining on pregnant women and their children.
This section examines whether early-life mortality was higher in coal-mining areas and whether it was associated with occupational hazards for female miners or with pollution.
I first present a simple framework for mortality selection \textit{in utero}, then describe the empirical strategy, and finally present the results.

\subsection{Theoretical Framework: Mortality Selection before Birth}\label{sec:sec71}

Health shocks during pregnancy can affect fetuses through two mechanisms: scarring and selection \citep{Almond:2011jf}.
Scarring reduces the health of all fetuses, whereas selection culls weaker fetuses before birth, so that the surviving infants are, on average, healthier.\footnote{See \citet{Catalano:2006vv} and \citet{Bruckner:2014fs}. Studies testing the Trivers--Willard hypothesis \citep{Trivers:1973fd} use the same framework to examine how fetal health shocks affect the sex ratio at birth \citep[e.g.,][]{Valente:2015ci}.}

Let $x \sim \mathcal{N}(\mu,\,\sigma^{2})$ be the initial health endowment of a fetus \textit{in utero}, and assume that a fetus with an endowment below the survival threshold $x_{s}$ dies before birth.\footnote{The survival threshold is usually well below the mean of the distribution ($x_{s} < \mu$).}
First, the scarring mechanism shifts the mean of the distribution to the left ($\mu^{*} < \mu$), so that $x^{*} \sim \mathcal{N}(\mu^{*},\,\sigma^{2})$.
Because the survival threshold is fixed, the share of fetuses that die before birth increases, $F^{*}(x_{s}) > F(x_{s})$, where $F^{*}(\cdot)$ and $F(\cdot)$ are the cumulative distribution functions of the normal distributions with means $\mu^{*}$ and $\mu$, respectively.
The expected health endowment of surviving infants also declines:
\begin{eqnarray}\label{cef_scar}
\footnotesize{
\begin{split}
E[x^{*}|x^{*} > x_{s}] < E[x|x > x_{s}].
\end{split}
}
\end{eqnarray}
Second, the selection mechanism shifts the survival threshold to the right ($x_{s}^{*} > x_{s}$).
Because the mean of the distribution is fixed, the share of fetuses that die before birth also increases, $F(x_{s}^{*}) > F(x_{s})$, but the expected health endowment of surviving infants improves:\footnote{See Online Appendix~\ref{sec:ce_tnd} for the derivation of the conditional expectation of the truncated normal distribution.}
\begin{eqnarray}\label{cef_sel}
\footnotesize{
\begin{split}
E[x|x > x_{s}] < E[x|x > x_{s}^{*}].
\end{split}
}
\end{eqnarray}
These results can be summarized as follows.
These results imply that fetal health shocks increase fetal deaths under both mechanisms, whereas their effects on infant mortality differ: infant mortality increases under the scarring mechanism and decreases under the selection mechanism.
If both mechanisms operate, the effect on infant mortality depends on which mechanism dominates.
Therefore, higher fetal death and infant mortality rates in mining areas would indicate that the scarring mechanism dominates.

\subsection{Empirical Strategy}\label{sec:sec72_es}

I estimate three sets of regressions using the municipal-level fetal death rate (FDR), infant mortality rate (IMR), and child mortality rate (CMR) in 1933 and 1938 (Section~\ref{sec:sec_data_dv}).
First, to compare early-life mortality between mining and surrounding municipalities, I estimate equations~(\ref{eq1}) and~(\ref{eq2}) by two-stage least squares, using the same instrument and control variables as in Section~\ref{sec:sec_ea}.

Second, to examine the channels, I restrict the sample to municipalities with coal mines and estimate the following equation by OLS:
\begin{eqnarray}\label{eq_health}
\footnotesize{
\begin{split}
y_{i} = \lambda_{0} + \lambda_{f} \textit{FemaleMiner}_{i} + \lambda_{m} \textit{MaleMiner}_{i} + \mathbf{x}_{i}' \boldsymbol{\phi} + u_{i},
\end{split}
}
\end{eqnarray}
where $\textit{FemaleMiner}_{i}$ and $\textit{MaleMiner}_{i}$ are the numbers of female and male miners in the coal mines of municipality $i$ recorded in the NDFM, and $\mathbf{x}_{i}$ includes the control variables used in Section~\ref{sec:sec_ea}, including the distance to the nearest river.
If heavy work during pregnancy increased fetal and infant deaths, $\lambda_{f}$ should be positive.
Because the numbers of female and male miners are both included, $\lambda_{f}$ captures the association with female employment conditional on the scale of mining.

Third, to examine changes between 1933 and 1938, I pool the two years and estimate panel regressions with municipal and year fixed effects.
For the full sample, I estimate the difference-in-differences specification in equations~(\ref{pe_did1}) and~(\ref{pe_did2}), in which $\text{\textit{MineDeposit}}_{i}$ is interacted with an indicator for 1938 and instrumented with $\text{\textit{Stratum}}_{i}$ interacted with the same indicator.
For the sample of mining municipalities, I include the numbers of female and male miners and their interactions with the 1938 indicator.
Because the numbers of miners are recorded in different years of the NDFM, their coefficients are identified from changes within municipalities, and the interaction terms indicate whether these associations differed between 1933 and 1938.

The timing of the data deserves attention.
The 1933 and 1938 mortality data are matched to the NDFM of October 1931 and October 1936, respectively.
Fetal deaths in 1933 occurred among pregnancies that began in 1932 or 1933, and infants who died in 1933 were born in 1932 or 1933.
Their mothers were therefore exposed to working conditions in 1931--32, when the numbers of female miners recorded in October 1931 are likely to reflect their exposure.
By contrast, the 1938 data reflect working conditions after the enforcement, when the remaining female miners worked mainly in thin-layer pits and in coal selection outside the pits (Section~\ref{sec:sec23}).

\subsection{Results}\label{sec:sec72}

\subsubsection*{Occupational Hazards}

%------------------
%Table 5
%\begin{landscape}
\begin{table}[]
\def\arraystretch{0.93}
\begin{center}
\captionsetup{justification=centering}
\caption{Testing the Occupational Hazards and Pollution\\ using the Early-life Mortality Data}
\label{tab:r_health}
\scriptsize
\scalebox{0.86}[1]{
{\setlength\doublerulesep{2pt}
\begin{tabular}{lD{.}{.}{-2}D{.}{.}{-2}D{.}{.}{-2}D{.}{.}{-2}D{.}{.}{-2}D{.}{.}{-2}}
\toprule[1pt]\midrule[0.3pt]
\multicolumn{7}{l}{\textbf{Panel A: Coal Mining and Early-life Mortality in 1933}}\\
&\multicolumn{6}{c}{Dependent Variable}\\
\cmidrule(rrrrrr){2-7}
&\multicolumn{1}{c}{(1) IMR}&\multicolumn{1}{c}{(2) FDR}&\multicolumn{1}{c}{(3) FDR}&\multicolumn{1}{c}{(4) IMR}&\multicolumn{1}{c}{(5) CMR}&\multicolumn{1}{c}{(6) CMR}\\\hline

\textit{MineDeposit}
&36.766$**$	&20.668$**$	&                	&                &1.456            &    \\
&(15.968)		&(9.553)		&               	 &                &(2.799)        &    \\
\textit{Female Miner}
&			&			&0.038$***$	&0.096$***$    &                &0.004        \\
&			&                	&(0.013)		&(0.030)        &                &(0.004)    \\
\textit{Male Miner}
&            	  	  &                &-0.003$*$	        &-0.009$**$    &                &-0.000    \\
&            	  	  &                &(0.002)        &(0.004)        &                &(0.001)    \\
\textit{River accessibility}
&-2.119$**$    &0.368            &-2.994        &-0.253        &-0.635$***$    &0.563        \\
&(1.027)        &(0.686)        &(2.290)        &(3.672)        &(0.223)        &(0.470)    \\\hdashline[0.5pt/3pt]
Analytical sample
&\multicolumn{1}{c}{Full}&\multicolumn{1}{c}{Full}&\multicolumn{1}{c}{Mine}&\multicolumn{1}{c}{Mine}&\multicolumn{1}{c}{Full}&\multicolumn{1}{c}{Mine}\\
Observations
&\multicolumn{1}{c}{1,142}&\multicolumn{1}{c}{1,142}&\multicolumn{1}{c}{91}&\multicolumn{1}{c}{91}&\multicolumn{1}{c}{1,142}&\multicolumn{1}{c}{91}\\\hline
City and town indicators
&\multicolumn{1}{c}{Yes}&\multicolumn{1}{c}{Yes}&\multicolumn{1}{c}{Yes}&\multicolumn{1}{c}{Yes}&\multicolumn{1}{c}{Yes}&\multicolumn{1}{c}{Yes}\\
Control variables
&\multicolumn{1}{c}{Yes}&\multicolumn{1}{c}{Yes}&\multicolumn{1}{c}{Yes}&\multicolumn{1}{c}{Yes}&\multicolumn{1}{c}{Yes}&\multicolumn{1}{c}{Yes}\\
Estimator
&\multicolumn{1}{c}{IV}&\multicolumn{1}{c}{IV}&\multicolumn{1}{c}{OLS}&\multicolumn{1}{c}{OLS}&\multicolumn{1}{c}{IV}&\multicolumn{1}{c}{OLS}\\
First-stage $F$-statistic    &36.69&36.69&$--$&$--$&36.69&$--$\\
Sample Mean of the DV   &117.11    &43.54    &53.96    &148.79    &13.41    &16.15\\
Std. Dev. of the DV		&42.63        &27.37    &23.78    &43.07        &9.58    &5.54\\
&&&&&&\\
\multicolumn{7}{l}{\textbf{Panel B: Coal Mining and Early-life Mortality in 1938}}\\
&\multicolumn{6}{c}{Dependent Variable}\\
\cmidrule(rrrrrr){2-7}
&\multicolumn{1}{c}{(1) IMR}&\multicolumn{1}{c}{(2) FDR}&\multicolumn{1}{c}{(3) FDR}&\multicolumn{1}{c}{(4) IMR}&\multicolumn{1}{c}{(5) CMR}&\multicolumn{1}{c}{(6) CMR}\\\hline
\textit{MineDeposit}
&16.709		&7.910        &                &                &1.964        &            \\
&(12.889)		&(8.458)    &                &                &(2.647)    &            \\
\textit{Female Miner}
&            &            &0.033$**$    &0.073$**$    &            &0.006        \\
&            &            &(0.014)        &(0.028)        &            &(0.005)    \\
\textit{Male Miner}
&            &            &-0.001        &-0.001        &            &-0.000        \\
&            &            &(0.001)        &(0.002)        &            &(0.001)    \\
\textit{River accessibility}
&-2.022$*$    &0.378        &0.175        &-0.276        &-0.244    &-0.122    \\
&(1.034)        &(0.582)    &(2.041)    &(3.538)        &(0.216)    &(0.719)    \\\hdashline[0.5pt/3pt]
Analytical sample
&\multicolumn{1}{c}{Full}&\multicolumn{1}{c}{Full}&\multicolumn{1}{c}{Mine}&\multicolumn{1}{c}{Mine}&\multicolumn{1}{c}{Full}&\multicolumn{1}{c}{Mine}\\
Observations
&\multicolumn{1}{c}{1,316}&\multicolumn{1}{c}{1,316}&\multicolumn{1}{c}{110}&\multicolumn{1}{c}{110}&\multicolumn{1}{c}{1,316}&\multicolumn{1}{c}{110}\\\hline
City and town indicators
&\multicolumn{1}{c}{Yes}&\multicolumn{1}{c}{Yes}&\multicolumn{1}{c}{Yes}&\multicolumn{1}{c}{Yes}&\multicolumn{1}{c}{Yes}&\multicolumn{1}{c}{Yes}\\
Control variables
&\multicolumn{1}{c}{Yes}&\multicolumn{1}{c}{Yes}&\multicolumn{1}{c}{Yes}&\multicolumn{1}{c}{Yes}&\multicolumn{1}{c}{Yes}&\multicolumn{1}{c}{Yes}\\
Estimator
&\multicolumn{1}{c}{IV}&\multicolumn{1}{c}{IV}&\multicolumn{1}{c}{OLS}&\multicolumn{1}{c}{OLS}&\multicolumn{1}{c}{IV}&\multicolumn{1}{c}{OLS}\\
First-stage $F$-statistic	&40.00	&40.00	&$--$	&$--$	&40.00	&$--$\\
Sample Mean of the DV	&114.57	&37.00	&51.46	&137.80	&16.80	&20.63\\
Std. Dev. of the DV		&42.94	&25.75	&23.21	&36.15	&8.81	&8.14\\
&&&&&&\\
\multicolumn{7}{l}{\textbf{Panel C: The Early-life Mortality Trend between 1933 and 1938}}\\
&\multicolumn{6}{c}{Dependent Variable}\\
\cmidrule(rrrrrr){2-7}
&\multicolumn{1}{c}{(1) IMR}&\multicolumn{1}{c}{(2) FDR}
&\multicolumn{1}{c}{(3) FDR}&\multicolumn{1}{c}{(4) IMR}
&\multicolumn{1}{c}{(5) CMR}&\multicolumn{1}{c}{(6) CMR}\\\hline
$\textit{MineDeposit}\times I\text{(Year = 1938)}$			&-6.724	&-3.095	&			&			&-1.683	&		\\
						&(14.151)	&(8.822)	&			&			&(3.586)	&		\\
\textit{Female Miner}			&		&		&0.033$**$	&0.081$***$	&		&-0.005	\\
						&		&		&(0.013)		&(0.029)		&		&(0.009)	\\
$\textit{Female Miner}\times I\text{(Year = 1938)}$			&		&		&-0.014		&0.008		&		&0.008	\\
						&		&		&(0.014)		&(0.022)		&		&(0.005)	\\
\textit{Male Miner}			&		&		&-0.004		&-0.014$**$	&		&0.001	\\
						&		&		&(0.003)		&(0.006)		&		&(0.002)	\\
$\textit{Male Miner}\times I\text{(Year = 1938)}$		&		&		&0.003$*$		&0.004		&		&-0.000	\\
										&		&		&(0.001)		&(0.003)		&		&(0.001)	\\\hdashline[0.5pt/3pt]
Analytical sample
&\multicolumn{1}{c}{Full}&\multicolumn{1}{c}{Full}&\multicolumn{1}{c}{Mine}&\multicolumn{1}{c}{Mine}&\multicolumn{1}{c}{Full}&\multicolumn{1}{c}{Mine}\\
Observations
&\multicolumn{1}{c}{2,182}&\multicolumn{1}{c}{2,182}&\multicolumn{1}{c}{168}&\multicolumn{1}{c}{168}&\multicolumn{1}{c}{2,182}&\multicolumn{1}{c}{168}\\\hline
Control variables
&\multicolumn{1}{c}{Yes}&\multicolumn{1}{c}{Yes}&\multicolumn{1}{c}{Yes}&\multicolumn{1}{c}{Yes}&\multicolumn{1}{c}{Yes}&\multicolumn{1}{c}{Yes}\\
Municipal and Year FEs
&\multicolumn{1}{c}{Yes}&\multicolumn{1}{c}{Yes}&\multicolumn{1}{c}{Yes}&\multicolumn{1}{c}{Yes}&\multicolumn{1}{c}{Yes}&\multicolumn{1}{c}{Yes}\\
Estimator
&\multicolumn{1}{c}{IV}&\multicolumn{1}{c}{IV}&\multicolumn{1}{c}{OLS}
&\multicolumn{1}{c}{OLS}&\multicolumn{1}{c}{IV}&\multicolumn{1}{c}{OLS}\\
First-stage $F$-statistic    &38.599&38.599&$--$&$--$&38.599&$--$\\
Sample Mean of the DV	&140.23	&51.84	&54.56	&148.60	&15.76	&16.06	\\
Std. Dev. of the DV		&43.23	&26.21	&23.19	&40.65	&7.16	&5.62	\\\midrule[0.3pt]\bottomrule[1pt]
\end{tabular}
}
}
{\scriptsize
\begin{minipage}{430pt}
\setstretch{0.85}
***, **, and * represent statistical significance at the 1\%, 5\%, and 10\% levels, respectively.
Standard errors based on the heteroskedasticity-robust covariance matrix estimator are reported in parentheses.\\
Notes:
Panels A and B show the results for the early-life mortality measured from the 1933 and 1938 Vital Statistics samples, respectively.
The IMR, FDR, and CMR indicate the infant, fetal, and child mortality rates, respectively.
Columns (1); (2); and (5) show the results for the entire sample.
Columns (3); (4); and (6) show the results for the municipalities with coal mines.
Control variables include railway accessibility, port accessibility, and elevation.
Panel C presents the results for the panel data created from the 1933 and 1938 Vital Statistics.
Columns (1); (2); and (5) show the results for the ever-and never-treated panel data sample.
The 5km municipalities from coal mines that are consistently observed (treated) in both 1933 and 1938 are defined as the ever-treated municipalities.
Columns (3); (4); and (6) show the results for the panel data for the municipalities with coal mines that are consistently observed in both 1933 and 1938.
Control variables include railway and port accessibility.
\end{minipage}
}
\end{center}
\end{table}
%\end{landscape}

Panel A of Table~\ref{tab:r_health} presents the results for 1933.
Columns (1) and (2) show that both the IMR and the FDR were significantly higher in mining municipalities than in surrounding municipalities.
As discussed in Section~\ref{sec:sec71}, this pattern is consistent with the scarring mechanism, in which mining reduced the health of fetuses \textit{in utero}.

Columns (3) and (4) restrict the sample to the $91$ municipalities with coal mines.
The coefficients on the number of female miners are positive and statistically significant for both the FDR and the IMR, which supports the argument that occupational hazards for pregnant women increased fetal and infant deaths.\footnote{The estimate indicates that a one-standard-deviation increase in the number of female miners ($271.15$ miners) increases fetal deaths by roughly $10$ ($0.038 \times 271.15$) per $1,000$ births. This magnitude is comparable to the difference in the mean FDR between the treatment and control groups, approximately $9.4$ fetal deaths per $1,000$ births (Panel B of Online Appendix Table~\ref{tab:sum_outcome}).}
By contrast, the coefficients on the number of male miners are negative.
Because the numbers of female and male miners are positively correlated, the negative coefficients on male miners indicate that, for a given number of female miners, mortality was lower in municipalities with more male miners.
Thus, the higher early-life mortality was associated with the employment of women rather than with the scale of mining itself.

Panel B presents the results for 1938.
The estimates in Columns (1) and (2) remain positive but are smaller and no longer statistically significant.
The coefficients on female miners in Columns (3) and (4) remain positive and statistically significant, although they are smaller than those in 1933.\footnote{A one-standard-deviation increase in the number of female miners ($179.14$ miners) increases fetal deaths by approximately $5.9$ ($0.033 \times 179.14$) per $1,000$ births, roughly $60$\% of the magnitude in 1933.}
This suggests that the association between female employment and early-life mortality persisted after the full enforcement, when women continued to work in thin-layer pits and outside the pits.

Panel C examines whether early-life mortality changed differently in mining municipalities between 1933 and 1938.
The estimates in Columns (1) and (2) are negative but statistically insignificant.
In Columns (3) and (4), the coefficients on female miners remain positive and statistically significant even with municipal fixed effects, indicating that early-life mortality increased when the number of female miners in the same municipality increased.
The coefficients on the interactions between female miners and the 1938 indicator are close to zero.
Thus, early-life mortality in mining municipalities did not improve more than in surrounding municipalities between 1933 and 1938, and its association with female employment did not change significantly.
Because both years are at or after the full enforcement, these results do not evaluate the effects of the regulation itself.
The smaller estimates for 1938 may instead reflect the general decline in early-life mortality in Japan during the 1930s (Online Appendix Figure~\ref{fig:ts_imr_fdr}).

\subsubsection*{Pollution Channels}

Another potential channel is pollution.
If wastewater from coal sorting increased fetal and infant deaths, mortality should have been higher in mining municipalities closer to rivers (Section~\ref{sec:sec23}).
Because river accessibility is measured as the log distance to the nearest river, a negative coefficient indicates higher mortality closer to rivers.
In the full samples, the coefficients on river accessibility are negative and statistically significant for the IMR and CMR in 1933 (Columns 1 and 5 of Panel A) and for the IMR in 1938 (Column 1 of Panel B).
However, these estimates are based on all municipalities and may reflect general health risks near rivers, such as waterborne diseases, rather than pollution from mines.
Within mining municipalities, where wastewater from coal sorting would matter most, the coefficients on river accessibility are small and statistically insignificant (Columns 3 and 4 of Panels A and B).
Thus, I find no evidence that wastewater from coal mines increased early-life mortality, although this proxy may not capture the actual exposure to contaminated water.

Air pollution from the boilers of steam-powered winding engines may also have increased early-life mortality.
If this channel mattered, mortality should have increased with the scale of mines, measured by the number of male miners.
However, the coefficients on male miners are negative or close to zero in Columns (3) and (4) of Panels A and B.
Column (5) uses the CMR as an alternative outcome, because young children who play outdoors are likely to be exposed to air pollution.
The estimates are close to zero in both years, and the coefficients on the numbers of female and male miners in Column (6) are also close to zero.
These results are consistent with the electrification of winding engines during the interwar period (Section~\ref{sec:sec23}).

\subsubsection*{Summary}

Fetal and infant mortality were higher in coal-mining municipalities than in surrounding municipalities in 1933, consistent with the scarring mechanism \textit{in utero}.
Within mining municipalities, early-life mortality increased with the number of female miners, conditional on the number of male miners.
This association also holds within municipalities over time and did not change significantly between 1933 and 1938, suggesting that it persisted after the full enforcement of the regulation, when women continued to work in thin-layer pits and in coal selection outside the pits.
By contrast, I find little evidence that water or air pollution increased early-life mortality.
These results suggest that occupational hazards faced by pregnant women were an important channel through which coal mining affected the health of the next generation.

%-------------------------------------------------------------------------------
\section{Robustness}\label{sec:sec_rob}

This section summarizes additional analyses that examine the sensitivity of the main results.

First, I examine the definitions of the treatment and control groups.
Estimates by distance band show that the effects of coal mines are concentrated within $5$ km of mining municipalities, supporting the thresholds used to define the treatment and control groups (Online Appendix~\ref{sec:test_threshold}).

Second, I examine the influence of spatial correlation.
When standard errors are clustered at the county level, the statistical significance of the main results is materially unchanged (Online Appendix~\ref{sec:app_crve}).
This is important because the instrument is based on geological formations that extend over wide areas.

Third, I examine regional differences in the organization of mining labor.
Coal mines in Hokkaido relied more on male miners from rural T\=ohoku and employed fewer women in the pits than those in Ky\=ush\=u \citep{Nishinarita1985}.
The results are materially similar when an indicator for Hokkaido is included in the regressions (Online Appendix~\ref{sec:app_r_hokkaido}).

Fourth, I use an alternative source of variation.
Because the locations of most coal mines were stable over the sample period, the main analysis relies on cross-sectional variation (Section~\ref{sec:sec_ea_as}).
As an alternative, I estimate a difference-in-differences model for two coal-mining areas in the T\=ohoku region where mines opened during the sample period.
The results are consistent with the main findings (Online Appendix~\ref{sec:app_did}).

Finally, I compare the results for coal mines with those for gold, silver, and copper mines.
Because these mines employed far fewer workers than coal mines, their effects on local population should be smaller if the main results reflect the employment created by mining rather than unobserved characteristics of mining areas.
The estimated effects for these mines are indeed smaller (Online Appendix~\ref{sec:secd_hm}).

%-------------------------------------------------------------------------------
\section{Conclusion}\label{sec:sec_conclusion}

This paper examines how coal mining shaped local demographic change in interwar Japan, focusing on the role of women's work.
In prewar Japanese coal mines, married couples worked underground as teams, and women earned substantially more in the pits than in other industries.
The mechanization of haulage and the revision of labor regulations enforced in 1933 removed most women from the pits.
Using municipal-level census and vital statistics linked to the locations of coal mines, and exploiting the distribution of coal-bearing geological strata as an instrumental variable, I examine how coal mining affected population, marriage, fertility, and mortality.

The results yield three main findings.
First, coal mining considerably enlarged local populations, largely because of the inflow of workers, whereas low marriage and fertility rates kept natural increase small before the regulation.
Second, once women were removed from the pits, the differences in fertility and marriage between mining and surrounding areas became smaller.
This pattern is in line with the prediction that women's earning opportunities raise the opportunity cost of childrearing.
Third, fetal and infant mortality was higher in mining areas and rose with the number of female miners, while there is little evidence that pollution contributed to early-life mortality.

These findings suggest that the demographic consequences of resource extraction depended on how mining employment was organized.
Studies of resource booms have emphasized the inflow of workers attracted by higher wages.
This paper shows that the employment of women in mining also shaped local demography: it lowered fertility and increased early-life mortality in mining areas.
The findings also shed light on the consequences of protective labor legislation for women.
Previous studies have focused on its political economy and employment effects.
In interwar Japan, the regulation that removed women from the pits was followed by the convergence of fertility in mining areas toward that of surrounding areas.

Several limitations remain.
Because the regulation coincided with mechanization and with firms' efforts to encourage miners to form families, I cannot fully distinguish the effects of women's earning opportunities from those of other changes in mining communities.
In addition, the lack of migration statistics prevents a direct analysis of who moved into mining areas.
Individual-level records of mining households, such as company records and household registers, would help address these questions in future research.

%-------------------------------------------------------------------------------
%\clearpage
%-------------------------------------------------------------------------------
% FIGURES
%-------------------------------------------------------------------------------
%-------------------------------------------------------------------------------
% TABLES
%-------------------------------------------------------------------------------
%-------------------------------------------------------------------------------
% REFERENCES
%-------------------------------------------------------------------------------
%\clearpage
\bibliographystyle{plainnat}%tfcse
\bibliography{reference.bib}
%-------------------------------------------------------------------------------
% Appendix
%-------------------------------------------------------------------------------
\clearpage
\thispagestyle{empty}

\begin{center}
\qquad

\qquad

\qquad

\qquad

\qquad

\qquad

{\LARGE \textbf{
%Online
Appendices
%: For online publication only (Supplemental materials for review)
}}
\end{center}
%-------------------------------------------------------------------------------
% Appendix A
%-------------------------------------------------------------------------------
\clearpage
\appendix
\def\thesection{Appendix~\Alph{section}}
\def\thesubsection{\Alph{section}.\arabic{subsection}}
\setcounter{page}{1}

\section{Background Appendix}\label{sec:seca}
\setcounter{figure}{0} \renewcommand{\thefigure}{A.\arabic{figure}}
\setcounter{table}{0} \renewcommand{\thetable}{A.\arabic{table}}

\subsection{Influence of Great Depression in Japan}\label{sec:seca_gd}

The impact of the Great Depression was not as substantial in Japan as in other countries (Miyamoto 2008, pp.~56--57; Abe 2008, p.~106; Blumenthal 1981, pp.~46).
Blumenthal (1981, p. 43) shows that Japan's index of industrial production did not decline dramatically between 1929 (=100) and 1930 (=97).
In contrast, the US and the UK experienced much larger reductions (81 in the US and 92 in the UK) during the same period.
Blumenthal (1981, p.~45) also reports that the unemployment rates in the US, the UK, and Japan between 1930 and 1935 were approximately 27\%, 15\%, and 6\%, respectively.
Consequently, Japan's GNP growth rate (from 1929 to 1930) was positive at $1.1$\%, whereas it was $-7.7$ and $-0.1$, respectively, in the US and UK.
This evidence suggests that the Depression did not cause a prolonged disruption of economic activity in Japan.
In the coal mining sector, however, the Depression caused a temporary contraction around 1930 and 1931.
Coal production declined in the early 1930s (Figure~\ref{fig:ts_coal}), and several mines in the Chikuh\=o coalfield temporarily ceased operations (Compilation Committee of the Chronology of Chikuh\=o Coal Mining Industry 1973).
The NDFM of October 1931 also lists fewer coal mines than the 1922 and 1936 editions (Figure~\ref{fig:hist_mines}), which may reflect a temporary decline in the number of miners in small mines below the threshold of $50$ miners; excluding small mines does not change the results (Online Appendix~\ref{sec:r_inf}).
Because the contraction was temporary and production recovered by the mid-1930s, it is unlikely to affect the cross-sectional comparisons in 1925 and 1935.
However, it may have temporarily affected outcomes in mining areas around 1930, and these effects should be largely captured by the time-fixed effect in my panel regressions.
Overall, the Depression mainly affected prices rather than production, and price declines were smaller for food and fuel than for textiles, metals, and construction materials (Nakamura and Odaka 1989, p.~53).

\subsection{Female Miners in Coal Mine}\label{sec:seca0}

%--------------
%Figure A1
\begin{figure}[htbp]
\centering
\captionsetup{justification=centering}
\subfloat[\textit{Saki-yama} (left) and \textit{Ato-yama} (right)]{\label{fig:ato_sakiyama}\includegraphics[width=0.404\textwidth]{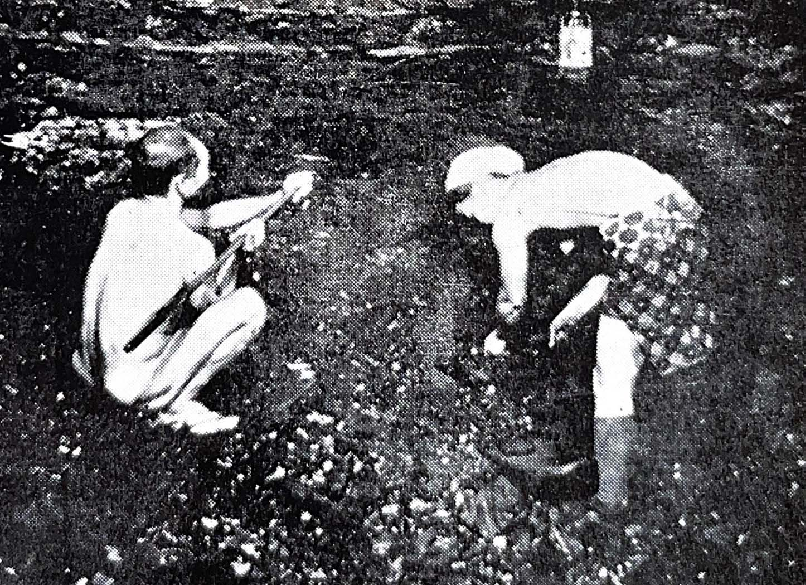}}
\quad
\subfloat[\textit{Ato-yama} using \textit{Sura}]{\label{fig:sura}\includegraphics[width=0.42\textwidth]{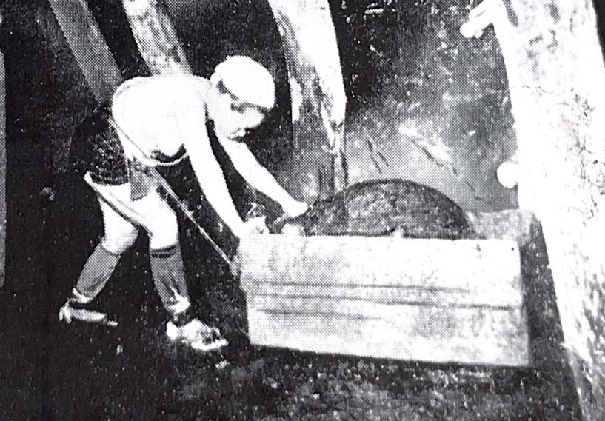}}
\caption{Male (\textit{saki-yama}) and Female Miners (\textit{ato-yama})}
\label{fig:photo1}
\scriptsize{\begin{minipage}{450pt}
\setstretch{0.9}
Notes:
Figure~\ref{fig:ato_sakiyama} shows the skilled male miner (\textit{saki-yama}) extracting coal and the female miner (\textit{ato-yama}) supporting her husband at the face.
Figure~\ref{fig:sura} illustrates an \textit{ato-yama} bringing coal using a bamboo basket (\textit{sura}) at the gangway through the steep pits.
Both figures were taken in the early Taish\=o period.
Source: Fumoto (1961, p.~2).
\end{minipage}}
\end{figure}
%--------------

By the end of the 19th century, many mines in the Chikuh\=o coalfields had completed mechanization of coal haulage.
However, this did not mean that the miner's workload related to hauling operations was reduced.
Instead, this mechanization in the main shafts had increased the workload of miners extracting and transporting coal around the coalface (Sumiya 1968, pp.~310--312).
Figure~\ref{fig:ato_sakiyama} shows a photograph of miners at the mine face (\textit{kiriha}) in the 1910s.
A male skilled miner (\textit{saki-yama}) extracted coal, whereas a female miner (\textit{ato-yama}) brought the coal to the surface using a bamboo basket (\textit{sura}).
Figure~\ref{fig:sura} shows that a female miner tied a rope around her body to prevent the heavy \textit{sura} from slipping off in the steep pits.
The environment in the pit was generally poor.
Moreover, rockfalls frequently caused accidents.\footnote{Available statistics indicate that the average number of accidents per year that occurred in the pits between 1917 and 1926, causing injuries and deaths of miners, was 142,595, of which 61,112 were from falling rocks. (Japan Mining Association, 1928, p.~table.6). Accidents caused by explosions, poisoning, and suffocation were fewer than those caused by falling rocks, but they still occurred 311 times per year on average between 1917 and 1926 (Japan Mining Association 1928, p.~table.6).}.

%--------------
%Figure A2
\begin{figure}[htb]
\centering
\captionsetup{justification=centering}
\subfloat[Longwall Method]{\label{fig:longwall}\includegraphics[angle=90, height = 0.28\textwidth, width=0.45\textwidth]{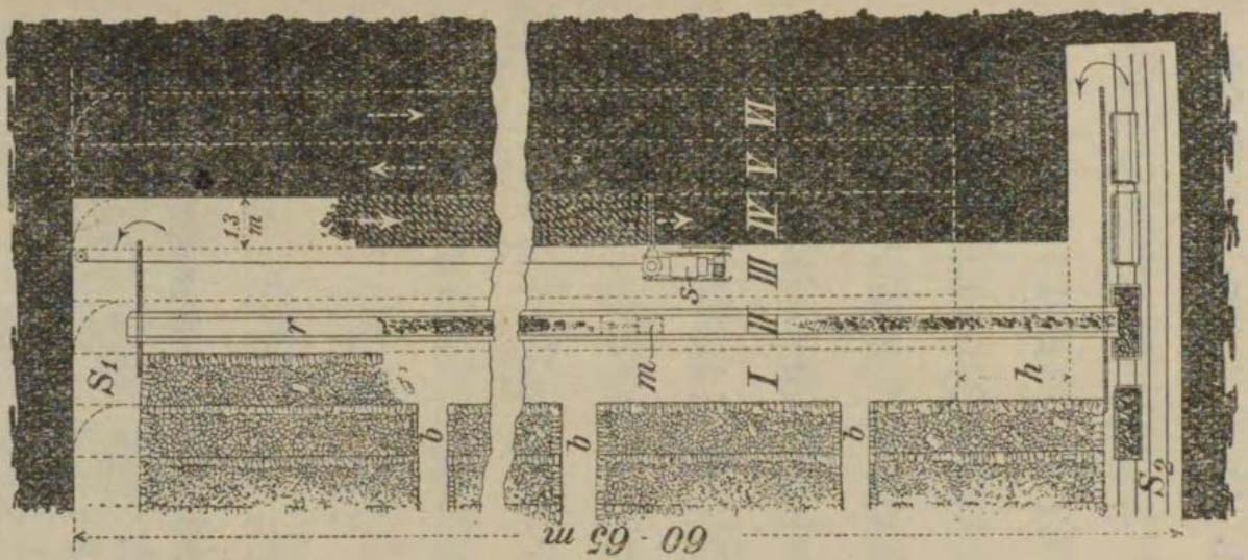}}
\quad
\subfloat[Mechanical Coal Extraction]{\label{fig:coal_cutter_conveyor}\includegraphics[width=0.4\textwidth]{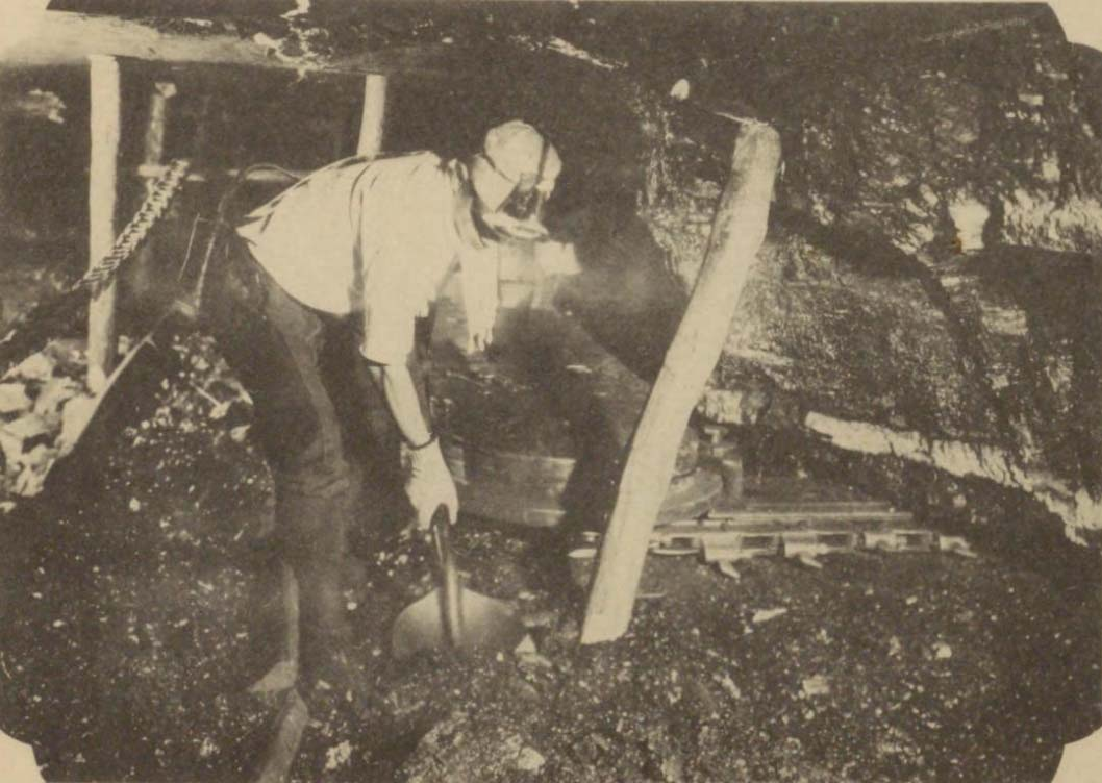}}
\caption{Longwall Method, Coal Cutter, and Conveyor}
\label{fig:photo2}
\scriptsize{\begin{minipage}{450pt}
\setstretch{0.9}
Notes:
Figure~\ref{fig:longwall} shows the system at the coal face used in the single-sided longwall method. This system includes a coal cutter (row III), positioned in front of the coal face (row IV), and a conveyor located behind the coal face (row II). The figure also shows a wagon in the main shaft on the right side.
Figure~\ref{fig:coal_cutter_conveyor} shows the labor pattern in the coal face under the longwall method, where a male minder uses a scoop to transfer coal extracted by the coal cutter to the conveyor behind him.
Sources: Mikawa (1939, p.~274) and Chikuh\=o Coal Mining Association (1935, p.~88).
\end{minipage}}
\end{figure}
%--------------

\begin{figure}[htb]
\centering
\captionsetup{justification=centering}
\subfloat[Wooden Baskets]{\label{fig:wagon_sura}\includegraphics[width=0.41\textwidth]{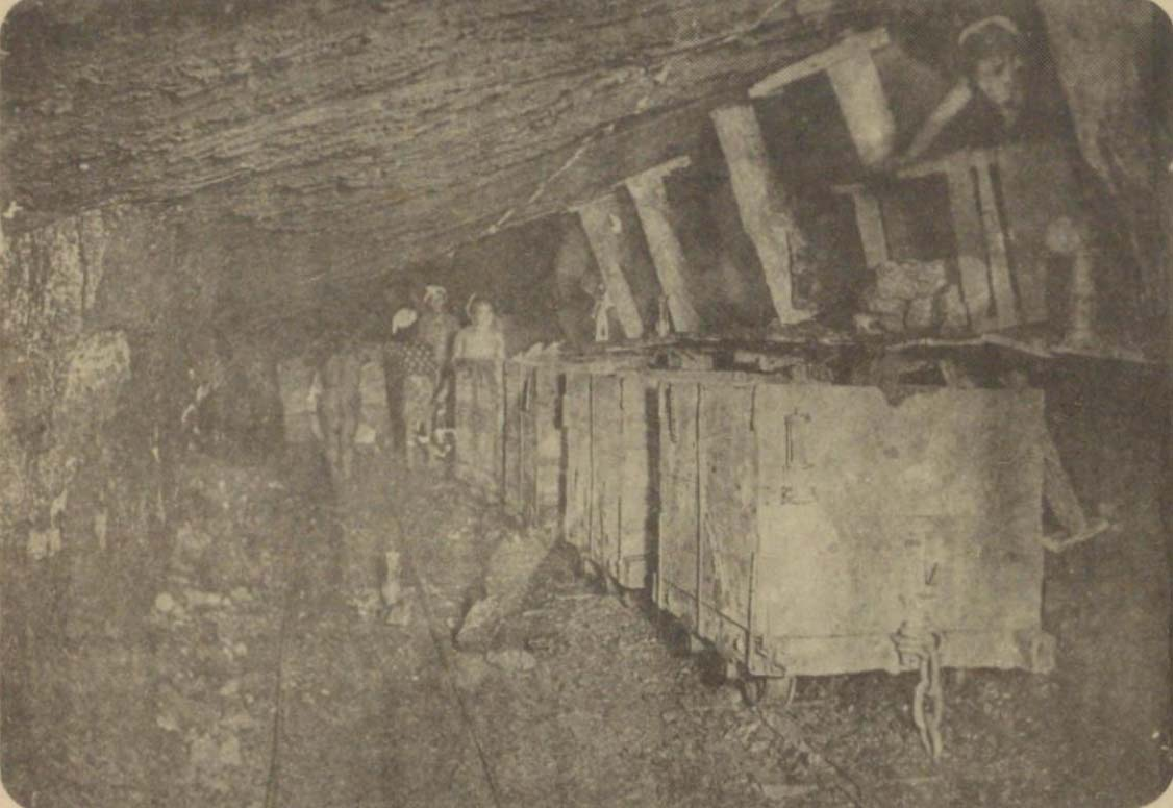}}
\quad
\subfloat[Conveyor]{\label{fig:wagon_conveyor}\includegraphics[width=0.4\textwidth]{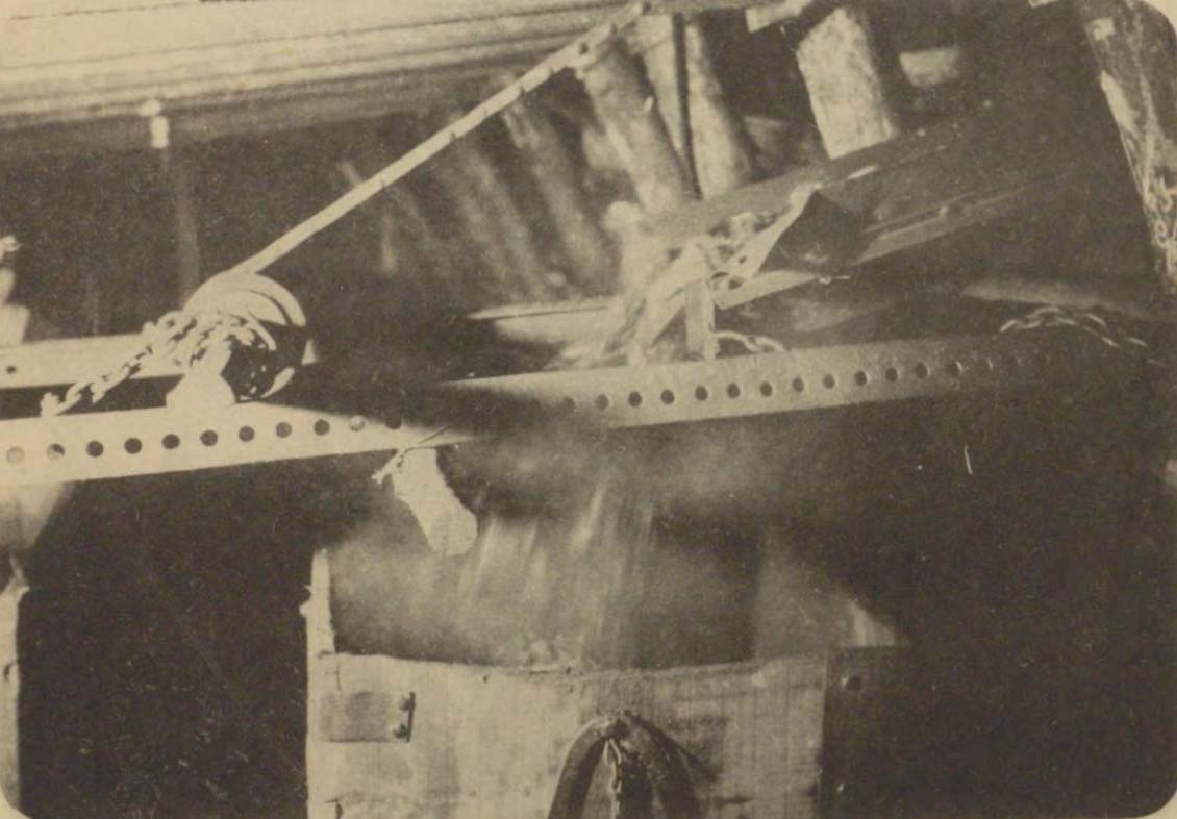}}
\caption{Changes in the Method of Transporting Extracted Coal\\ from the Coal Face to the Main Shaft}
\label{fig:photo3}
\scriptsize{\begin{minipage}{450pt}
\setstretch{0.9}
Notes:
Figure~\ref{fig:wagon_sura} shows the traditional coal mining method, where miners use a wooden basket (\textit{sura}) and must navigate through small, narrow tunnels to reach the main shaft. The same figure also shows the mechanical coal mining method, where a conveyor automatically deposits the extracted coal into a wagon.\\
Sources: Chikuh\=o Coal Mining Association (1935, p.~35).
\end{minipage}}
\end{figure}

The first International Labor Conference in 1919 motivated the government to improve harsh working conditions, especially for female miners.
In the early 1920s, the government consulted with mining companies to revise the Miners' Labor Assistance Regulation of 1916 to prevent females from working in coal mines.
Initially, the companies opposed this proposal.
However, a shift in mining techniques altered their perspective.
By the late 1920s, many coal mines had transitioned from the traditional room-and-pillar method to the longwall method.
The longwall method, compared with the room-and-pillar method, offers a more extensive space for coal extraction, which accommodates larger machinery such as coal cutters and conveyors, as shown in Figure~\ref{fig:longwall}.
This provided a strong incentive for mining companies to exclude female miners from the pits (Tanaka and Ogino 1977, p.~79; Nishinarita 1985, p.~91).\footnote{See also Tanaka (1984, Chapter~4) for a comprehensive summary of the relationships between the mechanization process and the changes in the labor patterns in the pits.}
Finally, the Revised Miners' Labor Assistance Regulations of September 1928 prohibited certain types of underground and late-night work, with a grace period for enforcement that lasted until September 1933.
During this period of mechanization from the late 1920s to the early 1930s, the number of mines employing both male and female workers decreased.
However, the apparent reduction was observed among female miners working in the pits (Table~\ref{tab:nminers}).
The photograph in Figure~\ref{fig:coal_cutter_conveyor} clearly shows how the work of coal extraction at the coal face changed with the introduction of mechanical coal extraction using the longwall method.
Additionally, the photographs in Figure~\ref{fig:photo3} show that the conveyors transported the extracted coal more efficiently than the miners could.

%--------------
%Figure A3
%--------------

Importantly, however, the revised regulations did not eliminate female miners altogether.
Female miners were allowed to work in the pits of the thin layer and were primarily responsible for coal selection and other out-of-pit labor.
More than 15,000 women still worked in the mines in the late 1930s, representing approximately 10\% of all the miners measured.

\subsection{Mechanization and  Female Miners}\label{sec:seca_tech}

%-------------
%Figure A4
\begin{figure}[htb]
\centering
\captionsetup{justification=centering,margin=1.5cm}
\includegraphics[width=0.5\textwidth]{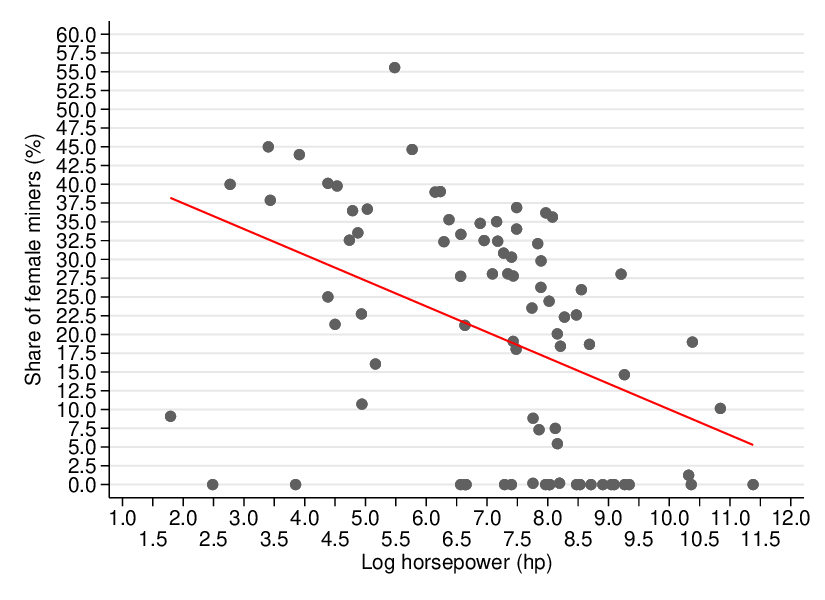}
\caption{Relationship between Share of Female Miners (\%) and Log-Horsepower (hp) in Coal Mines in Fukuoka}
\label{fig:scat_fminers}
\scriptsize{\begin{minipage}{400pt}
\setstretch{0.85}
Notes: This figure illustrates the raw relationship between the share of female miners and log-transformed horsepower in 82 coal mines in Fukuoka prefecture in 1930.
The share of female miners is the number of female miners per 100 total miners.
The mean and standard deviation of the share of female miners are $20.1$ and $15.6$, respectively.
The horsepower is the total horsepower of the machines using steam, electrical, or water power.
The mean and standard deviation of the log-transformed horsepower are $7.1$ and $2.0$, respectively.
The red line represents the fitted values for a linear regression model.
The Tobit estimate is reported in the text.
Although the Fukuoka Prefectural Statistics originally listed 90 coal mines, 8 were dropped due to missing horsepower data.
Source:
Created by the author.
Data are obtained from the official report published by Fukuoka Prefecture (1932).
\end{minipage}}
\end{figure}
%-------------

The history illustrated in Online Appendix~\ref{sec:seca0} suggests that female miners were more likely to be replaced via mechanization.
If this anecdote is accurate, mines that deployed more machines were less likely to rely on female miners.
To assess this point, I examine the relationship between the degree of mechanization and the use of female miners in coal mines.
Systematic data capturing mechanization in coal mines is unavailable.
Fortunately, the Fukuoka Prefectural Statistics document the horsepower and the number of miners in Fukuoka's coal mines, including the Chikuh\=o coalfield.
Figure~\ref{fig:scat_fminers} illustrates the raw relationship between the share of female miners and log-transformed horsepower in the 82 coal mines in the prefecture in 1930.
Overall, the figure shows a negative correlation between the two variables.
The estimate from the Tobit model, which accounts for the censored observations at zero in Figure~\ref{fig:scat_fminers}, is $-4.197$ and is statistically significant at the $1$\% level.
This estimate implies that a one-standard-deviation increase in log horsepower ($2.0$) is associated with a decline in the latent share of female miners of approximately $8.4$ percentage points, more than half of the standard deviation of the share of female miners ($15.6$).
This result is consistent with the view that mechanization replaced female miners in coal mines.

\subsection{Age Distribution of Female Miners}\label{sec:seca_age_dist}

%-------------
%Figure A5
\begin{figure}[htb]
\centering
\captionsetup{justification=centering,margin=1.5cm}
\includegraphics[width=0.5\textwidth]{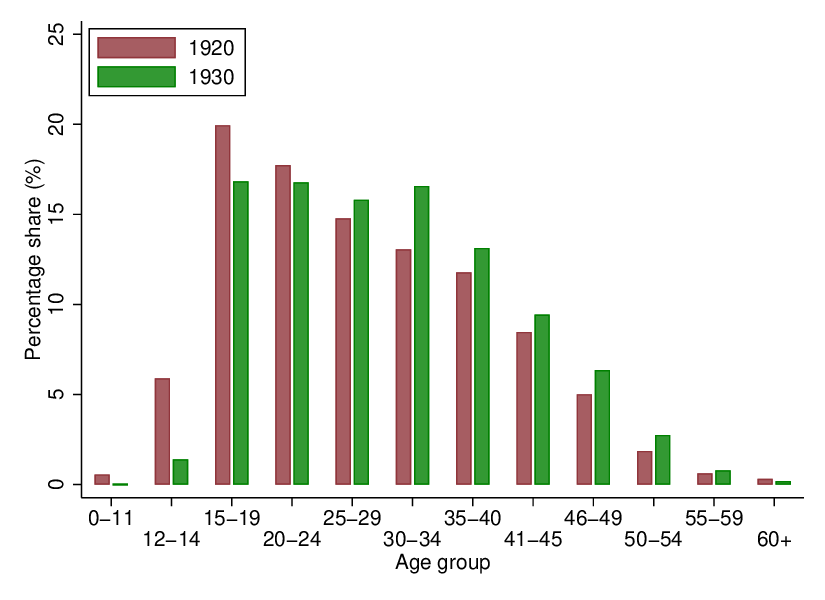}
\caption{Age Distribution of Female Workers in the Coal Mining Sector, as Measured in the 1920 and 1930 Population Censuses}
\label{fig:hist_age}
\scriptsize{\begin{minipage}{400pt}
\setstretch{0.85}
Notes:
This figure shows the age distribution of female workers in the mining sector, as recorded in the 1920 (red) and 1930 (green) Population Censuses. The percentage of female workers represents the number of female workers per 100 total workers. In the 1920 Population Census, individuals who reported their principal occupation (\textit{hongy\=o}) are classified as workers.
In the 1930 Population Census, individuals who reported their industry are classified as workers. I test the equality of the age distributions using the Pearson chi-squared statistic (Pearson 1900). The hypothesis of equal distribution is not rejected, with a $p$-value of $0.832$.
Data on occupations from the 1925 and 1935 Population Censuses are not available.
Sources:
Created by the author.
Data are obtained from the Statistics Bureau of the Cabinet (1929, pp.~206--207; 1935b, pp.~160--161).
\end{minipage}}
\end{figure}
%-------------

Figure~\ref{fig:hist_age} compares the age distributions of female workers in the coal mining sector as recorded in the 1920 and 1930 Population Censuses. The 1930 distribution shows a slight shift to the right, which may reflect the effect of the revised labor regulation established in 1928 that prohibited most in-pit work by female and child miners. However, the overall age distribution remains similar to that of the 1920 Population Census. 
The Pearson chi-squared test does not reject the null hypothesis of equal distribution, with a $p$-value of $0.832$.
Moreover, more than $70$\% of female miners were married in both the 1920 and 1930 censuses (Statistics Bureau of the Cabinet 1929, pp.~206--207; 1935b, pp.~160--161).
These results suggest that the composition of female miners changed little in the early stage of the regulation.
Because occupational statistics are unavailable in the 1935 Population Census, I cannot examine the composition after the full enforcement.

\subsection{Shares of Coal and GSC Mines}\label{sec:seca2}

%-------------
%Figure A6
\begin{figure}[htbp]
\centering
\captionsetup{justification=centering}
\subfloat[Coal and Oil Mines]{\label{fig:hist_fuel}\includegraphics[width=0.42\textwidth]{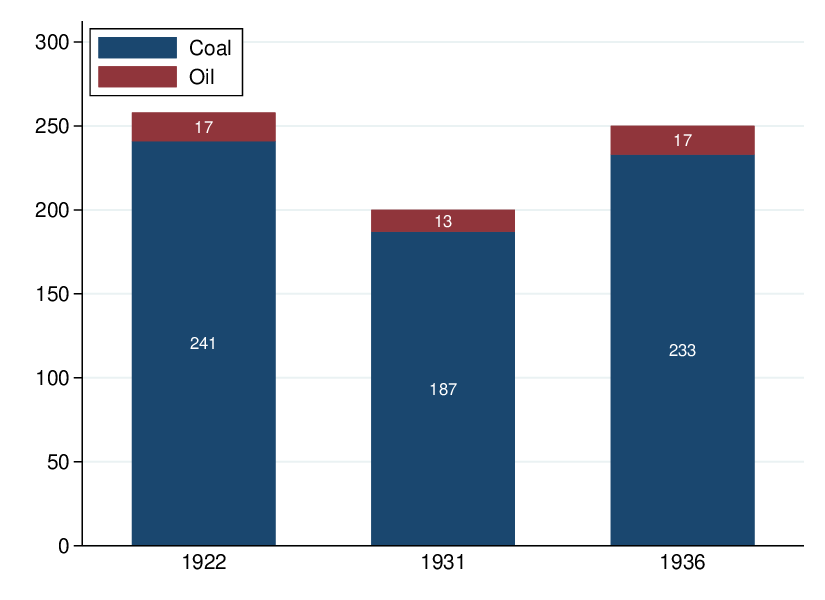}}
\subfloat[Metal Mines]{\label{fig:hist_metal}\includegraphics[width=0.42\textwidth]{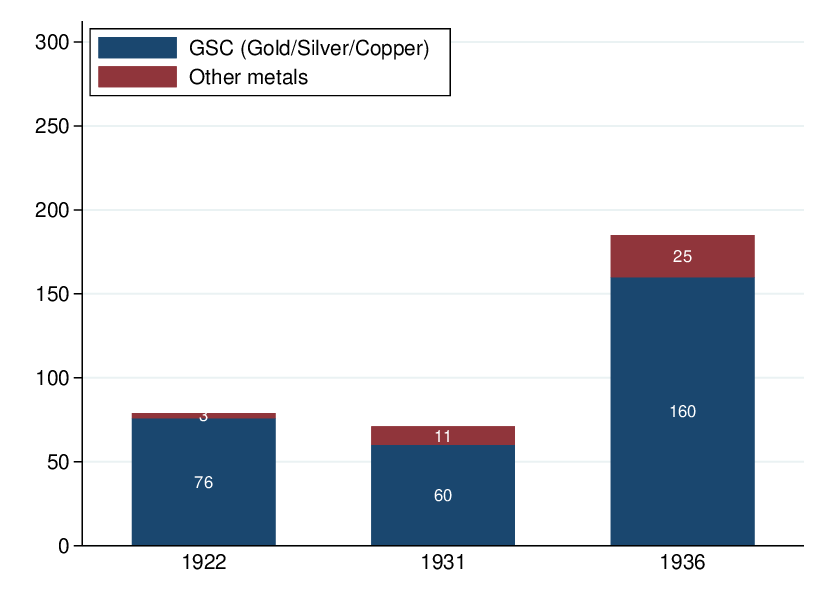}}
\caption{The Number of Coal, Oil, GSC, and Other Metal Mines\\ in 1922, 1931, and 1936}
\label{fig:hist_mines}
\scriptsize{\begin{minipage}{450pt}
\setstretch{0.9}
Notes:
This figure shows the number of mines in December 1922, October 1931, and October 1936 measured in the NDFM.
Figure~\ref{fig:hist_fuel} and Figure~\ref{fig:hist_metal} show the number of coal and oil mines and heavy metal mines, respectively.
GSC mines indicate heavy-metal mines that mine gold, silver, and/or copper.
A few smelting and oil refinery places are not included in the data.
Sources: Created by the author using the Cooperation Society (1924; 1932; 1937).
\end{minipage}}
\end{figure}
%-------------

Figure~\ref{fig:hist_fuel} shows the number of coal and oil mines documented in the NDFM.
Figure~\ref{fig:hist_metal} shows the numbers of GSC and other metal mines reported in the same document.
As shown, roughly 93 (85)\% of mines extracting fuels (metals) were coal (GSC) mines.
Note that Figure~\ref{fig:hist_mines} shows the number of mines, rather than the number of municipalities.
Since there were some cases where a few mines coexist in a municipality, the number of municipalities with mines reported in the main text is smaller than the number of mines.

%-------------------------------------------------------------------------------
% Appendix B
%-------------------------------------------------------------------------------
%\clearpage
\section{Conceptual Framework Appendix}\label{sec:secb}
\setcounter{figure}{0} \renewcommand{\thefigure}{B.\arabic{figure}}
\setcounter{table}{0} \renewcommand{\thetable}{B.\arabic{table}}

The regulations governing female miners decreased their relative wage rate (Section~\ref{sec:sec22}).
The purpose of this section is to provide a conceptual framework for how their fertility changes in response to the female relative wages.
I consider a simplified case of a collective model proposed by Siegel (2017) that illustrates optimal fertility based on the couple-household's optimization.

\subsubsection*{Setting}

Males and females derive utilities from consumption ($c$) and having children ($b$).
For a spouse $g \in \{m, f\}$, additively separable utility function is specified as follows:
\begin{eqnarray}\label{utility}
\footnotesize{
\begin{split}
u_{g} (c_{g}, b) = \log c_{g} + \tau \log b.
\end{split}
}
\end{eqnarray}
Home production required to raise children can be expressed as:
\begin{eqnarray}\label{hp_b}
\footnotesize{
\begin{split}
y(b) = \pi b,
\end{split}
}
\end{eqnarray}
where $\pi > 0$. The home production is accomplished using labor-saving technologies available in the markets ($s$) and parents' total home labor input ($\Omega$):
\begin{eqnarray}\label{hp_agents}
\footnotesize{
\begin{split}
y(b) = s^{\gamma} \Omega^{1-\gamma},
\end{split}
}
\end{eqnarray}
where $0 \leq \gamma < 1$ represents the availability of childcare services in the economy.
Let $\kappa_{g}$ be the productivity of housework including childcare, and $h_{g} \geq 0$ be the home labor input of each spouse.
The parents' total home labor input is then specified in the constant elasticity of substitution function as:
\begin{eqnarray}\label{hp_agents}
\footnotesize{
\begin{split}
\Omega = \left( \kappa_{m} h_{m}^{1-\rho} + \kappa_{f} h_{f}^{1-\rho} \right)^{\frac{1}{1-\rho}},
\end{split}
}
\end{eqnarray}
where $\rho > 0$ is the inverse of the elasticity of substitution.
A spouse earns wage $w_{g}$ for labor input $n_{g} \geq 0$.
The labor income is used for their consumption and for purchasing home labor-saving services.

\subsubsection*{Optimization}

The representative household maximizes the sum of utilities:
\begin{eqnarray}\label{max_prob}
\footnotesize{
\begin{split}
\theta u_{m} (c_{m}, b) + (1 - \theta) u_{f} (c_{f}, b)
\end{split}
}
\end{eqnarray}
subject to the budget and time constraints:
\begin{eqnarray}\label{constraints}
\footnotesize{
\begin{split}
c_{m} + c_{f} + s = w_{m} n_{m} + w_{f} n_{f},\\
n_{g} + h_{g} = 1~\text{for}~g \in \{m, f\}.
\end{split}
}
\end{eqnarray}
The optimal home labor inputs and fertility are given by:
\begin{eqnarray}\label{foc_hm}
\footnotesize{
\begin{split}
h_{m} = \left(\frac{\kappa_{m}}{w_{m}} \right)^{\frac{1}{\rho}}
 \left(\frac{1- \gamma}{\gamma} \right)^{\gamma}
 \{\kappa_{m}^{\frac{1}{\rho}} w_{m}^{\frac{1}{\rho-1}} + \kappa_{f}^{\frac{1}{\rho}} w_{f}^{\frac{1}{\rho-1}} \}^{\frac{\gamma \rho - 1}{1-\rho}}
  y_{b},
\end{split}
}
\end{eqnarray}
\begin{eqnarray}\label{foc_hf}
\footnotesize{
\begin{split}
h_{f} = \left(\frac{\kappa_{f}}{w_{f}} \right)^{\frac{1}{\rho}}
 \left(\frac{1- \gamma}{\gamma} \right)^{\gamma}
 \{\kappa_{m}^{\frac{1}{\rho}} w_{m}^{\frac{1}{\rho-1}} + \kappa_{f}^{\frac{1}{\rho}} w_{f}^{\frac{1}{\rho-1}} \}^{\frac{\gamma \rho - 1}{1-\rho}}
  y_{b},
\end{split}
}
\end{eqnarray}
\begin{eqnarray}\label{foc_b}
\footnotesize{
\begin{split}
b = \frac{\tau}{\pi}
 \left(\frac{c_{m}}{\theta} \right)
 \gamma^{\gamma} \left(1 - \gamma \right)^{1 - \gamma}
 \{\kappa_{m}^{\frac{1}{\rho}} w_{m}^{\frac{1}{\rho-1}} + \kappa_{f}^{\frac{1}{\rho}} w_{f}^{\frac{1}{\rho-1}} \}^{\frac{\rho (1-\gamma)}{1-\rho}}.
\end{split}
}
\end{eqnarray}
Equations~\ref{foc_hm} and~\ref{foc_hf} characterize the optimal ratio of home labor inputs as $\frac{h_{m}}{h_{f}}=\left( \frac{\kappa_{m}}{\kappa_{f}}\frac{w_{f}}{w_{m}} \right)^{\frac{1}{\rho}}$.
Substitute the optimal consumption and optimal time-allocation choice into budget constraints to yield:
\begin{eqnarray}\label{opt_constraint}
\footnotesize{
\begin{split}
\frac{c_{m}}{\theta} =
(w_{m} + w_{f}) -
\left( \frac{1}{\gamma} \right)^{\gamma}
\left( \frac{1}{1 - \gamma} \right)^{1 - \gamma}
 \{\kappa_{m}^{\frac{1}{\rho}} w_{m}^{\frac{1}{\rho-1}} + \kappa_{f}^{\frac{1}{\rho}} w_{f}^{\frac{1}{\rho-1}} \}^{\frac{\rho (1-\gamma)}{\rho-1}}y_{b}.
\end{split}
}
\end{eqnarray}
This implies that equation~\ref{foc_b} can be written in closed form:
\begin{eqnarray}\label{opt_fertility}
\footnotesize{
\begin{split}
b = \frac{\tau}{\pi (1 + \pi)}
 \gamma^{\gamma} \left(1 - \gamma \right)^{1 - \gamma}
 \left(w_{m} + w_{f} \right)
 \{\kappa_{m}^{\frac{1}{\rho}} w_{m}^{\frac{1}{\rho-1}} + \kappa_{f}^{\frac{1}{\rho}} w_{f}^{\frac{1}{\rho-1}} \}^{\frac{\rho (1-\gamma)}{1-\rho}}.
\end{split}
}
\end{eqnarray}
This is independent of the bargaining weight ($\theta$).
Finally, equation~\ref{opt_fertility} suggests that $\frac{\partial b}{\partial w_{f}}$ becomes negative under following condition:
\begin{eqnarray}\label{opt_fertility_sign}
\footnotesize{
\begin{split}
\gamma \kappa_{f}^\frac{1}{\rho} w_{f}^\frac{\rho-1}{\rho}
+ w_{m} \left[\left(\frac{\kappa_{m}}{w_{m}} \right)^{\frac{1}{\rho}} -  (1-\gamma)\left(\frac{\kappa_{f}}{w_{f}} \right)^{\frac{1}{\rho}} \right] < 0.
\end{split}
}
\end{eqnarray}

\subsubsection*{Prediction}

The inequality~\ref{opt_fertility_sign} indicates that $\frac{\partial b}{\partial w_{f}}$ is negative under the following condition:
\begin{equation}\label{opt_fertility_sign_mod}
\footnotesize{
\frac{w_{f}}{w_{m}} < \frac{1-\gamma}{\gamma} - \frac{1}{\gamma}\left(\frac{\kappa_{m}}{\kappa_{f}}\frac{w_{f}}{w_{m}}\right)^{\frac{1}{\rho}}.
}
\end{equation}
Since prewar Japan was a paternalistic society under the Old Civil Code, the most plausible setting for housework productivity at that time is $\kappa_{m} \simeq 0$, so that the second term on the right-hand side is negligible.
The condition then reduces to $w_{f}/w_{m} < (1-\gamma)/\gamma$.
Because the female-to-male wage ratio in coal mining was below one throughout the period (Table~1), the condition holds whenever $\gamma \leq 0.5$.

%------
%Table B1
%\begin{landscape}
\begin{table}[htbp]
\def\arraystretch{1.0}
\begin{center}
\captionsetup{justification=centering}
\caption{Nursery at Yamauchi Coal Mine in 1915}
\label{tab:nursery_1915}
\footnotesize
\scalebox{1.0}[1]{
{\setlength\doublerulesep{2pt}
\begin{tabular}{ccccc}
\toprule[1pt]\midrule[0.3pt]
\multicolumn{1}{c}{(a) Ave. years of}
&\multicolumn{1}{c}{(b) Ave. years of}
&\multicolumn{1}{c}{(c) Ave. days using}
&\multicolumn{1}{c}{(d) Ave. days}
&\multicolumn{1}{c}{(e) Utilization}\\
\multicolumn{1}{c}{service of parent}
&\multicolumn{1}{c}{use of nursery ($\bar{p}_{c}$)}
&\multicolumn{1}{c}{nursary ($\bar{n}_{c}$)}
&\multicolumn{1}{c}{of operation}
&\multicolumn{1}{c}{rate (\%)}\\\hline
2.4&1.3&85.2&320&20.0\\
(1.3)&(1.0)&(57.5)&&\\\midrule[0.3pt]\bottomrule[1pt]
\end{tabular}
}
}
{\scriptsize
\begin{minipage}{420pt}
\setstretch{0.85}
Notes: Columns (a) to (c) are from the survey report on $88$ children using the nursery at Yamauchi Coal Mine in Chikuh\=o coalfield in 1915.
The figures in parentheses are the median values.
Column (d) shows the average number of operating days of coal mines in 1924, surveyed in the mining census.
The average number of operating days is calculated as 365 minus the average number of days closed ($45.5$).
Column (e) shows the utilization rate, which is defined as:
$100\left(\sum_{c=1}^{88} n_{c}\right)/\left(\sum_{c=1}^{88} p_{c} \times (365-45.5)\right)$.
\\
I excluded several individuals whose statistics were unavailable or unreliable.
Three individuals whose length of service exceeded the number of years they had used the nursery are excluded.
Six individuals had more days in care than days in operation.
These individuals are also excluded because of suspected listing and transcription errors.
Finally, I excluded three individuals for whom statistics on childcare days were unavailable.
Consequently, this table uses statistics for the 88 individuals among the $100$ originally listed.
Sources: Noyori 2010, pp.~80--82.
The average number of days closed is obtained from the Statistics Bureau of the Cabinet 1926, p.~18.
\end{minipage}
}
\end{center}
\end{table}
%\end{landscape}
%------

To what extent were childcare services available to female miners at that time?
As described in Section~\ref{sec:sec2}, female miners could use the nurseries attached to some coal mines.
Fortunately, Noyori (2010) provides a comprehensive study of nurseries in coal mines, drawing on a unique archive of nursery surveys from a few mines in the Chikuh\=o coalfield.
According to the nursery's regulations, employees can use the nursery during working hours (Noyori 2010, p.~78).
However, evidence shows that children over 5 years old were unlikely to use the nurseries in mines (Noyori 2010, pp.~76--77).
Moreover, the number of days using the nursery was quite small.
Table~\ref{tab:nursery_1915} summarizes the utilization of the nursery based on the survey report on 88 children at the Yamauchi Coal Mine in 1915.
Column (a) shows that mothers' average number of working years is 2.4, whereas their average years of nursery use is 1.3.
Column (c) indicates that the average number of annual days used was roughly 85 days (median is $57.5$ days), whereas the average number of annual operating days available from the mining census is 320 (Column (d)).
The years spent in daycare ($p_{c}$) can be considered the number of years the mother worked after the child was born.
Thus, one can calculate the utilization rate for each child ($c$) as follows:
\begin{eqnarray}\label{utilization}
\footnotesize{
\begin{split}
\frac{\sum_{c=1}^{88} n_{c}}{\sum_{c=1}^{88} p_{c} \times 320} \times 100,
\end{split}
}
\end{eqnarray}
where $n_{c}$ is the number of days using the nursery.
Column (e) indicates that this rate is estimated to be $20$\%.
Note that several mothers used the nursery for more than 200 days, indicating that the nursery must have been open on all operating days.
This rejects the possibility that the nursery could not be opened due to a lack of staff.
Despite this, it is possible that miners might not work fully on operating days.
Even if one assumes that miners worked only one-third of the operating days (i.e., $107$ days), the utilization rate is still $60$\%.

In addition, childcare fees were usually deducted from miners' paychecks (Noyori 2010, p.~109).
This implies that some miners who could not use the nursery may use alternative childcare institutions.
Another possible informal institution may be hiring babysitting labor (\textit{komori}).
Noyori (2010, pp.~47--48) shows that these babysitters were usually children from other households.
The young babysitter was expected to carry the infant on her back up to the mine entrance.
The mother would then climb up and down multiple times like a ``mouse'' from the pits to the entrance to feed her baby.
This anecdotal evidence suggests that the quality of babysitting was not high enough to compensate for the costs of childrearing.

Overall, while the nursery was a useful institution for female miners, it replaced only a small part of their childcare work.
Although babysitting was available, it offered an incomplete service, covering only a limited portion of mothers' childcare needs.
This meant that the marketization parameter $\gamma$ would be considerably low in inequality~\ref{opt_fertility_sign_mod}.
Therefore, optimal fertility should be negatively correlated with the relative wage ($w_{f}/w_{m}$).
This predicts the relationship that the fertility rate would be lower before the female labor regulation but could be higher after the regulation. This is because the relative wage declined substantially after the prohibition of female miners' work in the pits (Section~\ref{sec:sec22}).

%-------------------------------------------------------------------------------
% Appendix C
%-------------------------------------------------------------------------------
%\clearpage
\section{Data Appendix}\label{sec:secc}
\setcounter{figure}{0} \renewcommand{\thefigure}{C.\arabic{figure}}
\setcounter{table}{0} \renewcommand{\thetable}{C.\arabic{table}}

\subsection{Mine Deposit}\label{sec:secc1}

Data on the location and type of mines are obtained from the National Directory of Factories and Mines (\textit{Zenkoku K\=ojyo K\=ozan Meibo}) published by the Cooperation Society (1924, 1932, 1937).
They document all mines with more than $50$ miners in December 1922, October 1931, and October 1936.
The documents list three types of mines: metal, coal, and oil.\footnote{The metal mine includes minerals such as gold, silver, copper, blister copper, pig iron, steel, iron sulfide, lead, zinc, bismuth, arsenious acid, tin, mercury, chromium, manganese, and sulfide minerals. A smelting mine is also included in the metal mine category. Coal mines include a few mines mining lignite. Oil mines include oil, crude oil, gas, volatile oil, kerosene, light oil, machine oil, and heavy oil.}
In these documents, the mine's location is indicated by the municipality's name.
Hence, the municipality name was used during geocoding to match the mine dataset with the other variable datasets.
The shapefile used for geocoding was obtained from the official database of the Ministry of Land, Infrastructure, Transport and Tourism (available at: \url{https://nlftp.mlit.go.jp/ksj/gml/datalist/KsjTmplt-N03-v2_2.html} (accessed on 23rd February 2021).

Several municipalities were consolidated during the sample period.
Thus, the municipality boundaries defined in 1938 were used, and aggregate statistics for consolidated municipalities were calculated.
A minimal number of municipalities were divided; these were excluded from the original dataset.
To check all consolidations and divisions of municipalities, \textit{Zenkoku Shich\=osonmei Hensen S\=oran} (Municipal Autonomy Research Association 2006), was used because it is the most comprehensive book on the history of municipalities' names.

\begin{figure}[htbp]
\centering
\captionsetup{justification=centering}
\subfloat[December 1922]{\label{fig:mine_1922}\includegraphics[width=0.35\textwidth]{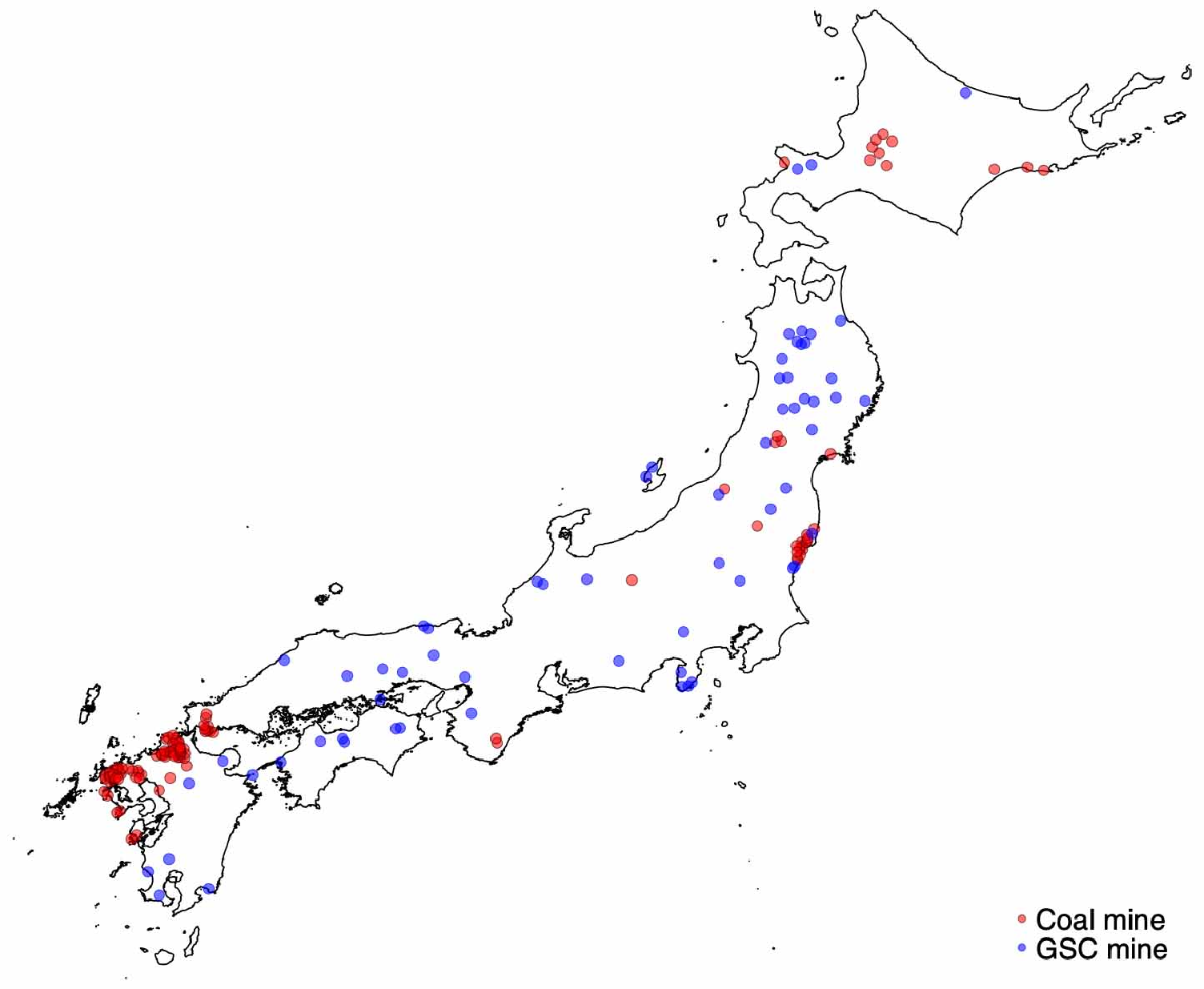}}
\subfloat[October 1931]{\label{fig:mine_1931}\includegraphics[width=0.35\textwidth]{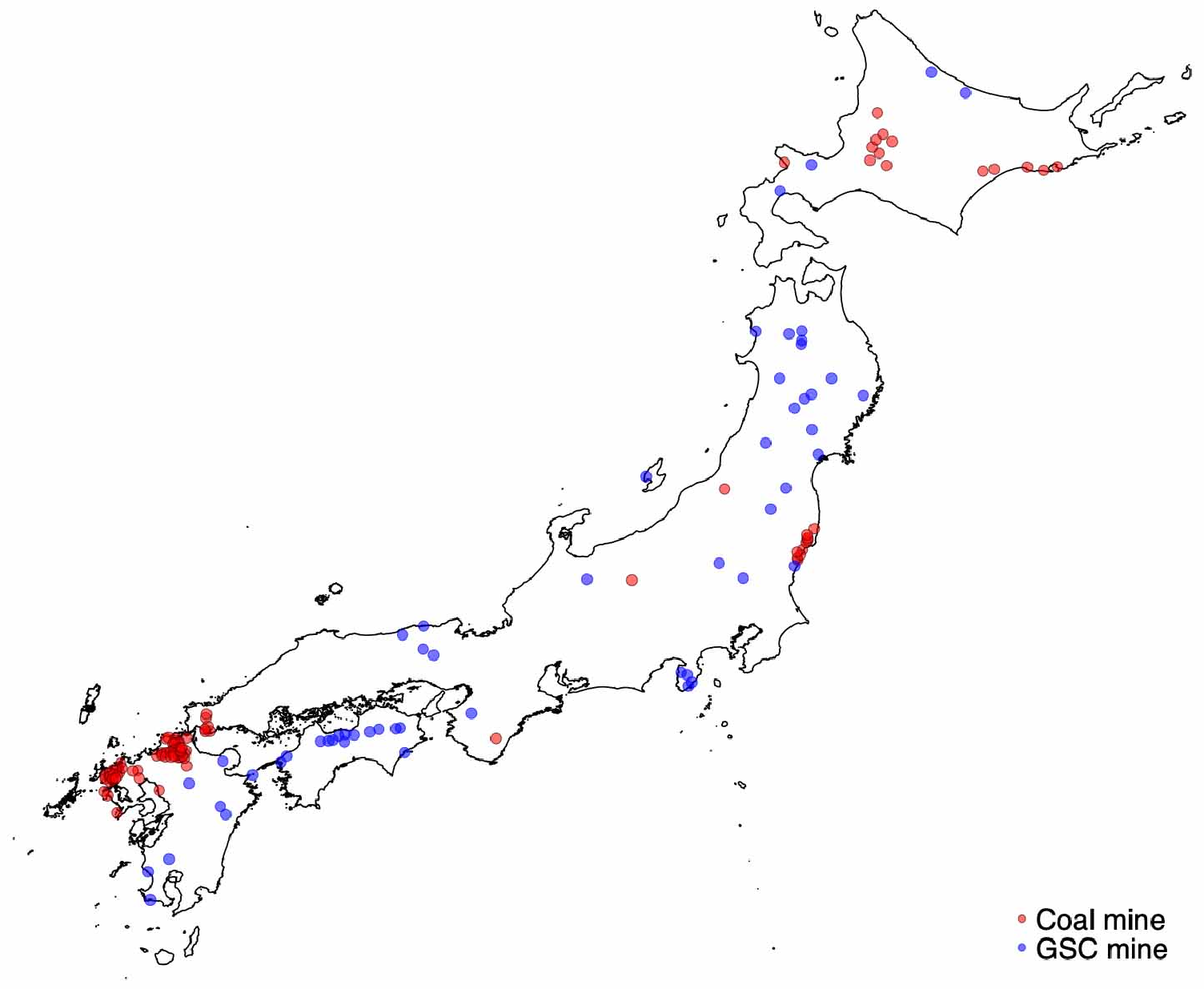}}
\subfloat[October 1936]{\label{fig:mine_1936}\includegraphics[width=0.35\textwidth]{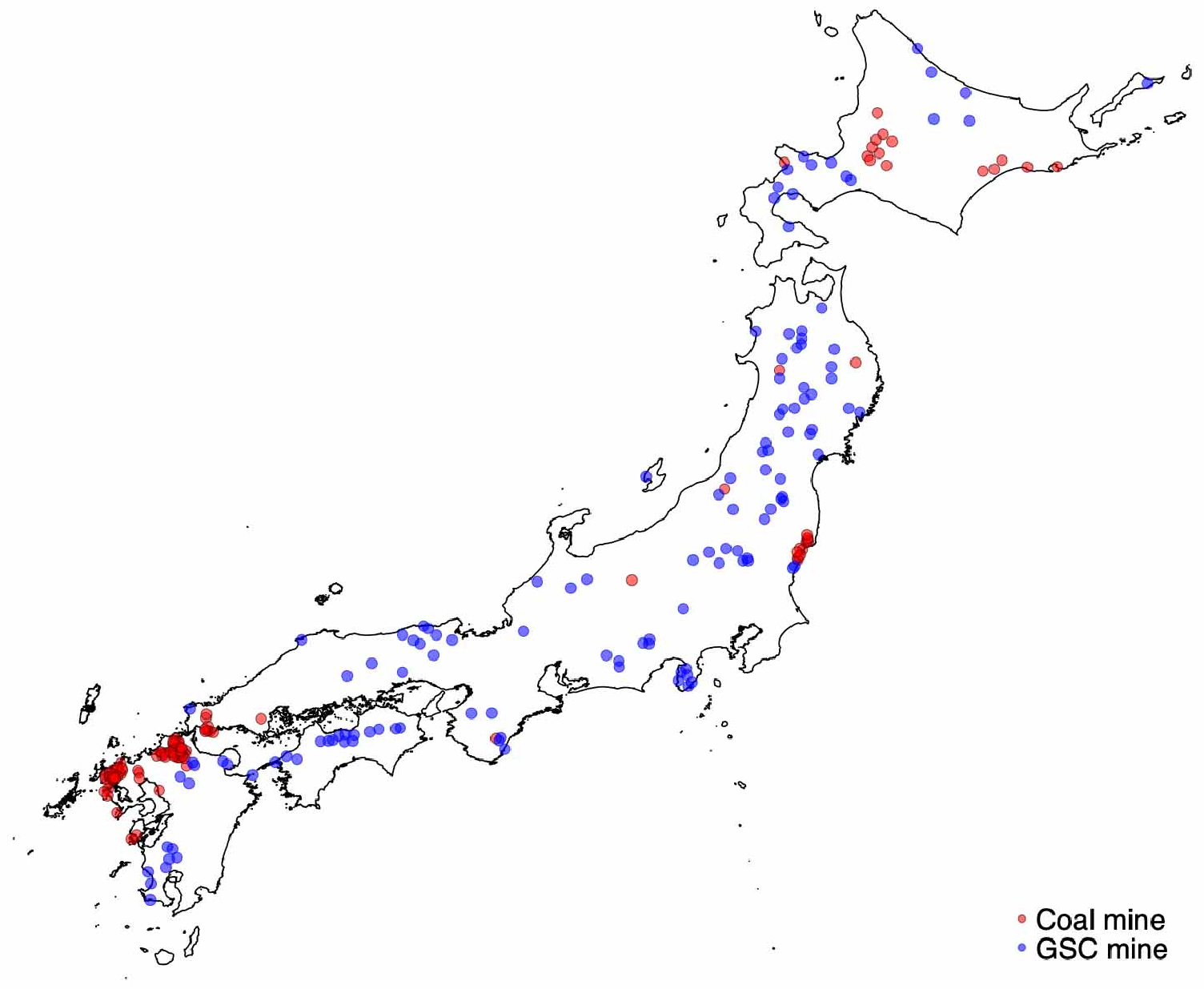}}
\caption{Location of Coal and GSC mines in 1922 and 1936}
\label{fig:map_mines}
\scriptsize{\begin{minipage}{450pt}
\setstretch{0.9}
Notes:
This figure illustrates the location of mines measured in the NDFM.
The red and blue circles indicate the locations of coal and GSC (gold, silver, and copper) mines, respectively.\\
Source: Created by the author.
Data on the location of mines are from the Cooperation Society (1932, 1937).
\end{minipage}}
\end{figure}

The initial dataset included both coal and heavy metal mines.
The GSC mines were the most representative metal mines of their time (Online Appendix~\ref{sec:secd_hm}).
Generally, the GSC mines extracted combinations of these two or three metals.
The other metal mines, such as zinc and tin, were few and negligible in scale and number of miners.
To prepare the analytical samples, I therefore retained only the coal and the GSC mines.
Figure~\ref{fig:map_mines} illustrates the location of coal and GSC mines considered.

%----------
%Figure C2
%----------

Table~\ref{tab:dup} summarizes how my analytical samples are obtained from all $11,142$ municipalities based upon the 1938 boundaries.
I first retained the municipalities within 30km of a mine.
The duplications were then excluded to better identify the impact of coal mines.
For example, among $1,416$ municipalities within 30km of coal mines, there are $14$ duplications with the municipalities within 5km of GSC mines (Column (1) in Panel A-1).
Thus, the 1925 census sample finally includes $1,402$ municipalities.
Figure~\ref{fig:map_coal_1922} illustrates the spatial distribution of this 1925 census sample.
The same trimming methods were applied for the other cross-sectional samples as listed in Columns (2) and (3).

Panel B of the same table summarizes similar but slightly different trimming steps used to create the ever- and never-treated panel data samples.
I first retained the mines that were consistently observed in the sample years.
I then excluded duplicates from the other mines that were not consistently measured in the sample years.
This trimming created some unbalanced units, so I finally balanced the panel by removing them.
For example, to create the 1925-1935 panel data sample, I first retained the coal mines measured in all years.
I then excluded the municipalities within a $5$ km radius of the other coal mines and GSC mines ($80$ and $42$ observations, respectively, out of $3,369$ observations).
Finally, I trimmed the unbalanced municipalities due to duplication, yielding a final $3,117$ observations.
The sample methods were applied to the 1933-1938 vital statistics panel data sample in Column (2).

%----------
%Table C1
\begin{landscape}
\begin{table}[htbp]
\def\arraystretch{1.0}
\begin{center}
\captionsetup{justification=centering}
\caption{Generating Process of the Analytical Sample}
\label{tab:dup}
\scriptsize
\scalebox{1.0}[1]{
{\setlength\doublerulesep{2pt}
\begin{tabular}{lrrr}
\toprule[1pt]\midrule[0.3pt]
&\multicolumn{3}{c}{Analytical Sample}\\
\cmidrule(rrr){2-4}
&(1) 1925 Census	&(2) 1930 Census /	&(3) 1935 Census / \\
\textbf{Panel A: Cross-Sectional Sample}&	&1933 Vital Statistics	&1938 Vital Statistics\\\hline
&&&\\
\hspace{10pt}\textbf{Panel A-1: Coal Mine Sample} &&&\\
\hspace{10pt}Number of municipalities				&11,142	&11,142	&11,142   \\
\hspace{20pt}Municipalities within 30km of coal mines	&1,416	&1,142	&1,316        \\
\hspace{30pt}Municipalities within 5 km of GCS mines	&14		&4		&28\\
\hspace{40pt}Number of municipalities in the final sample	&1,402	&1,142	&1,316\\
\hspace{50pt}Municipalities in the TG				&197		&166		&169\\
&&&\\
\hspace{10pt}\textbf{Panel A-2: GSC Mine Sample}		&&&\\
\hspace{10pt}Number of municipalities				&11,142	&11,142	&11,142	\\
\hspace{20pt}Number of municipalities within 30km of coal mines	&3,280	&2,281	&3,973	\\
\hspace{30pt}Number of municipalities within 5 km of GCS mines	&34		&11		&37		\\
\hspace{40pt}Number of municipalities in the final sample	&3,246	&2,270	&3,936	\\
\hspace{50pt}Municipalities in the TG				&131		&109		&225\\
&&&\\
&&\multicolumn{2}{c}{Analytical Sample}\\
\cmidrule(rr){3-4}
\textbf{Panel B: Panel Data Sample}		&&(1) 1925-1935 Census	&(2) 1933-1938 Vital Statistics\\\hline
&&&\\
\hspace{10pt}\textbf{Panel B-1: Coal Mine Sample} &&&\\
\hspace{10pt}Number of observations								&&33,426	&22,284\\
\hspace{20pt}Panels within 30km of ever-treated coal mines				&&3,369	&2,276\\
\hspace{30pt}[A] Observations within 5km radius from the other coal mines	&&80	&21\\
\hspace{30pt}[B] Observations within 5km radius from the GSC mines		&&42	&30\\
\hspace{30pt}Unbalanced observations due to [A] and [B]					&&130	&43\\
\hspace{40pt}Final balanced panels			 						&&3,117	&2,182\\
\hspace{50pt}Ever-treated panels									&&426	&306\\
&&&\\
\hspace{10pt}\textbf{Panel B-2: GSC Mine Sample}						&&&\\
\hspace{10pt}Number of observations								&&33,426	&22,284\\
\hspace{20pt}Panels within 30km of ever-treated coal mines				&&4,914	&3,874\\
\hspace{30pt}[A] Observations within 5km radius from the other GSC mines	&&102	&69\\
\hspace{30pt}[B] Observations within 5km radius from the coal mines		&&41	&23\\
\hspace{30pt}Unbalanced observations due to [A] and [B]					&&157	&66\\
\hspace{40pt}Final balanced panels			 						&&4,614	&3,716\\
\hspace{50pt}Ever-treated panels									&&171	&166\\
\midrule[0.3pt]\bottomrule[1pt]
\end{tabular}
}
}
{\scriptsize
\begin{minipage}{550pt}
\setstretch{0.85}
Notes:
Panels A-1 and A-2 show the trimming steps of the cross-sectional coal and GSC mine samples, respectively.
Columns (1), (2), and (3) list the 1925 census sample, 1930 census and 1933 vital statistics samples, and 1935 and 1938 vital statistics samples, respectively.
The spatial distributions of these samples are illustrated in Figures~\ref{fig:map_coal_1922},~\ref{fig:map_coal_1931}, and~\ref{fig:map_coal_1936}.
Panels B-1 and B-2 show the trimming steps of the panel-data coal and GSC mine samples, respectively.
Columns (1) and (2) list the 1925-1935 census sample and 1933-1938 vital statistics sample, respectively.
Source: See Online Appendix~\ref{sec:secc1}.
\end{minipage}
}
\end{center}
\end{table}
\end{landscape}
%----------

\subsection{Geological Stratum}\label{sec:secc2}

The location information on geological strata (\textit{chis\=o}) obtained from the official database of the Ministry of Land, Infrastructure, Transport and Tourism was used (available at: \url{http://nrb-www.mlit.go.jp/kokjo/inspect/landclassification/download/index.html}, accessed on 27th February 2020).

%----------
%Figure C2
\begin{figure}[htbp]
\centering
\captionsetup{justification=centering,margin=1.5cm}
\includegraphics[width=0.5\textwidth]{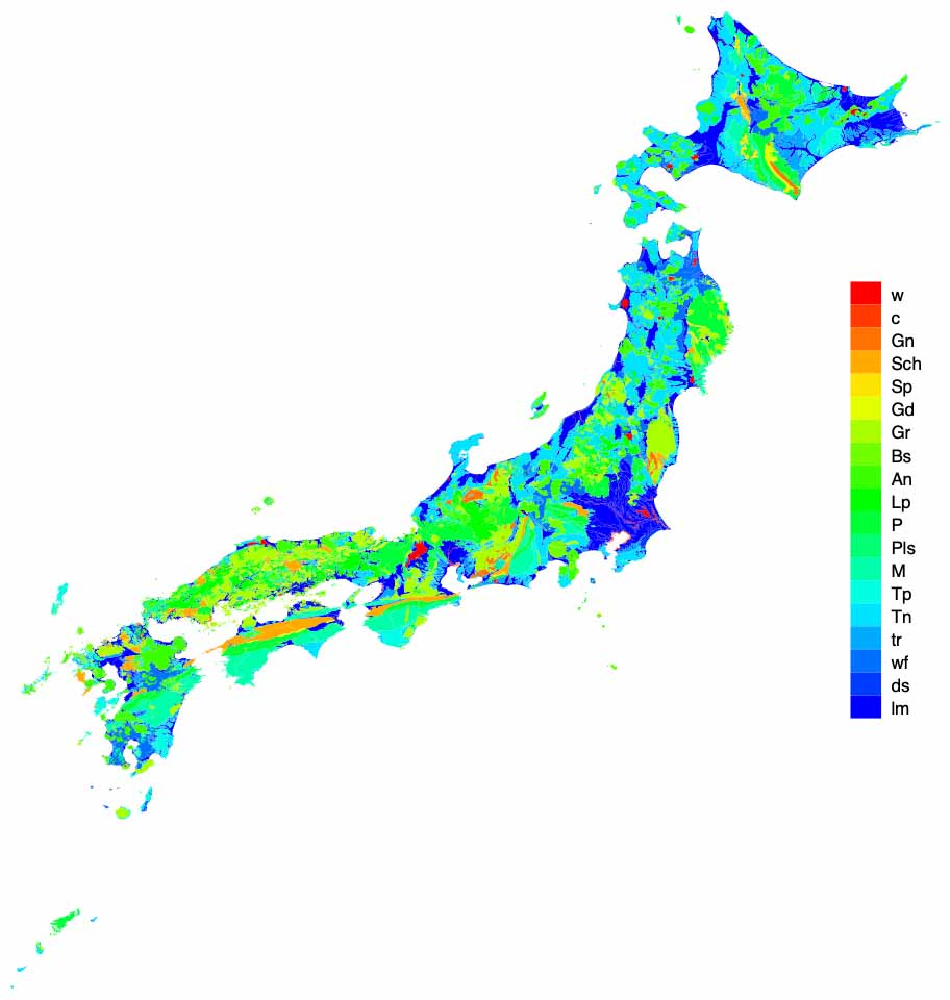}
\caption{Spatial Distribution of Geological Strata}
\label{fig:map_mine}
\scriptsize{\begin{minipage}{400pt}
\setstretch{0.85}
Notes:
This figure illustrates the locations of strata measured in the soil-boring investigations conducted by the Ministry of Land, Infrastructure, Transport and Tourism.
Sources:
Created by the author.
Data on the location of mines are obtained from the Cooperation Society (1932, 1937).
Data on the geological stratum are obtained from the Ministry of Land, Infrastructure, Transport and Tourism.
\end{minipage}}
\end{figure}
%----------

\begin{figure}[htbp]
\centering
\captionsetup{justification=centering,margin=1.5cm}
\includegraphics[width=0.5\textwidth]{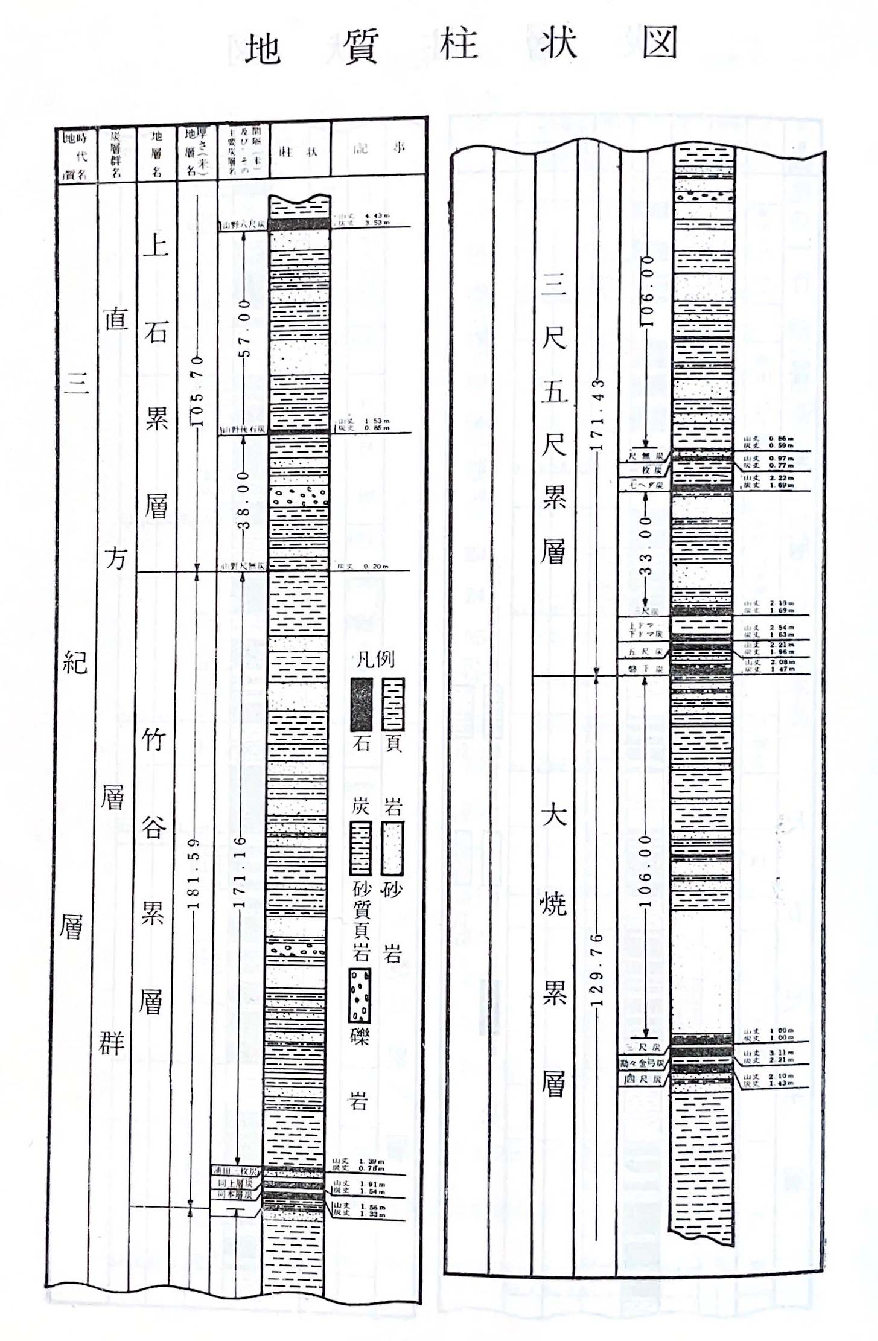}
\caption{Geological Columnar Section of the Iizuka Coal Mine}
\label{fig:columnar}
\scriptsize{\begin{minipage}{400pt}
\setstretch{0.85}
Notes:
This geological columnar section shows the strata around the Iizuka coal mine in the Chikuh\=o coalfield in the Ky\=ush\=u region.
The first through third columns list the names of layers formed during the Paleogene period.
The fourth to seventh columns show the lengths of each layer, the column, and the legend.
The black layers shown in the sixth column are coal layers: more than ten coal layers are included in the layers formed during the Paleogene period.
Source:
Fumoto (1961, p.~\textit{Chishitsu ch\=ujy\=o zu}).
\end{minipage}}
\end{figure}

First, each municipality was matched with the nearest stratum point ($11,142$ municipalities with $8,748$ stratum points).
The median distance from the municipality's centroid to the nearest stratum point is $3.38$ km.
Figure~\ref{fig:map_mine} illustrates the spatial distributions of the geological strata.
Second, I define the instrument based on the geology of Japanese coalfields.
Most of Japan's major coalfields, including Chikuh\=o, Miike, Ishikari, and J\=oban, lie in coal-bearing formations of the Paleogene period (\textit{ko daisan ki}) (Nagao 1975; Tanaka 1984).
I therefore use an indicator for the Tp stratum, which consists of sedimentary rocks formed during the Paleogene period of the Cenozoic era (\textit{shinseidai}).
Figure~\ref{fig:columnar} shows a geological columnar section of the Iizuka coal mine, a representative coal mine in the Chikuh\=o coalfield, which includes more than ten coal layers formed during the Paleogene period.

To confirm the relevance of this choice, I regress the indicator for the treatment group on indicators for each of the 19 types of strata, which consist of four categories of rock (and field) and six categories of era.
Table~\ref{tab:r_iv_def} shows that Tp has the largest and most precisely estimated coefficient in all three years.
The only other stratum with a significant positive coefficient is Tn, a sedimentary rock formed in the later Neogene period (\textit{shin daisan ki}), which includes some coal-bearing formations.
This pattern is consistent with the geological evidence that Japan's coal deposits are concentrated in the Paleogene formations.
Table~\ref{tab:r_iv_def} shows the summary statistics for the mine and stratum variables by type of mine and NDFM year.

%------------------
%Table C2
\begin{landscape}
\begin{table}[htbp]
\def\arraystretch{1.0}
\begin{center}
\captionsetup{justification=centering}
\caption{Defining Instrumental Variables for the Coal Mine Subsample}
\label{tab:r_iv_def}
\footnotesize
\scalebox{0.95}[1]{
{\setlength\doublerulesep{2pt}
\begin{tabular}{lD{.}{.}{-2}D{.}{.}{-2}D{.}{.}{-2}D{.}{.}{-2}D{.}{.}{-2}D{.}{.}{-2}D{.}{.}{-2}D{.}{.}{-2}D{.}{.}{-2}}
\toprule[1pt]\midrule[0.3pt]
&\multicolumn{3}{c}{December 1922}&\multicolumn{3}{c}{October 1931}&\multicolumn{3}{c}{October 1936}\\
\cmidrule(rrr){2-4}\cmidrule(rr){5-7}\cmidrule(rr){8-10}
Stratum
&\multicolumn{1}{c}{Coeff.}&\multicolumn{1}{c}{Std. Err.}&\multicolumn{1}{c}{$R$-squared}
&\multicolumn{1}{c}{Coeff.}&\multicolumn{1}{c}{Std. Err.}&\multicolumn{1}{c}{$R$-squared}
&\multicolumn{1}{c}{Coeff.}&\multicolumn{1}{c}{Std. Err.}&\multicolumn{1}{c}{$R$-squared}\\\hline

\hspace{10pt}Sedimentary rock    &&&&&&&&&\\
\hspace{20pt}M		&-0.001		&(0.050)&0.000		&0.069		&(0.073)    &0.001    &0.023		&(0.051)        &0.000\\
\hspace{20pt}P		&-0.030		&(0.077)&0.000		&0.037		&(0.122)    &0.000    &-0.026		&(0.058)        &0.000\\
\hspace{20pt}Pls	&0.026		&(0.167)&0.000		&0.021		&(0.167)    &0.000    &0.038		&(0.167)        &0.000\\
\hspace{20pt}Tn	&0.107$***$	&(0.034)&0.010		&0.135$***$	&(0.040)    &0.015    &0.179$***$	&(0.042)        &0.024\\\hdashline
\hspace{20pt}Tp	&0.284$***$	&(0.046)&0.052		&0.324$***$	&(0.049)    &0.072    &0.309$***$	&(0.046)        &0.069\\\hdashline
\hspace{20pt}al		&-0.061$***$	&(0.023)&0.004		&-0.035		&(0.031)    &0.001    &-0.036		&(0.025)        &0.001\\
\hspace{20pt}ds	&-0.141$***$	&(0.009)&0.001		&-0.146$***$	&(0.010)    &0.001    &-0.130$***$	&(0.009)        &0.001\\
\hspace{20pt}lm	&0.026		&(0.167)&0.000		&-0.003		&(0.143)    &0.000    &0.038		&(0.167)        &0.000\\
\hspace{20pt}tr		&-0.067$***$	&(0.020)&0.006		&-0.104$***$	&(0.020)    &0.015    &-0.075$***$	&(0.019)        &0.008\\
\hspace{20pt}wf	&-0.098$***$	&(0.033)&0.002		&-0.121$***$	&(0.030)    &0.004    &-0.107$***$	&(0.026)        &0.003\\
\hspace{10pt}Metamorphic rock    &&&&&&&&&\\
\hspace{20pt}Gn	&-0.141$***$	&(0.009)&0.000		&-0.146$***$	&(0.010)    &0.000    &-0.131$***$	&(0.009)        &0.003\\
\hspace{20pt}Sch	&0.007		&(0.047)&0.000		&-0.013		&(0.046)    &0.000    &-0.022		&(0.040)        &0.000\\
\hspace{10pt}Body of water        &&&&&&&&&\\
\hspace{20pt}w		&0.060		&(0.092)&0.000		&-0.146$***$    &(0.010)    &0.001    &-0.129$***$    &(0.009)        &0.001\\
\hspace{10pt}Igneous rock        &&&&&&&&&\\
\hspace{20pt}An	&-0.093$***$	&(0.024)&0.005		&-0.107$***$	&(0.025)    &0.006    &-0.104$***$	&(0.021)        &0.007\\
\hspace{20pt}Bs	&0.006		&(0.052)&0.000		&-0.030		&(0.051)    &0.000    &-0.020		&(0.047)        &0.000\\
\hspace{20pt}Gd	&0.128		&(0.119)&0.001		&0.135		&(0.109)    &0.002    &0.108		&(0.106)        &0.001\\
\hspace{20pt}Gr	&-0.024		&(0.025)&0.001		&-0.058$**$	&(0.025)    &0.004    &-0.054$***$	&(0.021)        &0.004\\
\hspace{20pt}Lp	&-0.145$***$	&(0.010)&0.005		&-0.148$***$	&(0.011)    &0.004    &-0.132$***$	&(0.009)        &0.004\\
\hspace{20pt}Sp	&0.060		&(0.200)&0.000		&0.055		&(0.200)    &0.000    &0.015		&(0.143)        &0.000\\\hline
Observations	&\multicolumn{3}{c}{1,402}&\multicolumn{3}{c}{1,142}&\multicolumn{3}{c}{1,316}\\\midrule[0.3pt]\bottomrule[1pt]
\end{tabular}
}
}
{\scriptsize
\begin{minipage}{630pt}
\setstretch{0.85}
***, **, and * represent statistical significance at the 1\%, 5\%, and 10\% levels, respectively.
Standard errors based on the heteroskedasticity-robust covariance matrix estimator (Horn et al.~1975) are reported in parentheses.\\
Notes:
This table summarizes the results of the linear probability models for the $19$ strata types.
The dependent variable is an indicator variable for the treated municipalities (i.e., those within 5km of coal mines), and the independent variable is an indicator variable for each stratum.
This means that $19 \times 3 = 57$ regressions were run in total.
Source: Created by author.
\end{minipage}
}
\end{center}
\end{table}
\end{landscape}
%------------------
%Figure C3
%------------------

%------------------
%Table C3
%\begin{landscape}
\begin{table}[htbp]
\def\arraystretch{0.95}
\begin{center}
\caption{Summary Statistics: Mine Variables and IV}
\label{tab:sum_key}
\scriptsize
\scalebox{1.0}[1]{
{\setlength\doublerulesep{2pt}
\begin{tabular}{lcD{.}{.}{2}D{.}{.}{2}D{.}{.}{2}D{.}{.}{2}D{.}{.}{2}D{.}{.}{2}D{.}{.}{2}}
\toprule[1pt]\midrule[0.3pt]
&Edition of&\multicolumn{3}{c}{Coal Mines}&\multicolumn{3}{c}{GSC Mines}\\
\cmidrule(rrr){3-5}\cmidrule(rrr){6-8}
Mine Variables and IV&the NDFM
&\multicolumn{1}{c}{Mean}&\multicolumn{1}{c}{Std. Dev.}&\multicolumn{1}{c}{Obs.}
&\multicolumn{1}{c}{Mean}&\multicolumn{1}{c}{Std. Dev.}&\multicolumn{1}{c}{Obs.}\\\hline
Mine Deposit	&1922        &0.141        &0.348        &1402    &0.040        &0.197        &3246    \\
			&1931        &0.145        &0.353        &1142    &0.048        &0.214        &2270    \\
			&1936        &0.128        &0.335        &1316    &0.057        &0.232        &3936    \\
Female Miners	&1922        &598.16 	   &932.48    &110    &167.59        &234.36        &64    \\
			&1931        &210.12    &271.15    &91    &117.36        &203.92        &56    \\
			&1936        &140.03    &179.14    &110    &82.19        &151.56        &129    \\
Male Miners	&1922        &1641.23    &2407.85    &110    &1269.11    &2196.83    &64    \\	
			&1931        &1374.43    &1724.95    &91    &1205.66    &2358.63    &56    \\
			&1936        &1385.49    &1808.70    &110    &893.47    &1794.48    &129    \\
Stratum		&(1922)   	 &0.086        &0.280        &1402    &$--$        &$--$        &$--$    \\
			&(1931)   	 &0.094        &0.292        &1142    &$--$        &$--$        &$--$    \\
			&(1936)   	 &0.089        &0.285        &1316    &$--$        &$--$        &$--$    \\\midrule[0.3pt]\bottomrule[1pt]
\end{tabular}
}
}
{\scriptsize
\begin{minipage}{400pt}
\setstretch{0.85}Notes:
This table presents summary statistics for the treatment-group indicator variable, the number of miners, and the Tp stratum indicator variable.
The treatment (control) group is 0--5 (5--30) km from the centroid of a municipality that has a mine.
The geological stratum is a time-constant variable: ``Year'' in parentheses indicates the matched year.
GSC mines indicate gold, silver, and copper mines.\\
Sources:
Data on the location of mines are from the Cooperation Society (1924; 1932; 1937).
\end{minipage}
}
\end{center}
\end{table}
%\end{landscape}
%------------------

\subsection{Dependent Variables}\label{sec:secc3}

%------------------
%Table C4
%\begin{landscape}
\begin{table}[htbp]
\def\arraystretch{0.95}
\begin{center}
\caption{Summary Statistics: Dependent Variables}
\label{tab:sum_outcome}
\scriptsize
\scalebox{1.0}[1]{
{\setlength\doublerulesep{2pt}
\begin{tabular}{lcD{.}{.}{2}D{.}{.}{2}D{.}{.}{2}D{.}{.}{2}D{.}{.}{2}D{.}{.}{2}}
\toprule[1pt]\midrule[0.3pt]
&\multirow{2}{*}{Year}&\multicolumn{3}{c}{Treated}&\multicolumn{3}{c}{Controlled}\\
\cmidrule(rrr){3-5}\cmidrule(rrr){6-8}
&&\multicolumn{1}{c}{Mean}&\multicolumn{1}{c}{Std. Dev.}&\multicolumn{1}{c}{Obs.}
&\multicolumn{1}{c}{Mean}&\multicolumn{1}{c}{Std. Dev.}&\multicolumn{1}{c}{Obs.}\\\hline
					&&&&&&&\\
\multicolumn{8}{l}{\textbf{Panel A: Dependent Variables used in Sections~\ref{sec:sec_ea} and~\ref{sec:sec_pe}}}\\
ln(Population)			&1925&8.68&0.92&197&8.16&0.63&1205\\
					&1930&8.87&0.97&166&8.19&0.64&976\\
					&1935&8.88&0.97&169&8.18&0.65&1147\\			
Crude Marriage Rate		&1925&8.06&2.31&197&9.83&2.33&1205\\
					&1930&7.43&2.19&166&8.79&2.23&976\\
					&1935&7.93&2.06&169&9.23&2.42&1147\\
Crude Birth Rate		&1925&34.47&6.38&197&37.42&5.92&1205\\
					&1930&31.76&5.64&166&34.46&5.40&976\\
					&1935&32.25&5.01&169&34.47&5.48&1147\\
Marital Fertility Rate		&1925&170.76&33.26&197&186.11&31.48&1205\\
					&1930&161.95&30.67&166&176.24&29.00&976\\
					&1935&168.76&28.11&169&178.52&29.80&1147\\
Crude Death Rate (male)	&1925&20.43&3.86&197&20.72&4.59&1205\\
					&1930&20.07&4.62&166&20.06&4.45&976\\
					&1935&17.93&4.00&169&18.46&4.15&1147\\
Crude Death Rate (female)&1925&18.27&4.14&197&19.39&4.37&1205\\
					&1930&18.21&3.88&166&18.66&4.39&976\\
					&1935&16.38&3.83&169&17.04&4.26&1147\\
Sex Ratio (male/female)	&1925&1.02&0.11&197&0.98&0.07&1205\\
					&1930&1.03&0.10&166&0.98&0.07&976\\
					&1935&1.03&0.08&169&0.98&0.07&1147\\
Household Size			&1925&5.04&0.49&197&5.35&0.62&1205\\
					&1930&5.18&0.41&166&5.37&0.47&976\\
					&1935&5.20&0.44&169&5.35&0.58&1147\\
					&&&&&&&\\
\multicolumn{8}{l}{\textbf{Panel B: Dependent Variables used in Section~\ref{sec:sec_aa}}}\\
Infant Mortality Rate			&1933&139.27&44.67&166&113.35&41.13&976\\
						&1938&131.55&37.24&169&112.06&43.18&1147\\
Fetal Death Rate$^{\star}$	&1933&51.54&26.32&166&42.18&27.33&976\\
						&1938&47.15&25.83&169&35.51&25.41&1147\\
Child Mortality Rate$^{\star}$	&1933&15.54&7.13&166&13.05&9.89&976\\
						&1938&19.15&8.48&169&16.45&8.81&1147\\\midrule[0.3pt]\bottomrule[1pt]
\end{tabular}
}
}
{\scriptsize
\begin{minipage}{380pt}
\setstretch{0.85}
Notes:
%$\ddagger$ and $\dagger$ in Panels A--C indicate that the mean difference between treatment and control groups is statistically significant at the 1\% and 5\% levels, respectively.
Panel A summarizes the summary statistics for the dependent variables in the cross-sectional municipal-level datasets used in Sections~\ref{sec:sec_ea} and~\ref{sec:sec_pe}.
The treatment (control) group is 0--5 (5--30) km from the centroid of a municipality that has a mine.
The crude marriage rate is the number of marriages per $1,000$ people.
The crude birth rate is the number of live births per $1,000$ people.
The marital fertility rate is the number of live births per $1,000$ married females.
The crude death rate is the number of male (female) deaths per $1,000$ males (females).
The sex ratio is the number of males divided by the number of females.
The household size is the total number of people divided by the number of households.\\
Panel B summarizes the summary statistics for the dependent variables in the cross-sectional municipal-level datasets used in Section~\ref{sec:sec_aa}.
The infant mortality rate is the number of infant deaths per $1,000$ live births.
The fetal death rate is the number of fetal deaths per $1,000$ births.
The child mortality rate is the number of deaths of children aged 1--4 per $1,000$ children aged 1--5.
The variables with $\star$ have a set of censored observations; the results from the Tobit model are materially similar to the main results.\\
Sources:
Data on the number of people and households are from the Statistics Bureau of the Cabinet (1926, 1935b, 1939).
Data on marriage, live births, and deaths are from the Statistics Bureau of the Cabinet (1927, 1932, 1938).
Data on births, fetal deaths, infant deaths, and child deaths are from Aiikukai (1935); Social Welfare Bureau (1941).
\end{minipage}
}
\end{center}
\end{table}
%\end{landscape}
%------------------

\subsubsection{Population, Marriage, Fertility, and Mortality}\label{sec:secc31}

The official reports of the Population Censuses published by the Statistics Bureau of the Cabinet document the number of people (by sex) and households.
The number of marriages, live births, and deaths used to calculate the crude marriage rate, birth rate, and death rate are obtained from the Municipal Vital Statistics published by the Statistics Bureau of the Cabinet.
The number of married women used to calculate the marital fertility rate is obtained from the Population Censuses.

Data on the number of fetal, infant, and child deaths are digitized using the Municipal Vital Statistics of 1933 and 1938 published by the Aiikukai and Social Welfare Bureau.
Data on the number of births and live births used to calculate fetal death rate and infant mortality rate are obtained from the same report.
Data on the number of children aged 1--5 used for the child mortality rate are obtained from the 1930 Population Census.

Live births and deaths are recorded in the municipality where they occurred, and fetal deaths in the municipality where the delivery took place.
Marriages are recorded in the municipality where the notification was filed under the Family Registration Law, which is generally where the couple resided.
Births and deaths of miners' families are therefore counted in mining areas rather than in their places of origin.
Table~\ref{tab:sum_outcome} presents the summary statistics of the demographic outcomes by year and treatment status.

\subsubsection{Labor Supply}\label{sec:secc32}

%Table C5
%\begin{landscape}
\begin{table}[htbp]
\def\arraystretch{0.95}
\begin{center}
\caption{Summary Statistics: Labor Statistics}
\label{tab:sum_ls}
\scriptsize
\scalebox{1.0}[1]{
{\setlength\doublerulesep{2pt}
\begin{tabular}{lcD{.}{.}{2}D{.}{.}{2}D{.}{.}{2}D{.}{.}{2}D{.}{.}{2}D{.}{.}{2}}
\toprule[1pt]\midrule[0.3pt]
&\multirow{2}{*}{Year}&\multicolumn{3}{c}{Treated}&\multicolumn{3}{c}{Controlled}\\
\cmidrule(rrr){3-5}\cmidrule(rrr){6-8}
&&\multicolumn{1}{c}{Mean}&\multicolumn{1}{c}{Std. Dev.}&\multicolumn{1}{c}{Obs.}
&\multicolumn{1}{c}{Mean}&\multicolumn{1}{c}{Std. Dev.}&\multicolumn{1}{c}{Obs.}\\\hline
Labor force participation rate	&1930&45.76	&6.12	&166		&49.52	&5.76	&976\\
\hspace{10pt}Male            		&1930&56.87	&3.61	&166		&57.18	&3.24	&976\\
\hspace{10pt}Female        		&1930&34.08	&11.02	&166		&41.95	&10.16	&976\\
Mining sector$^{\star}$		&1930&17.23	&17.65	&166		&0.28	&1.62	&976\\
\hspace{10pt}Male$^{\star}$	&1930&19.43	&19.17	&166		&0.39	&2.00	&976\\
\hspace{10pt}Female$^{\star}$	&1930&11.93	&14.32	&166		&0.11	&0.99	&976\\
Agricultural sector			&1930&45.73	&25.90	&166		&69.30	&20.94	&976\\
\hspace{10pt}Male			&1930&39.04	&24.96	&166		&64.09	&21.32	&976\\
\hspace{10pt}Female		&1930&57.80	&25.21	&166		&76.85	&19.99	&976\\
Manufacturing sector			&1930&12.31	&6.01	&166		&10.59	&8.20	&976\\
\hspace{10pt}Male			&1930&16.71	&7.06	&166		&13.41	&8.08	&976\\
\hspace{10pt}Female		&1930&3.97	&3.54	&166		&6.26	&9.34	&976\\
Commerce sector			&1930&10.35	&6.52	&166		&8.18	&7.16	&976\\
\hspace{10pt}Male			&1930&8.91	&5.63	&166		&7.66	&6.52	&976\\
\hspace{10pt}Female		&1930&14.13	&10.49	&166		&9.41	&9.85	&976\\
Domestic sector			&1930&2.05	&1.05	&166		&1.72	&1.34	&976\\
\hspace{10pt}Male$^{\star}$	&1930&0.28	&0.25	&166		&0.38	&0.57	&976\\
\hspace{10pt}Female		&1930&5.74	&3.71	&166		&3.94	&3.91	&976\\\midrule[0.3pt]\bottomrule[1pt]
\end{tabular}
}
}
{\scriptsize
\begin{minipage}{380pt}
\setstretch{0.85}
Notes:
This table reports the summary statistics for labor force participation rates and employment shares.
The treatment (control) group is 0--5 (5--30) km from the centroid of a municipality with a mine.
Labor force participation is the number of workers per $100$ people.
The employment share is the number of workers in each sector per $100$ workers.
The variables with $\star$ have a set of censored observations; the results from the Tobit model are materially similar to the main results.\\
Source:
Data on the labor force participation rates and employment shares are from the Statistics Bureau of the Cabinet (1935a).
\end{minipage}
}
\end{center}
\end{table}
%\end{landscape}

In Online Appendix~\ref{sec:test_lfpr}, the labor force participation rate, defined as the number of workers per $100$ people, is considered to investigate the impacts on local labor supplies.
To analyze structural shifts, the employment share of workers in the mining, agricultural, manufacturing, commercial, and domestic sectors is used.
These are calculated as the number of workers in each industry per $100$ workers.
These variables are considered for both sexes to understand the potential gender bias in the shifts.
These labor statistics are obtained from the prefectural part of the 1930 Population Census.\footnote{The prefectural part comprises $47$ reports (one for each prefecture), which I collected and digitized. For simplicity, those reports are cited as one document, Statistics Bureau of the Cabinet (1935a). The full population census is conducted every ten years in Japan. The 1930 census surveyed all labor statistics. Thus, similar statistics were unavailable in the 1935 Population Census. The 1940 Population Census was also disorganized because of the Second World War.}
Table~\ref{tab:sum_ls} presents the summary statistics of these variables by treatment status and census year.\footnote{The employment share in the mining sector has a set of censored observations because mining is a localized economic activity. As a robustness check, it was confirmed that the estimates from the Tobit estimator Tobin (1958) are similar to the main results (not reported).}
Similar to the demographic outcomes, the mean differences are statistically significant for most variables.

\subsection{Control Variables}\label{sec:secc4}

I include three measures of transportation accessibility (distances to the nearest railway station, seaport, and river) and one geographical characteristic (average elevation).
I used spatial joins to link these geospatial datasets with municipal boundaries and aggregated them to the municipal level.
Table~\ref{tab:sum_control} presents the summary statistics of the control variables.
The following subsections examine each control variable in detail.

%------------------
%Table C6
%\begin{landscape}
\begin{table}[htbp]
\def\arraystretch{0.95}
\begin{center}
\caption{Summary Statistics: Control Variables}
\label{tab:sum_control}
\footnotesize
\scalebox{1.0}[1]{
{\setlength\doublerulesep{2pt}
\begin{tabular}{lcD{.}{.}{2}D{.}{.}{2}D{.}{.}{2}D{.}{.}{2}D{.}{.}{2}D{.}{.}{2}}
\toprule[1pt]\midrule[0.3pt]
&Edition of&\multicolumn{3}{c}{Coal Mine Sample}&\multicolumn{3}{c}{GSC Mine Sample}\\
\cmidrule(rrr){3-5}\cmidrule(rrr){6-8}
Control Variables&the NDFM
&\multicolumn{1}{c}{Mean}&\multicolumn{1}{c}{Std. Dev.}&\multicolumn{1}{c}{Obs.}
&\multicolumn{1}{c}{Mean}&\multicolumn{1}{c}{Std. Dev.}&\multicolumn{1}{c}{Obs.}\\\hline
Distance to station	&1922	&7.72	&7.94	&1402    &8.01		&8.54	&3246\\
				&1931	&5.39	&5.84	&1142    &6.27		&7.98	&2498\\
				&1936	&5.75	&6.27	&1316    &5.51		&6.61	&3293\\
Distance to port		&1922	&41.10	&37.68	&1402    &41.14	&28.85	&3246\\
				&1931	&26.16	&22.19	&1142    &30.73	&21.89	&2498\\
				&1936	&24.09	&21.68	&1316    &31.77	&22.81	&3293\\
Distance to river	&(1922)	&0.99	&1.40	&1402    &0.83		&0.97	&3246\\
				&(1931)	&0.99	&1.59	&1142    &0.83		&0.97	&2498\\
				&(1936)	&1.02	&1.57	&1316    &0.84		&1.78	&3293\\
Elevation			&(1922)	&187.58	&236.36	&1402    &202.40	&219.68	&3246\\
				&(1931)	&187.37	&241.53	&1142    &191.33	&206.00	&2498\\
				&(1936)	&183.73	&234.38	&1316    &227.35	&234.20	&3293\\\midrule[0.3pt]\bottomrule[1pt]
\end{tabular}
}
}
{\scriptsize
\begin{minipage}{430pt}
\setstretch{0.85}Notes:
This table presents summary statistics for the control variables by mine type.
GSC mines indicate gold, silver, and copper mines.
Distances to the nearest station, port, and river are in kilometers.
Elevation is in meters.
The distance to the river and elevation are time-constant variables: ``Year'' in parentheses indicates the matched year.\\
Sources:
See Online Appendices~\ref{sec:secc41} to~\ref{sec:secc44} for the data on the distance and elevation variables.
Data on the location of mines are from the Cooperation Society (1924, 1932, 1937).
\end{minipage}
}
\end{center}
\end{table}
%\end{landscape}

\subsubsection{Railway Stations}\label{sec:secc41}

\begin{figure}[htbp]
\centering
\captionsetup{justification=centering}
\subfloat[1922]{\label{fig:map_station_1922}\includegraphics[width=0.35\textwidth]{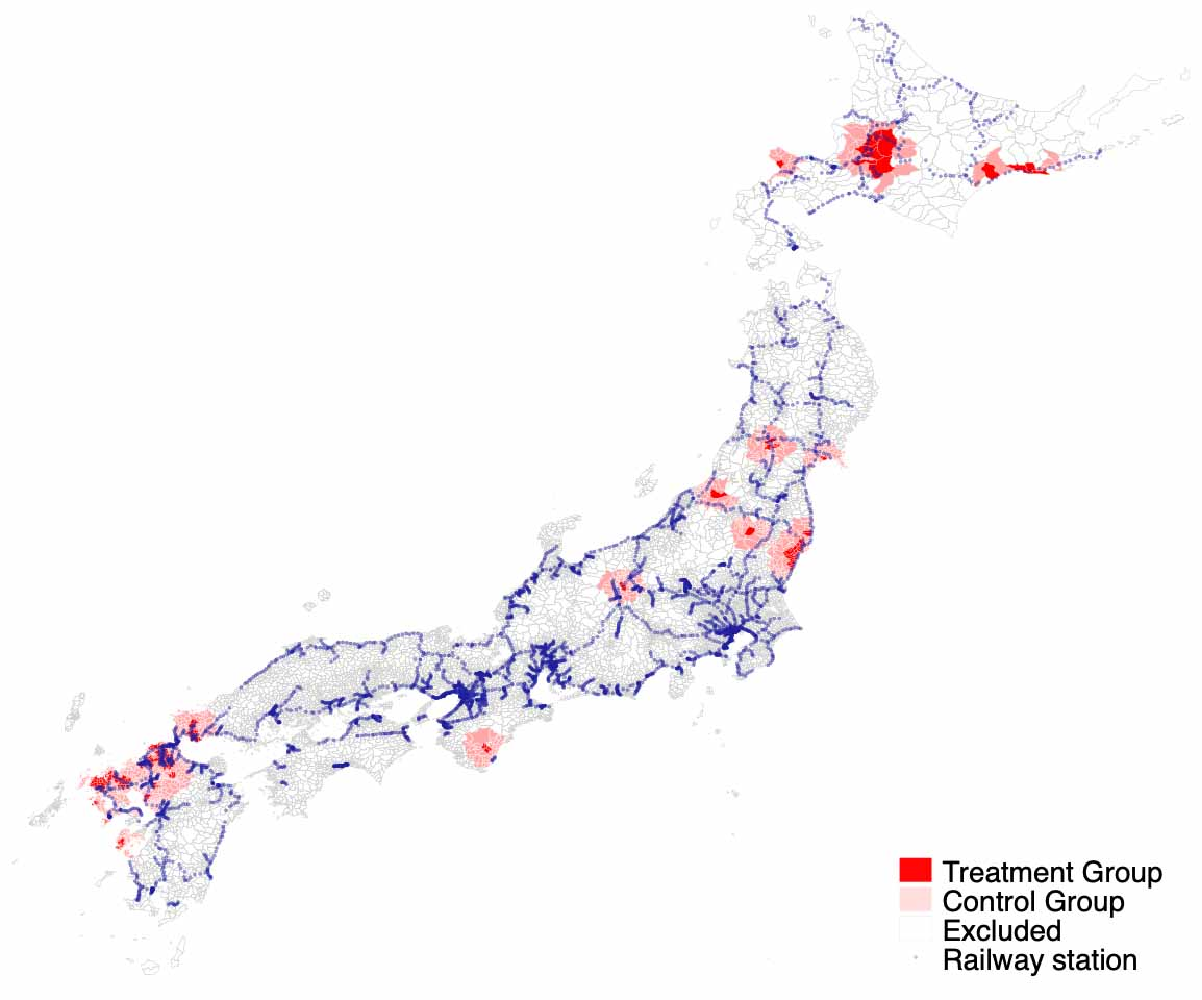}}
\subfloat[1931]{\label{fig:map_station_1931}\includegraphics[width=0.35\textwidth]{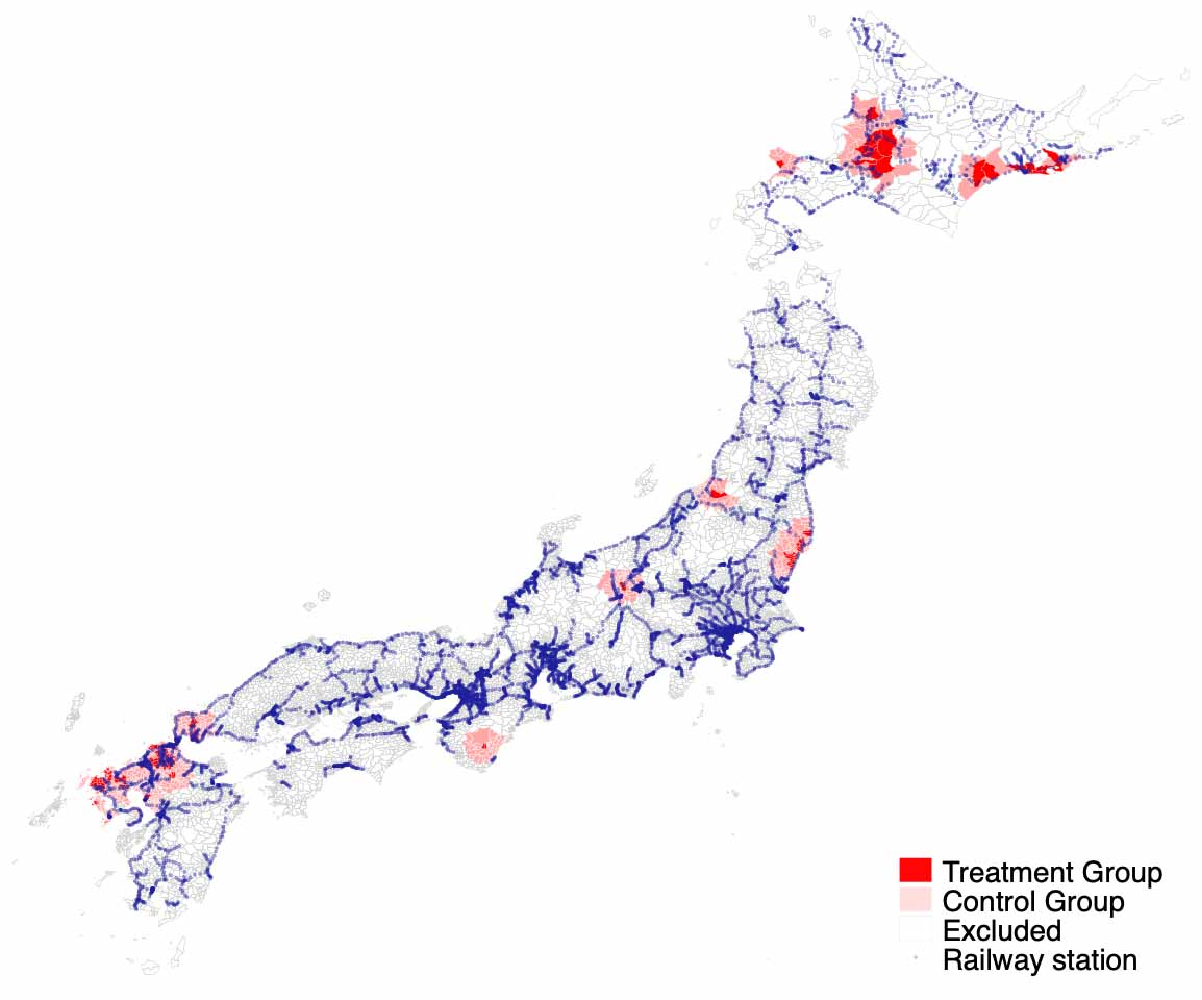}}
\subfloat[1936]{\label{fig:map_station_1936}\includegraphics[width=0.35\textwidth]{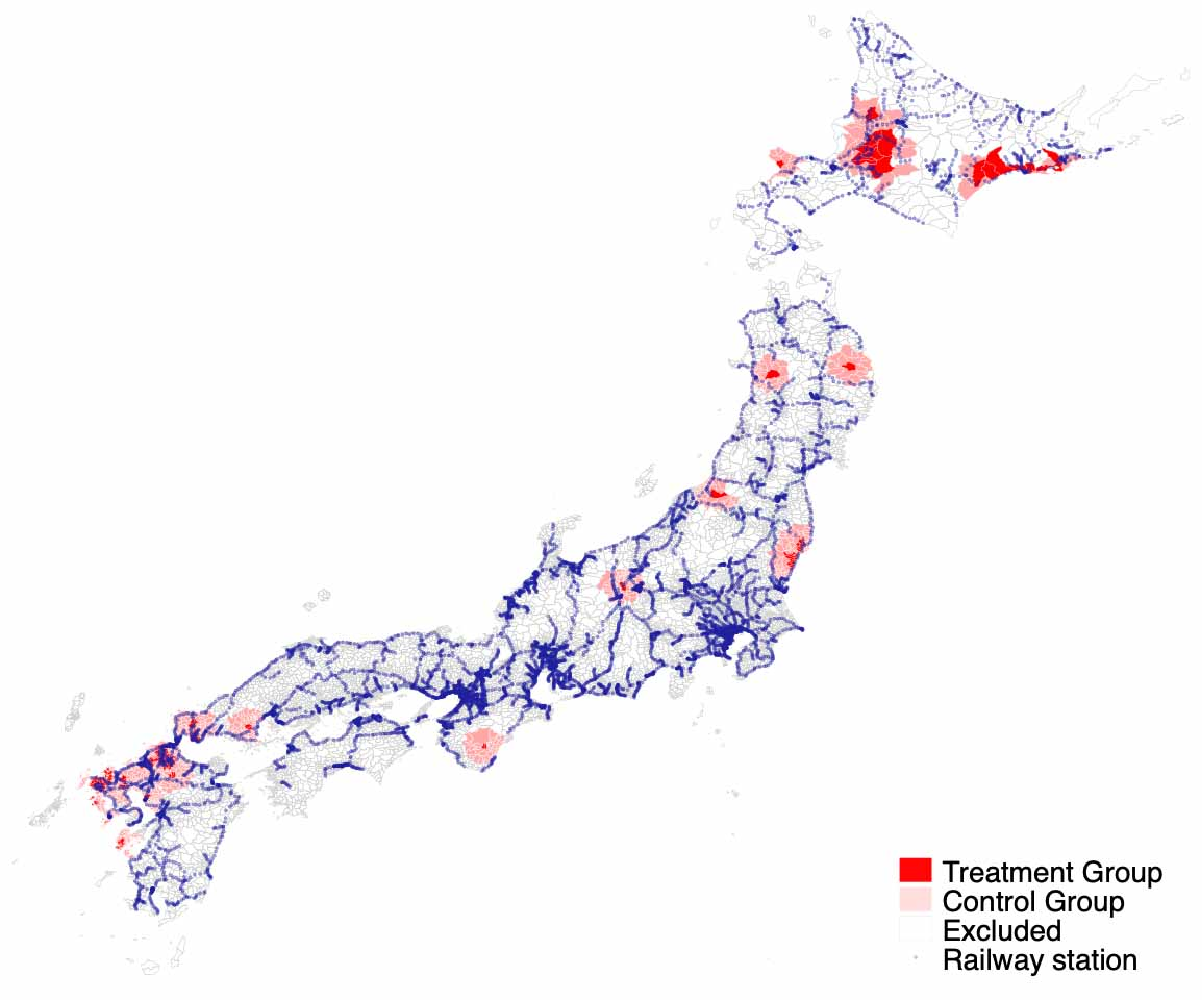}}
\caption{Spatial Distribution of Railway Stations}
\label{fig:map_station}
\scriptsize{\begin{minipage}{450pt}
\setstretch{0.9}
Notes:
The blue circles indicate the location of railway stations.
The base map used in this figure is Figure~\ref{fig:map_coal_mines}.
The treatment group highlighted in red includes the municipalities within 5 kilometers of a mine.
The control group, highlighted in pink, includes municipalities located between 5 and 30 kilometers from a mine.
The excluded municipalities are shown as empty lattices in the figures.
Source: Created by the author.
The location data on the railway stations are from the Ministry of Land, Infrastructure, Transport and Tourism (website).
\end{minipage}}
\end{figure}

The location information of railway stations was obtained from the official shapefile file created by the Ministry of Land, Infrastructure, Transport and Tourism (\url{https://nlftp.mlit.go.jp/ksj/gml/datalist/KsjTmplt-N05-v1_3.html}, accessed 29th July 2022).
The railway stations used to match the mining dataset included all stations (i.e., national, public, and private) within the railway sections.
Figure~\ref{fig:map_station} illustrates the distribution of stations.
The shortest distance to the nearest neighboring station is used as a railway accessibility variable.
Overall, Figure~\ref{fig:map_station} shows that railway stations are not concentrated in the mining areas.
This is consistent with evidence that accessibility to railways does not influence the results.

%----------
%Figure C4

\subsubsection{Seaports}\label{sec:secc42}

\begin{figure}[htbp]
\centering
\captionsetup{justification=centering}
\subfloat[1922]{\label{fig:map_port_1922}\includegraphics[width=0.35\textwidth]{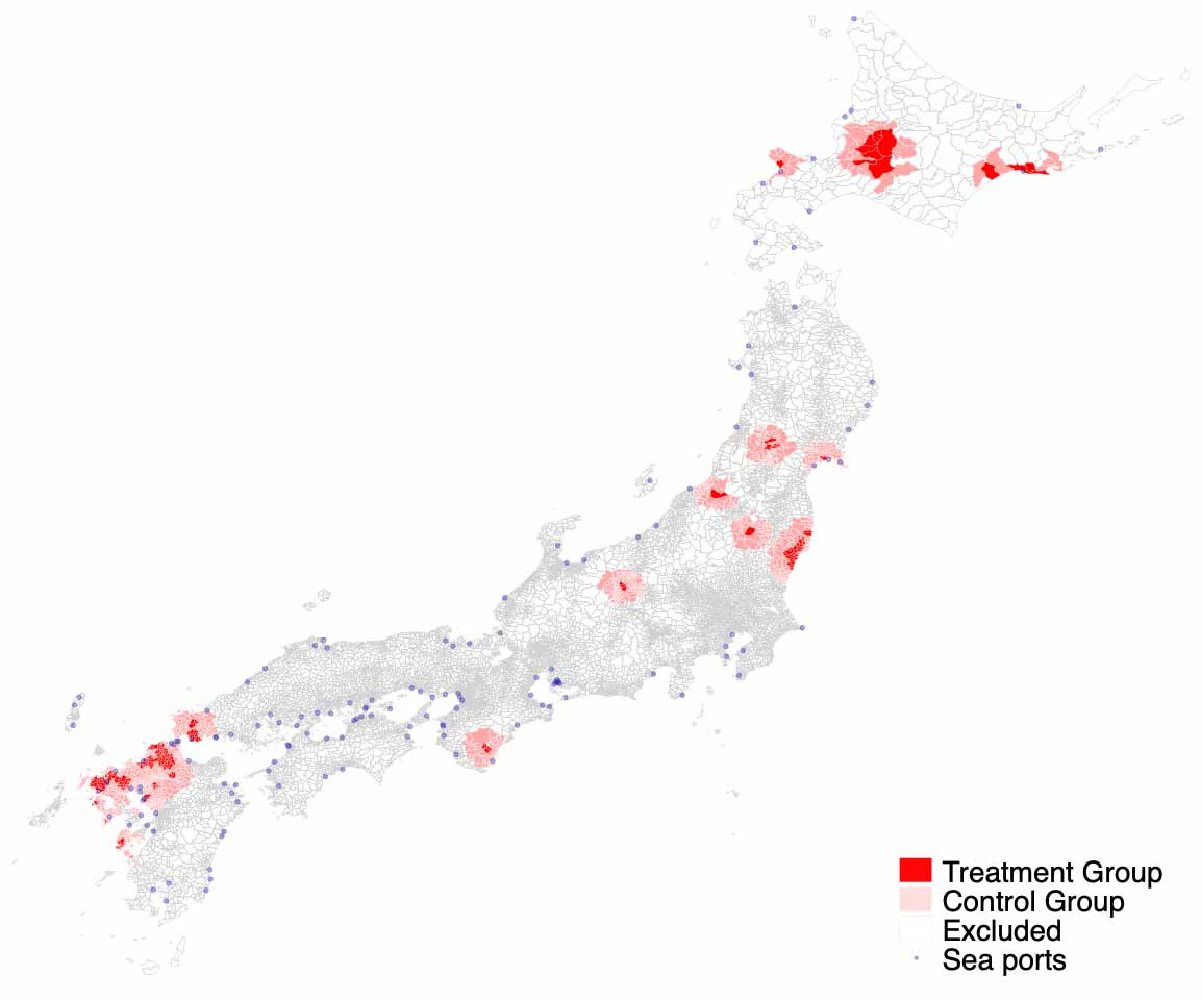}}
\subfloat[1931]{\label{fig:map_port_1931}\includegraphics[width=0.35\textwidth]{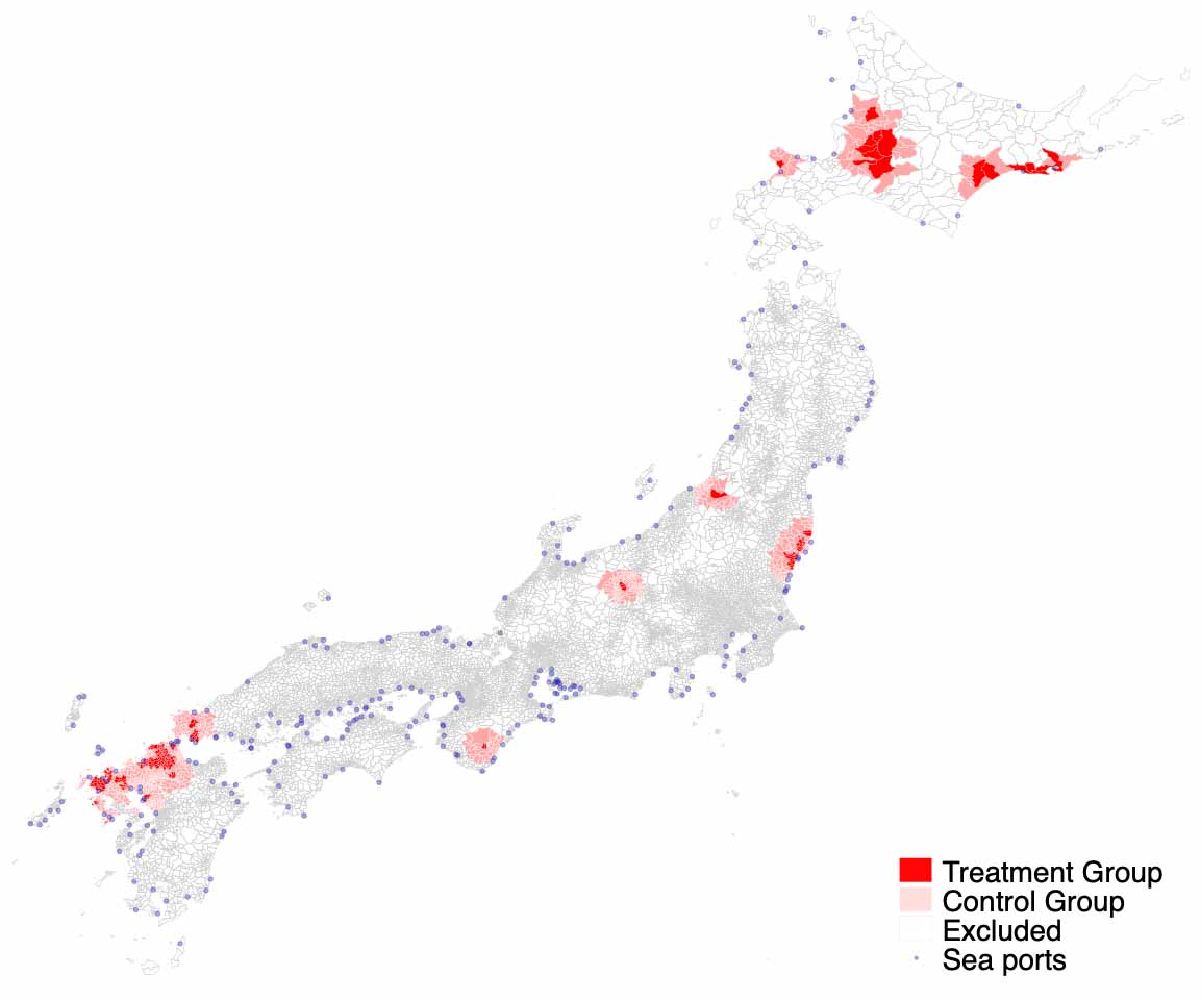}}
\subfloat[1936]{\label{fig:map_port_1936}\includegraphics[width=0.35\textwidth]{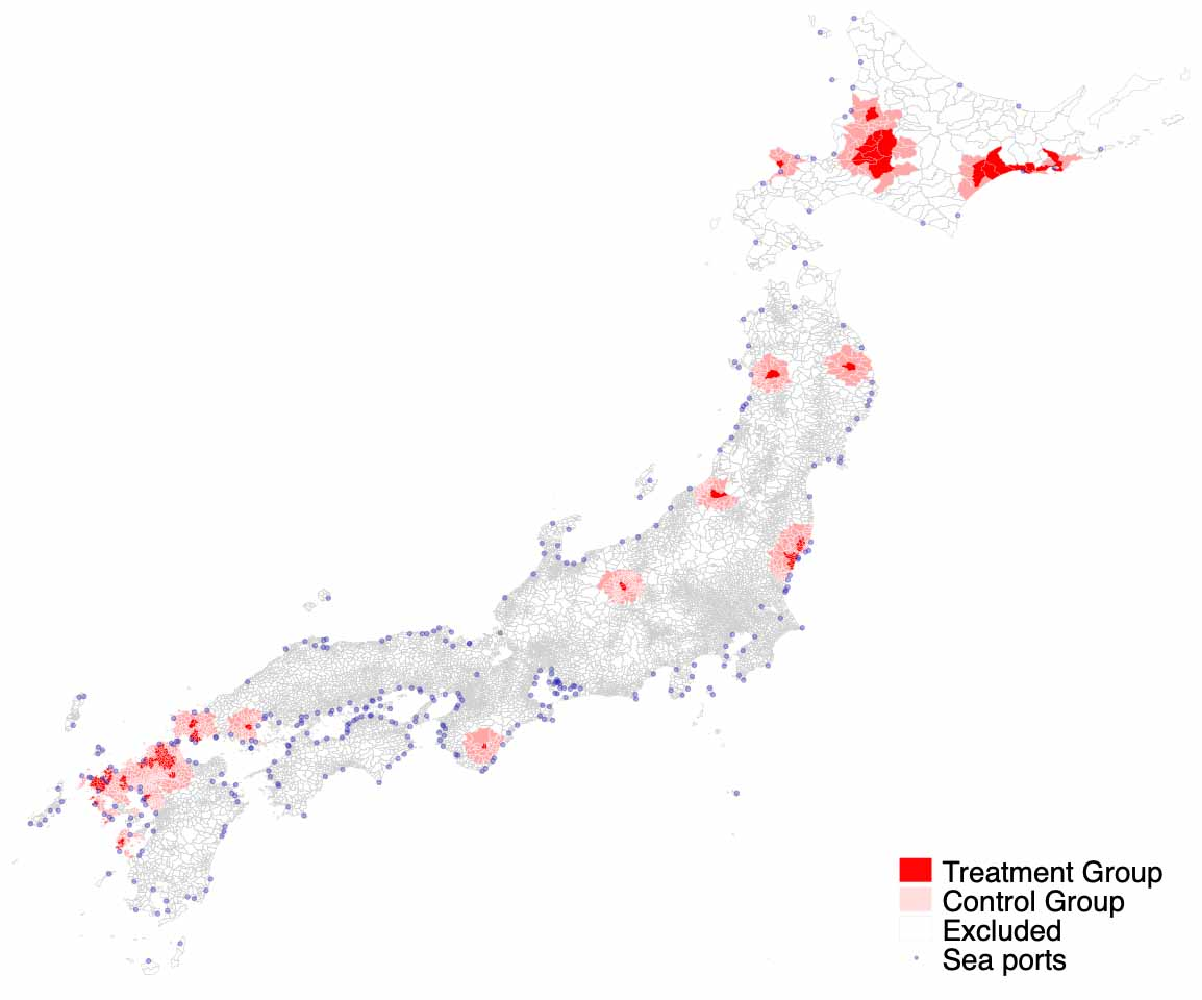}}
\caption{Spatial Distribution of Sea Ports}
\label{fig:map_port}
\scriptsize{\begin{minipage}{450pt}
\setstretch{0.9}
Notes:
The blue circles indicate the location of the seaports.
The ports in Okinawa Prefecture and a few ports on islands far from the main island are not shown in the figures.
The basemap used in this figure is the same as in Figure~\ref{fig:map_coal_mines}.
The treatment group highlighted in red includes the municipalities within 5 kilometers of a mine.
The control group, highlighted in pink, includes municipalities located between 5 and 30 kilometers from a mine.
The excluded municipalities are shown as empty lattices in the figures.
Source: Created by the author.
The location data on the seaports are from the Ministry of Land, Infrastructure, Transport and Tourism (website).
\end{minipage}}
\end{figure}

Seaport locations are obtained from the official shapefile created by the Ministry of Land, Infrastructure, Transport and Tourism (\url{https://nlftp.mlit.go.jp/ksj/gmlold/datalist/gmlold_KsjTmplt-C02.html}, accessed 1st October 2022).
The seaports used to match the mining dataset included all seaports (both commercial and fishery) listed in the Harbor Statistics of the Great Empire of Japan (\textit{Dainihon Teikoku K\=owan T\=okei}).
The ports listed in the Harbor Statistics but not included in the original shapefile taken from the Ministry of Land, Infrastructure, Transport, and Tourism were added manually.
Figure~\ref{fig:map_port} illustrates the distribution of seaports.

%----------
%Figure C5

\subsubsection{Rivers}\label{sec:secc43}

\begin{figure}[htbp]
\centering
\captionsetup{justification=centering,margin=1.5cm}
\includegraphics[width=0.6\textwidth]{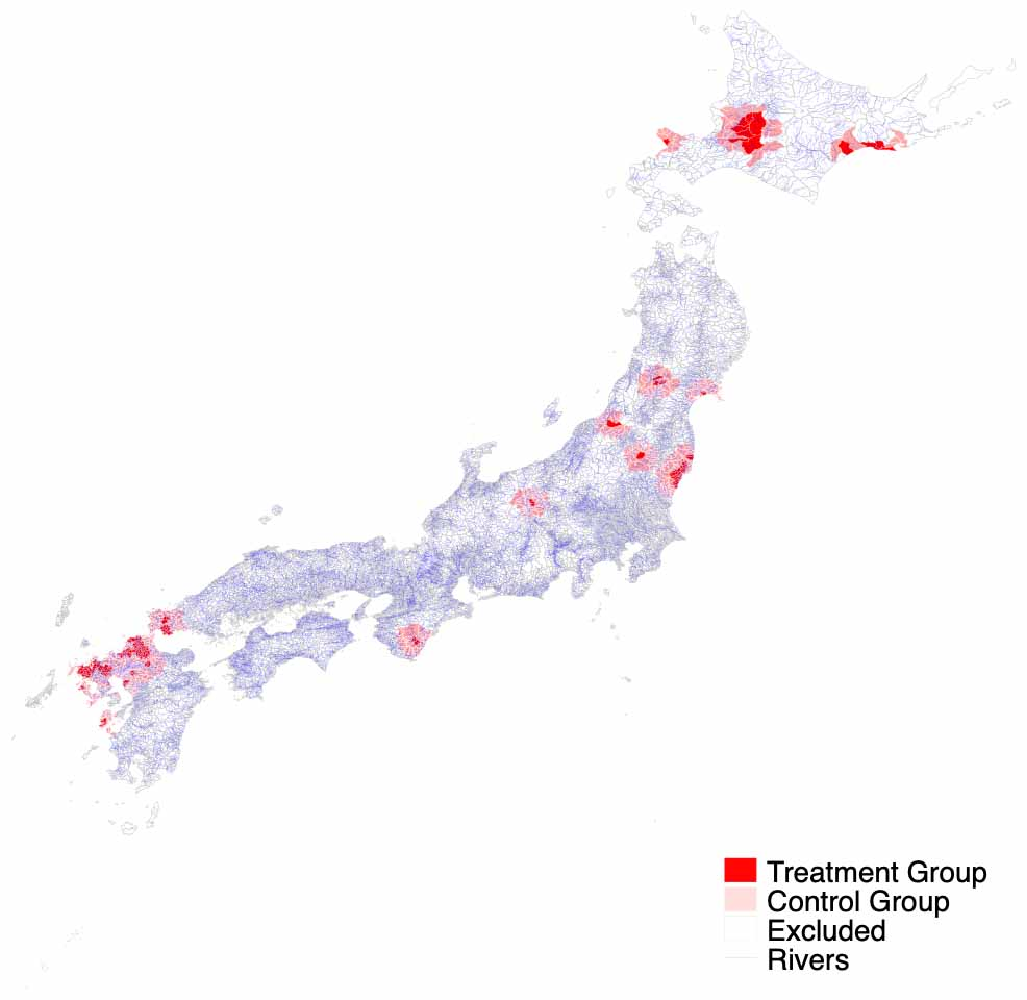}
\caption{Spatial Distribution of Rivers}
\label{fig:map_river}
\scriptsize{\begin{minipage}{450pt}
\setstretch{0.85}
Notes:
The blue lines indicate river locations.
The rivers used in the mining dataset include those managed by the government or by prefectures.
Rivers in Okinawa Prefecture and a few points on islands far from the main island are not shown in the figures.
River locations are time-constant, so the December 1922 coal mines map is used as an example basemap. (Figure~\ref{fig:map_coal_mines}).
The treatment group, highlighted in red, includes municipalities within 5 kilometers of a mine.
The control group, highlighted in pink, includes municipalities located between 5 and 30 kilometers from a mine.
Excluded municipalities are shown as empty lattices in the figures.
Source: Created by the author.
River location data are from the Ministry of Land, Infrastructure, Transport, and Tourism (website).
\end{minipage}}
\end{figure}

The location of rivers was obtained from the official shapefile provided by the Ministry of Land, Infrastructure, Transport, and Tourism (\url{https://nlftp.mlit.go.jp/ksj/jpgis/datalist/KsjTmplt-W05.html}, accessed August 21, 2018).
The rivers used to match the mining dataset included those managed either by the government or prefectures.
The location of rivers is time-constant.
Figure~\ref{fig:map_river} shows the distribution of the rivers matched to the 1922 coal mines map as an example.

%----------
%Figure C6
%----------

\subsubsection{Elevation}\label{sec:secc44}

\begin{figure}[htbp]
\centering
\captionsetup{justification=centering,margin=1.5cm}
\includegraphics[width=0.6\textwidth]{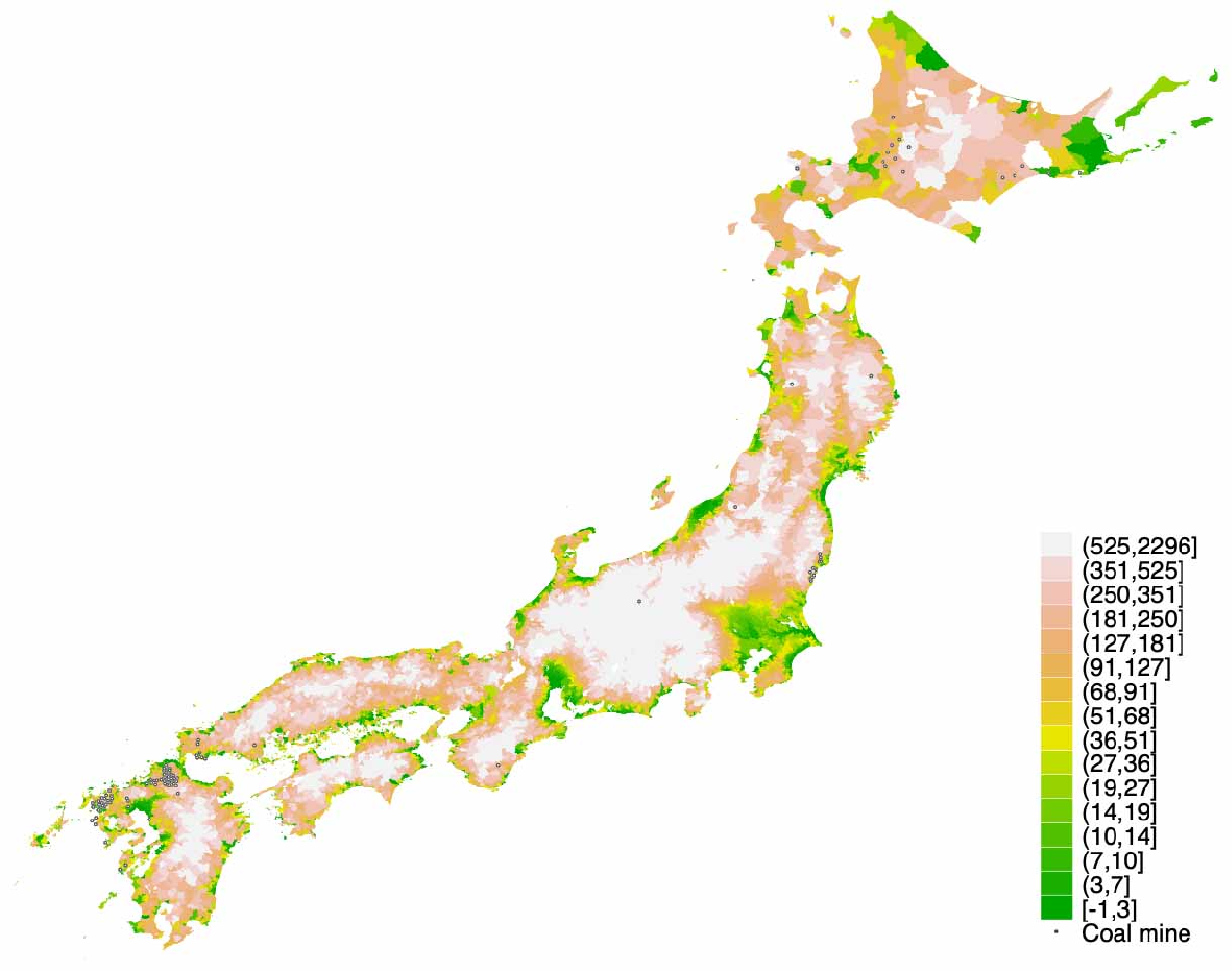}
\caption{Spatial Distribution of Elevation}
\label{fig:elevation}
\scriptsize{\begin{minipage}{400pt}
\setstretch{0.85}
Notes:
This figure shows the distribution of average elevation (in meters) across municipalities.
Rivers in Okinawa Prefecture and a few points on islands far from the main island are not shown in the figure.
The white circles mark the locations of coal mines in October 1936.
Source:
Created by the author.
The grid-cell elevation data are from the Ministry of Land, Infrastructure, Transport, and Tourism (website).
\end{minipage}}
\end{figure}

The grid cell elevation data are obtained from the official GIS database of the Ministry of Land, Infrastructure, Transport, and Tourism (\url{https://nlftp.mlit.go.jp/ksj/gml/datalist/KsjTmplt-G04-d.html}, accessed February 24, 2023).
The average elevation for a 1 km square grid cell is calculated using the elevations measured in the 250 square grid cells of a third mesh code initially assigned by the Ministry of Land, Infrastructure, Transport, and Tourism.
The nearest neighbor grid cell is then linked to each municipality.
The average distance to the nearest neighbor cell is $0.49$km, which is sufficiently close to represent each municipality's average elevation.
Figure~\ref{fig:elevation} shows the spatial distribution of the elevation.

%----------
%Figure C7
%----------

%-------------------------------------------------------------------------------
% Appendix D
%-------------------------------------------------------------------------------
%\clearpage
\section{Empirical Analysis Appendix}\label{sec:secd}
\setcounter{figure}{0} \renewcommand{\thefigure}{D.\arabic{figure}}
\setcounter{table}{0} \renewcommand{\thetable}{D.\arabic{table}}

\subsection{Assumptions for the Identification}\label{sec:app_as}

To simplify the discussion on the identification assumptions, I focus on the most straightforward structural equation, a dummy endogenous variable model taking the form:
\begin{eqnarray}\label{devm}
\footnotesize{
\begin{split}
y_{i} = \alpha + \beta \text{\textit{MineDeposit}}_{i} + e_{i},\\
\text{\textit{MineDeposit}}_{i} = \iota + \zeta \text{\textit{Stratum}}_{i} + \epsilon_{i}.
\end{split}
}
\end{eqnarray}
All the variables are defined in Section~\ref{sec:sec_ea_es}.
Note that the same Greek symbols are used as those used in equations~\ref{eq1} and~\ref{eq2} to simplify the expressions of the model, although they are different parameters from those considered in Section~\ref{sec:sec_ea_es}.

Five assumptions are made to identify the parameter of interest under this setting: the stable unit treatment value assumption (hereafter SUTVA), random assignments, exclusion restriction, nonzero average causal effect of the IV, and monotonicity (Angrist et al. 1996).
Section~\ref{sec:sec_ea_es} discusses the independence of the instrument and the exclusion restriction: the geological stratum, which was formed tens of millions of years ago, is unlikely to be related to local characteristics in the interwar period, and it is plausible that it affected the outcomes only through the presence of coal mines, conditional on the control variables.
This subsection then discusses the remaining three assumptions.

First, the SUTVA assumption holds that the potential outcomes for municipality $i$ in the system (\ref{devm}) depend only on its own treatment status.
In my empirical setting, which uses the municipal-level dataset, this assumption is less likely to be violated. This is because there are certain distances among different treated municipalities, even in the agglomerated area such as the Chikuh\=o coalfield.
Moreover, to minimize the risk of SUTVA violation, a defined threshold is considered to define the treatment group.
The results demonstrate that the baseline definition of the treatment distance is plausible for modeling potential spillover effects in the regression (Online Appendix~\ref{sec:test_threshold}).

Second, let $\textit{MineDeposit}_{i}(\textit{Stratum}_{i})$ be an indicator for whether $i$ is treated given the random assignments of the geological strata, under the SUTVA.
The nonzero average causal effect indicates that $E[\textit{MineDeposit}_{i}(1)-\textit{MineDeposit}_{i}(0)] \neq 0$.
This assumption clearly holds because the likelihood of having coal mines increases with the IV strata (Online Appendix~\ref{sec:secc2}).

Third, the monotonicity assumption suggests that $\forall i$, $\textit{MineDeposit}_{i}(1) \geq \textit{MineDeposit}_{i}(0)$.
In sum, a combination of the nonzero average causal effect and monotonicity assumptions argues that strong monotonicity should be satisfied at least in one unit as $\textit{MineDeposit}_{i}(1) > \textit{MineDeposit}_{i}(0)$ (Angrist et al.~1996).
Although this assumption is a potential outcome and untestable (Imbens and Angrist~1994), there is no fundamental reason for mining companies to avoid the stratum containing coal seams.
Thus, the instrument should affect the selection decision monotonically.

Under these assumptions, the IV estimate identifies the local average treatment effect for municipalities in which coal mines were located because of the Paleogene formations (Imbens and Angrist 1994; Angrist et al. 1996).

\subsection{Alternative Variance Estimator}\label{sec:app_crve}

\begin{figure}[h]
\centering
\captionsetup{justification=centering}
\subfloat[CS: Population]{\label{fig:vce_lnpop}\includegraphics[width=0.32\textwidth]{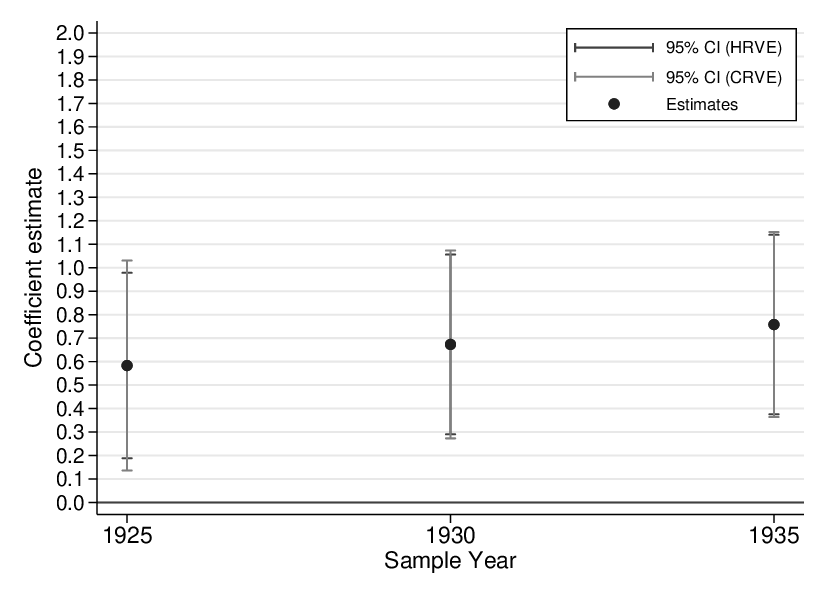}}
\subfloat[CS: MFR]{\label{fig:vce_mfr}\includegraphics[width=0.32\textwidth]{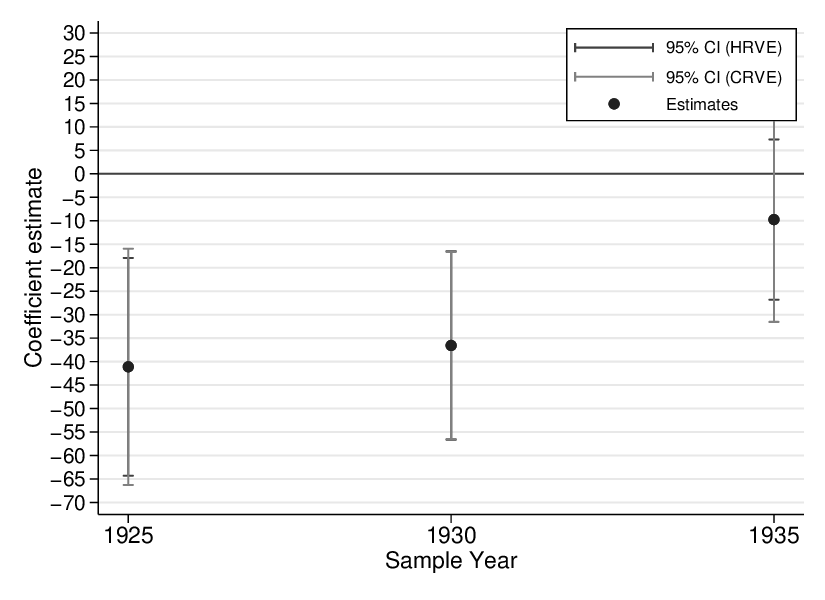}}
\subfloat[CS: CDR (Female)]{\label{fig:vce_cdrf}\includegraphics[width=0.32\textwidth]{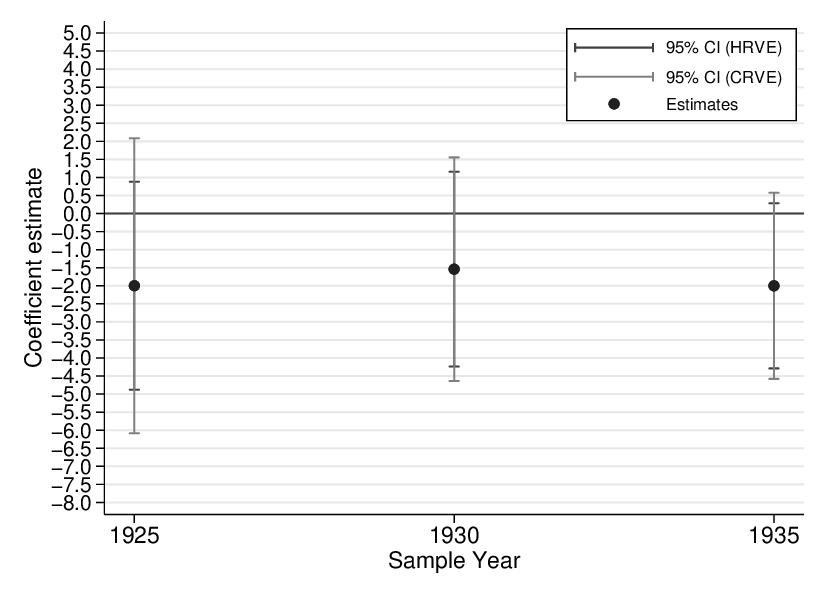}}\\
\subfloat[CS: Sex Ratio]{\label{fig:vce_sr}\includegraphics[width=0.32\textwidth]{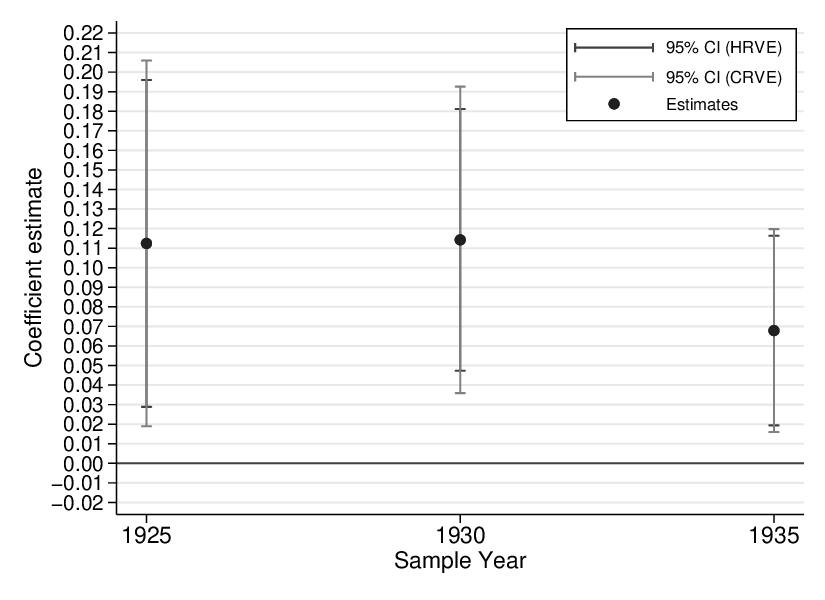}}
\subfloat[CS: IMR]{\label{fig:vce_imr}\includegraphics[width=0.32\textwidth]{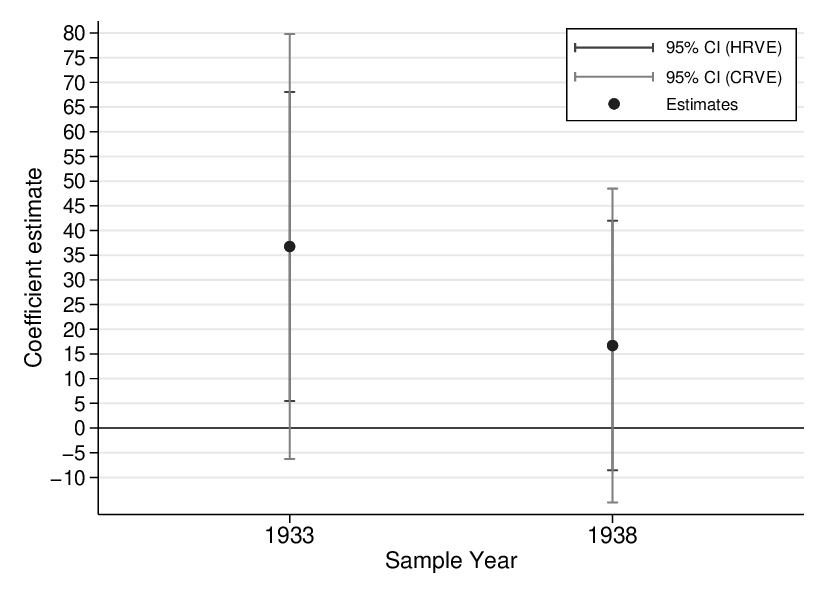}}
\subfloat[Panel Data]{\label{fig:vce_pd}\includegraphics[width=0.33\textwidth]{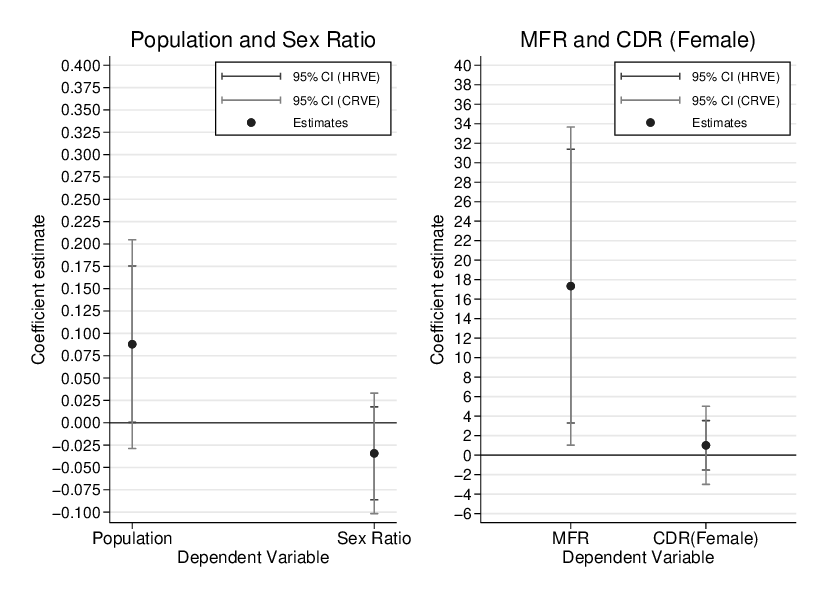}}
\caption{Robustness to the Alternative Variance-Covariance Matrix Estimators}
\label{fig:r_crve}
\scriptsize{\begin{minipage}{450pt}
\setstretch{0.9}
This figure shows the standard errors obtained from the heteroskedasticity-robust (HRVE) and cluster-robust (CRVE) variance estimators.
The darker (lighter) gray line indicates the 95\% confidence interval from the HRVE (CRVE).\\
Notes:
Figures~\ref{fig:vce_lnpop},~\ref{fig:vce_mfr},~\ref{fig:vce_cdrf},~\ref{fig:vce_sr}, and~\ref{fig:vce_imr} show the results for the local population (log-transformed), marital fertility rate, crude death rate among women, sex ratio, and infant mortality rate, respectively.
Figure~\ref{fig:vce_pd} shows the results for the panel data analysis using the ever- and never-treated samples (Sections~\ref{sec:sec_pe} and~\ref{sec:sec_aa}).
Source: Created by the author.
\end{minipage}}
\end{figure}

I use a cluster-robust covariance matrix estimator (Arellano 1987) and clustered standard errors at the broader county-level to assess potential local-scale spatial correlations in the errors.
The finite-sample adjustment proposed by Hansen (2007) is implemented in the estimator.
Figure~\ref{fig:r_crve} summarizes the results based on the cluster-robust variance-covariance matrix estimator, confirming that the results are materially similar to the baseline results in the main text.

%----------
%Figure D1
%----------

\subsection{Gender-Biased Shifts in the Labor Force Structure:~Evidence from the 1930 Census}\label{sec:test_lfpr}

%----------
%Table D1
%\begin{landscape}
\begin{table}[htb]
\def\arraystretch{1.0}
\begin{center}
\captionsetup{justification=centering}
\caption{Labor Force Participation and Structural Shifts: \\Evidence from 1930 Population Census}
\label{tab:r_labor}
\scriptsize
\scalebox{0.82}[1]{
{\setlength\doublerulesep{2pt}
\begin{tabular}{lD{.}{.}{-2}D{.}{.}{-2}D{.}{.}{-2}D{.}{.}{-2}D{.}{.}{-2}D{.}{.}{-2}}
\toprule[1pt]\midrule[0.3pt]
\multicolumn{7}{l}{\textbf{Panel A: Male Workers}}\\
&\multicolumn{6}{c}{Dependent Variable}\\
\cmidrule(rrrrrr){2-7}
&\multicolumn{1}{c}{}&\multicolumn{5}{c}{Employment Share}\\
\cmidrule(rrrrr){3-7}
&\multicolumn{1}{c}{(1) LFPR}&\multicolumn{1}{c}{(2) Mining}&\multicolumn{1}{c}{(3) Agricultural}&\multicolumn{1}{c}{(4) Manufacturing}&\multicolumn{1}{c}{(5) Commerce}&\multicolumn{1}{c}{(6) Domestic}\\\hline
\textit{MineDeposit}
&-0.306	&33.655$***$	&-31.794$***$	&1.551	&-2.716$*$	&-0.378$***$    \\
&(1.146)	&(4.803)		&(5.852)		&(2.000)	&(1.456)		&(0.131)        \\\hline
City and town indicators
&\multicolumn{1}{c}{Yes}&\multicolumn{1}{c}{Yes}&\multicolumn{1}{c}{Yes}&\multicolumn{1}{c}{Yes}&\multicolumn{1}{c}{Yes}&\multicolumn{1}{c}{Yes}\\
Railway accessibility
&\multicolumn{1}{c}{Yes}&\multicolumn{1}{c}{Yes}&\multicolumn{1}{c}{Yes}&\multicolumn{1}{c}{Yes}&\multicolumn{1}{c}{Yes}&\multicolumn{1}{c}{Yes}\\
Port accessibility
&\multicolumn{1}{c}{Yes}&\multicolumn{1}{c}{Yes}&\multicolumn{1}{c}{Yes}&\multicolumn{1}{c}{Yes}&\multicolumn{1}{c}{Yes}&\multicolumn{1}{c}{Yes}\\
River accessibility
&\multicolumn{1}{c}{Yes}&\multicolumn{1}{c}{Yes}&\multicolumn{1}{c}{Yes}&\multicolumn{1}{c}{Yes}&\multicolumn{1}{c}{Yes}&\multicolumn{1}{c}{Yes}\\
Elevation
&\multicolumn{1}{c}{Yes}&\multicolumn{1}{c}{Yes}&\multicolumn{1}{c}{Yes}&\multicolumn{1}{c}{Yes}&\multicolumn{1}{c}{Yes}&\multicolumn{1}{c}{Yes}\\
Observations
&\multicolumn{1}{c}{1,142}&\multicolumn{1}{c}{1,142}&\multicolumn{1}{c}{1,142}&\multicolumn{1}{c}{1,142}&\multicolumn{1}{c}{1,142}&\multicolumn{1}{c}{1,142}\\
Estimator
&\multicolumn{1}{c}{IV}&\multicolumn{1}{c}{IV}&\multicolumn{1}{c}{IV}&\multicolumn{1}{c}{IV}&\multicolumn{1}{c}{IV}&\multicolumn{1}{c}{IV}\\
First-stage $F$-statistic
&36.69&36.69    &36.69    &36.69&36.69&36.69    \\
Mean of the DV        &57.13    &3.16    &60.45    &13.89    &7.84&0.37\\
Std. Dev. of the DV    &3.30    &10.08    &23.59    &8.02    &6.41&0.54\\\hline
&&&&&&\\
\multicolumn{7}{l}{\textbf{Panel B: Female Workers}}\\
&\multicolumn{6}{c}{Dependent Variable}\\
\cmidrule(rrrrrr){2-7}
&\multicolumn{1}{c}{}&\multicolumn{5}{c}{Employment Share}\\
\cmidrule(rrrrr){3-7}
&\multicolumn{1}{c}{(1) LFPR}&\multicolumn{1}{c}{(2) Mining}&\multicolumn{1}{c}{(3) Agricultural}&\multicolumn{1}{c}{(4) Manufacturing}&\multicolumn{1}{c}{(5) Commerce}&\multicolumn{1}{c}{(6) Domestic}\\\hline
\textit{MineDeposit}
&-16.250$***$	&24.167$***$	&-21.836$***$	&-10.308$***$	&1.791	&4.183$***$    \\
&(3.611)		&(4.006)		&(5.487)		&(2.115)		&(2.498)	&(1.301)        \\\hline
City and town indicators
&\multicolumn{1}{c}{Yes}&\multicolumn{1}{c}{Yes}&\multicolumn{1}{c}{Yes}&\multicolumn{1}{c}{Yes}&\multicolumn{1}{c}{Yes}&\multicolumn{1}{c}{Yes}\\
Railway accessibility
&\multicolumn{1}{c}{Yes}&\multicolumn{1}{c}{Yes}&\multicolumn{1}{c}{Yes}&\multicolumn{1}{c}{Yes}&\multicolumn{1}{c}{Yes}&\multicolumn{1}{c}{Yes}\\
Port accessibility
&\multicolumn{1}{c}{Yes}&\multicolumn{1}{c}{Yes}&\multicolumn{1}{c}{Yes}&\multicolumn{1}{c}{Yes}&\multicolumn{1}{c}{Yes}&\multicolumn{1}{c}{Yes}\\
River accessibility
&\multicolumn{1}{c}{Yes}&\multicolumn{1}{c}{Yes}&\multicolumn{1}{c}{Yes}&\multicolumn{1}{c}{Yes}&\multicolumn{1}{c}{Yes}&\multicolumn{1}{c}{Yes}\\
Elevation
&\multicolumn{1}{c}{Yes}&\multicolumn{1}{c}{Yes}&\multicolumn{1}{c}{Yes}&\multicolumn{1}{c}{Yes}&\multicolumn{1}{c}{Yes}&\multicolumn{1}{c}{Yes}\\
Observations
&\multicolumn{1}{c}{1,142}&\multicolumn{1}{c}{1,142}&\multicolumn{1}{c}{1,142}&\multicolumn{1}{c}{1,142}&\multicolumn{1}{c}{1,142}&\multicolumn{1}{c}{1,142}\\
Estimator
&\multicolumn{1}{c}{IV}&\multicolumn{1}{c}{IV}&\multicolumn{1}{c}{IV}&\multicolumn{1}{c}{IV}&\multicolumn{1}{c}{IV}&\multicolumn{1}{c}{IV}\\
First-stage $F$-statistic	&36.69&36.69&36.69&36.69&36.69&36.69\\
Mean of the DV			&40.81&1.83&74.08&5.93&10.09&4.20\\
Std. Dev. of the DV		&10.65&6.92&21.88&8.77&10.08&3.93\\\midrule[0.3pt]\bottomrule[1pt]
\end{tabular}
}
}
{\scriptsize
\begin{minipage}{415pt}
\setstretch{0.85}
***, **, and * represent statistical significance at the 1\%, 5\%, and 10\% levels, respectively.
Standard errors based on the heteroskedasticity-robust covariance matrix estimator are reported in parentheses.\\
Notes:
Panels A and B show the results for male and female worker samples, respectively, from the 1930 Population Census (Table~\ref{tab:sum_ls}).
Column (1) shows the results for the labor force participation rates (\%), whereas Columns (2)--(6) show the results for the employment shares in each industrial sector (\%).
\end{minipage}
}
\end{center}
\end{table}
%\end{landscape}
%----------

I aim to evaluate how the coal mines influence the local industrial structure using labor supply statistics from the 1930 population census.

\subsubsection*{Male Workers}

Panel A of Table~\ref{tab:r_labor} presents the results for the male workers.
Column (1) shows that the estimate for the male LFP is close to zero.
This indicates that coal mines did not influence the labor supply of male workers in the local economy.
Columns (2)--(6) show the results for the employment share of male workers.
Column 2 indicates that coal mines increased the mining sector's employment share by approximately $33.7$ percentage points, leading to a similar decline in the agricultural sector's employment share (Column 3).
This suggests that the structural shift in male workers' employment occurred mainly in the agricultural sector.
In Column (4), the estimate for the manufacturing sector is positive, suggesting a marginal spillover effect of mines on manufacturing in mining areas.
However, this effect is not statistically significant.
Compared with the case of the 1970s U.S.,\footnote{Black et al. (2005) found that the 1970s coal boom in the U.S. increased employment and earnings with modest spillovers into the non-mining sectors.} the rapid mechanization may not have created the time window for generating spillover effects in the case of interwar Japan.
The estimates for the commercial and domestic sectors are moderately negative and close to zero (Columns (5)--(6), respectively).
This is consistent with male workers being less likely to work in these service sectors (Table~\ref{tab:sum_ls}).

\subsubsection*{Female Workers}

The results for female labor listed in Panel B demonstrate different responses.
Column (1) shows a statistically significant negative estimate, providing evidence that coal mines reduced female LFP.
The estimate suggests that coal mines decreased the female LFP rate by approximately $16.3$ percentage points.
This suggests that, outside mining employment, women in mining areas had few opportunities for paid work, and that women who did not work in the mines were more likely to stay at home.
Columns (2)--(3) show the results for the employment share of female workers.
Column (2) indicates that coal mines increase the mining sector's employment share by $24.2$ percentage points and decrease the agricultural sector's share by the same degree (Column (3)).
The estimated magnitude is smaller than that for male workers (Panel A).
This is logical because, on average, females were less likely to work in mines than males (Panel A of Table~\ref{tab:nminers}).
While the manufacturing sector's employment share decreases in the mining area (Column (3)), the estimates for the commercial and domestic sectors are moderately positive (Columns (5)--(6), respectively).
This appears to be consistent with the increased relative demand for personal services among mining workers in mining areas (Caselli and Coleman II 2001).

\subsubsection*{Summary}

Studies of resource booms show that resource extraction increases local employment and wages (e.g., Black et al. 2005; Allcott and Keniston 2018).
My results show that, in interwar Japan, the effects of coal mining on local labor supply differed by gender.
Coal mining did not affect male labor force participation, but it shifted male employment from agriculture to mining.
By contrast, female labor force participation was considerably lower in mining areas than in surrounding areas.
Because women in the surrounding agricultural areas commonly worked on family farms, the smaller agricultural sector in mining areas reduced women's opportunities for paid work, and mining jobs offset only part of this decline.
These results suggest that the employment opportunities created by mining for women were concentrated in the mines themselves, so that the removal of women from the pits would have further reduced their paid employment.

\subsection{Conditional Expectation of the Truncated Normal Distribution}\label{sec:ce_tnd}

Let $x \sim \mathcal{N}(\mu,\,\sigma^{2})$ be the initial health endowment of a fetus \textit{in utero}, $x_{s}$ be the survival threshold, and $f(\cdot)$ and $F(\cdot)$ be the density and cumulative distribution functions of $x$, respectively.
Because $f'(x) = -\frac{x-\mu}{\sigma^{2}} f(x)$, the conditional expectation of the truncated normal distribution can be derived as:
\begin{eqnarray*}
\footnotesize{
\begin{split}
E[x|x > x_{s}] &= \int_{x_{s}}^{\infty} x \frac{f(x)}{1-F(x_{s})}dx\\
&= \frac{-\sigma^{2}}{1-F(x_{s})} \int_{x_{s}}^{\infty} -\frac{(x-\mu)}{\sigma^{2}} f(x)dx + \frac{\mu}{1-F(x_{s})} \int_{x_{s}}^{\infty} f(x)dx\\
&= \mu + \sigma^{2} \frac{f(x_{s})}{1-F(x_{s})}.
\end{split}
}
\end{eqnarray*}
Let $\alpha = (x_{s}-\mu)/\sigma$ and $\lambda(\alpha) = \phi(\alpha)/[1-\Phi(\alpha)]$ be the inverse Mills ratio, where $\phi(\cdot)$ and $\Phi(\cdot)$ are the standard normal density and cumulative distribution functions.
Then $E[x|x > x_{s}] = \mu + \sigma\lambda(\alpha)$.
Because $0 < \lambda'(\alpha) < 1$, it follows that $\partial E[x|x > x_{s}]/\partial \mu = 1 - \lambda'(\alpha) > 0$ and $\partial E[x|x > x_{s}]/\partial x_{s} = \lambda'(\alpha) > 0$.
This implies that the ``scarring'' mechanism ($\mu^{*} < \mu$) shifts the conditional mean toward the left, whereas the ``selection'' mechanism ($x_{s}^{*} > x_{s}$) shifts it toward the right.

\subsection{Testing the Validity of Threshold}\label{sec:test_threshold}

%----------
%Figure D2
\begin{figure}[h]
\centering
\captionsetup{justification=centering}
\subfloat[Population]{\label{fig:het_lnpop}\includegraphics[width=0.32\textwidth]{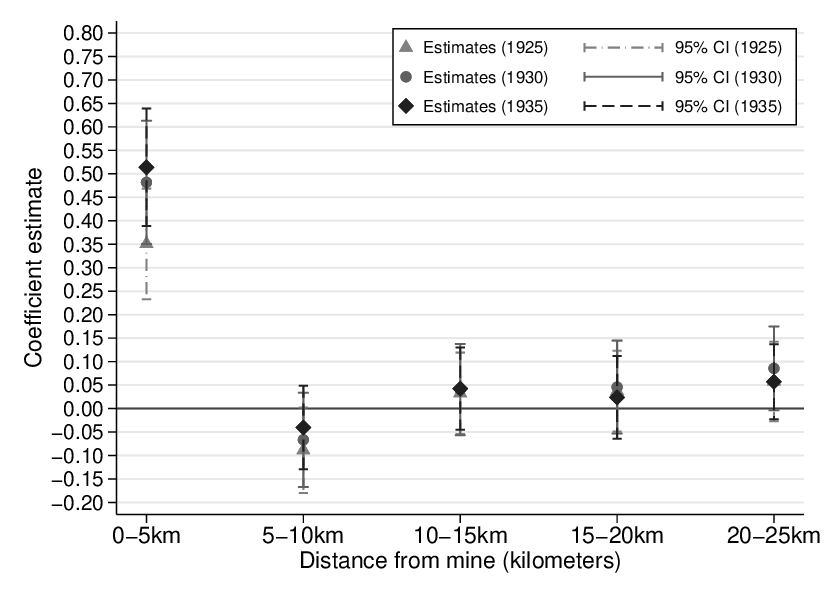}}
\subfloat[Fertility]{\label{fig:het_mfr}\includegraphics[width=0.32\textwidth]{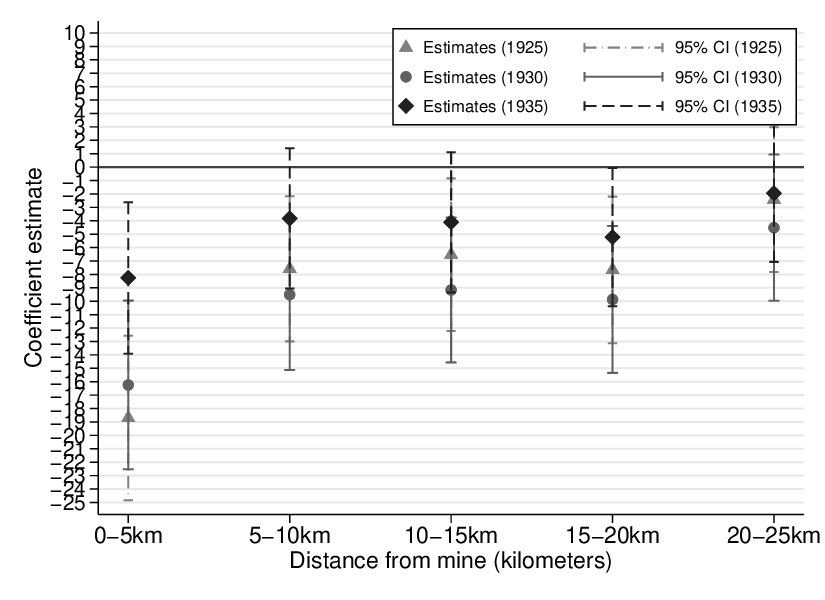}}
\subfloat[Mortality]{\label{fig:het_cdrf}\includegraphics[width=0.32\textwidth]{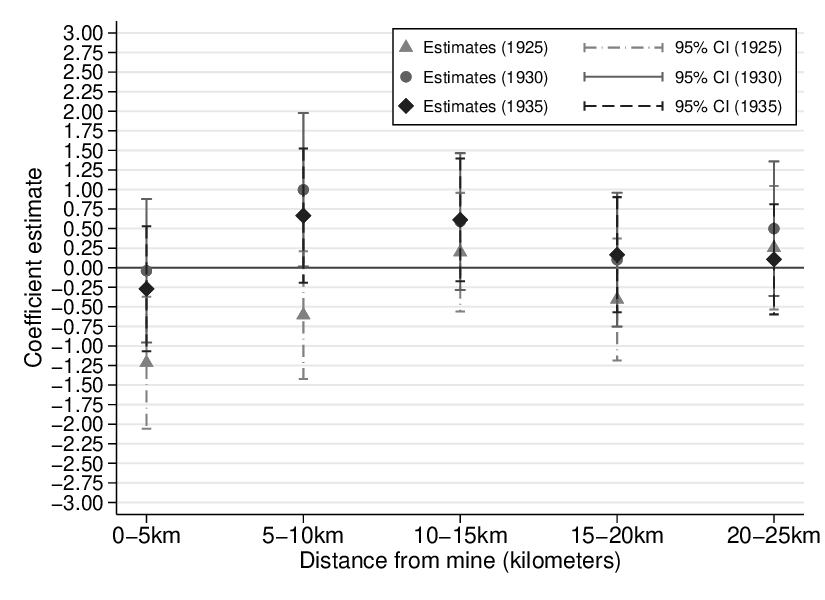}}\\
\subfloat[Sex Ratio]{\label{fig:het_sr}\includegraphics[width=0.32\textwidth]{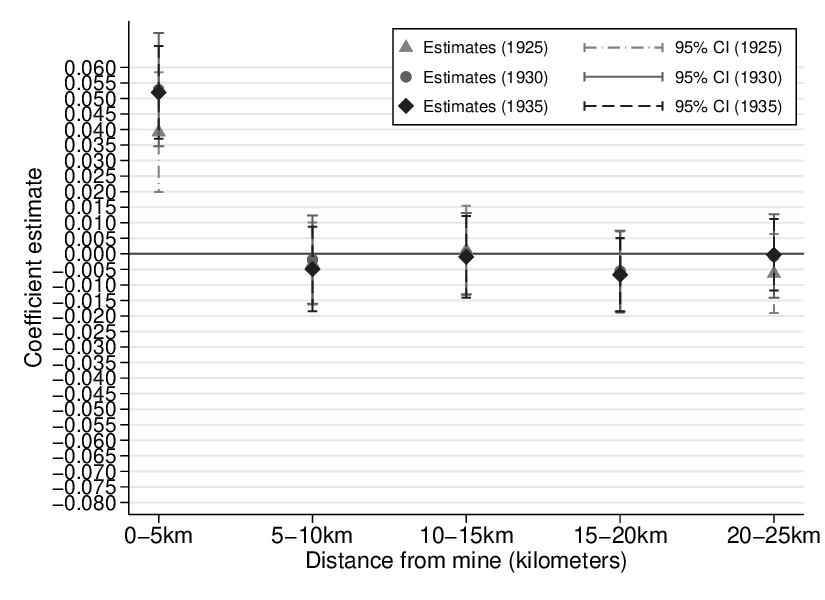}}
\subfloat[Household Size]{\label{fig:het_size}\includegraphics[width=0.32\textwidth]{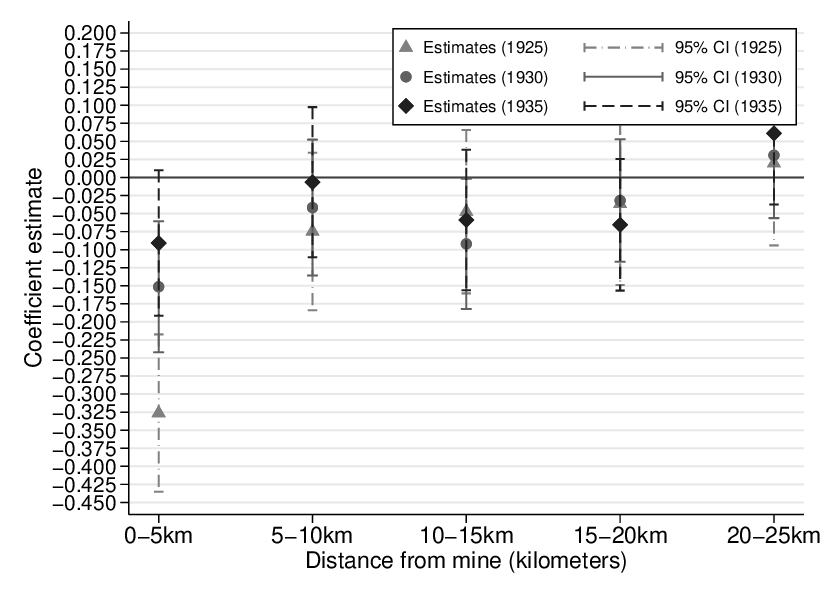}}
\subfloat[IMR]{\label{fig:het_imr}\includegraphics[width=0.32\textwidth]{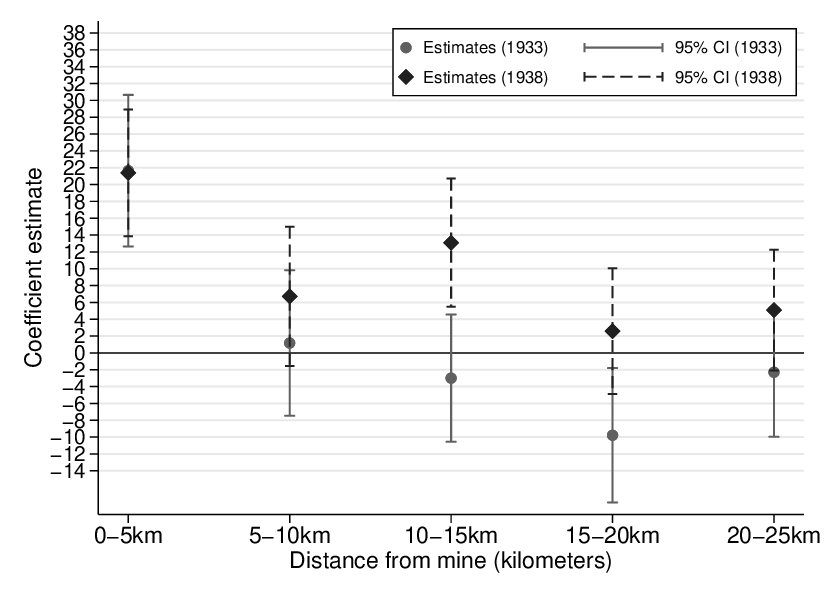}}
\caption{Heterogeneity in the Impact of Mines by Distance from the Mines}
\label{fig:het_dist}
\scriptsize{\begin{minipage}{450pt}
\setstretch{0.9}
The lines with caps show the estimates and their 95 percent confidence intervals, respectively.
The confidence intervals are calculated using the standard errors based on the heteroskedasticity-robust covariance matrix estimator.\\
Notes:
This figure shows the cross-sectional results from the expanded regressions of Equation~\ref{eq1} that include the five indicator variables for each 5 km bin from a mine to 25 km from the origin.
0--5 km distance from mines indicates the municipalities within 5 km of a mine.
5--10 (10--15, 15--20, 20--25) km distance from mines indicates the municipalities within $(5, 10]$ ($(10, 15], (15, 20], (20, 25]$) km of a mine.
The (25--30] km bin is used as a reference group.
Figures~\ref{fig:het_lnpop},~\ref{fig:het_mfr},~\ref{fig:het_cdrf},~\ref{fig:het_imr},~\ref{fig:het_sr}, and~\ref{fig:het_size} show the results for the local population (log-transformed), marital fertility rate, female mortality rate, infant mortality rate, sex ratio, and average household size, respectively.
Source: Created by the author.
\end{minipage}}
\end{figure}
%----------

The validity of the threshold for the exposure variable (\textit{MineDeposit} in equation~\ref{eq1}) is assessed using the main outcome variables.
Figure~\ref{fig:het_dist} provides the results from the expanded regression of equation~\ref{eq1} that includes the five indicator variables for each 5 km bin from a mine.
Figures~\ref{fig:het_lnpop},~\ref{fig:het_mfr},~\ref{fig:het_cdrf},~\ref{fig:het_imr},~\ref{fig:het_sr}, and~\ref{fig:het_size} show the results for the local population, marital fertility rate, female mortality rate, infant mortality rate, sex ratio, and average household size, respectively.
Note that these estimates are obtained from the OLS estimator, because the instrument cannot separately predict mine presence in each distance bin.
The estimates should therefore be interpreted as descriptive evidence on how the outcomes change with distance from mines, rather than as causal effects.
For outcomes with moderate effects, such as fertility and mortality, the estimates are imprecise.
In addition, the estimates shown in these figures are relative to the reference group, so statistical significance does not provide meaningful interpretation.
Therefore, the important point herein is that the shapes of the estimates for the dependent variables that captured clear effects in Section~\ref{sec:sec_ea} show a clear decreasing trend as distance from the mine increases.
Given this point, the results for the population (Figure~\ref{fig:het_lnpop}) and the sex ratio (Figure~\ref{fig:het_sr}) are useful for this test.\footnote{In other words, the other variables, such as fertility and mortality, are not useful herein because the main results do not have systematic effects. For example, the estimate for 1935 marital fertility is statistically insignificant in the main text (Table~\ref{tab:r_mech}). This means that estimates outside the 5 km threshold can be similar to those for the 0-5 km bin. Figure~\ref{fig:het_mfr} confirms this relationship.}

Overall, Figure~\ref{fig:het_dist} provides evidence that the estimates are larger in the absolute sense within $5$ km, whereas those for the areas outside the 5 km bin are close to zero in Figures~\ref{fig:het_lnpop} and~\ref{fig:het_sr}.
This provides a basis for defining the exposure variable, using $5$ km as the threshold.

A 10-15 km bin for the IMR in 1938 shown in Figure~\ref{fig:het_imr} shows a relatively clear hike compared with the other estimates for the bins outside the 5km threshold.
Since the main estimate for the 1938 IMR is not statistically significant (Table~\ref{tab:r_health}), it may be useful to analyze the reasons for this increase.
The descriptive analysis indicates that this increase in the IMR is mainly attributable to three observations in Iwate Prefecture.
The average IMR for these three Iwate municipalities is $163.1$, whereas that for the other municipalities is $119.9$.
This difference accounts for roughly one standard deviation ($42.6$) of the entire IMR.
Iwate Prefecture is a rural prefecture in the T\=ohoku region with lower sanitation standards, leading to greater fluctuations in infant mortality rates at that time.
My baseline results for the IMR and FDR, listed in Panel B of Table~\ref{tab:r_health}, remain unchanged when I exclude these three municipalities in Iwate Prefecture in 1938 (not reported).

\subsection{Sensitivity to Regional Differences in the Prevalence of Male Miners}\label{sec:app_r_hokkaido}

%----------
%Figure D3
\begin{figure}[h]
\centering
\captionsetup{justification=centering}
\subfloat[Population]{\label{fig:hkd_lnpop}\includegraphics[width=0.32\textwidth]{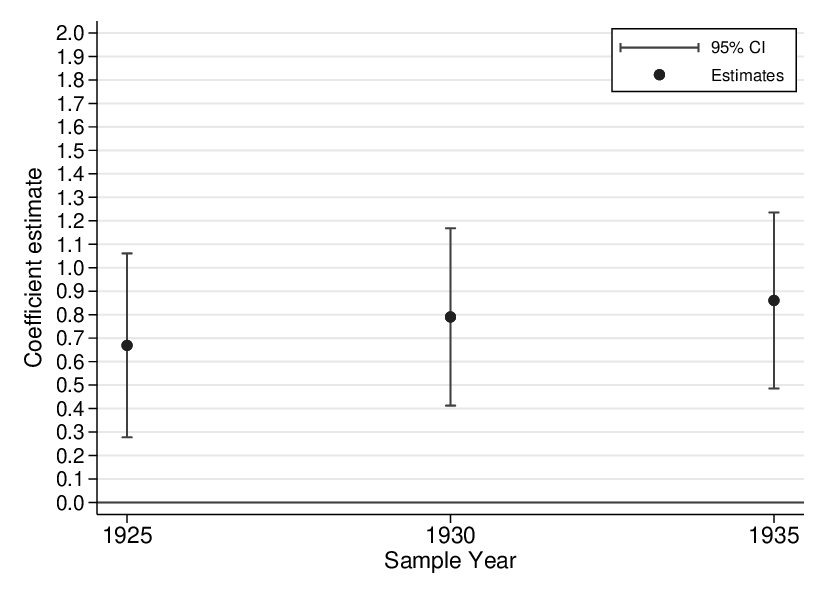}}
\subfloat[Fertility]{\label{fig:hkd_mfr}\includegraphics[width=0.32\textwidth]{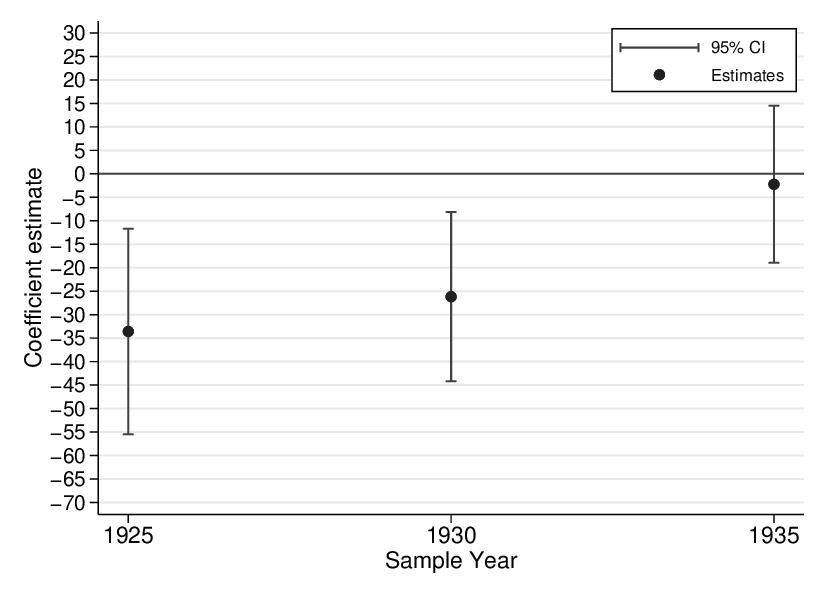}}
\subfloat[Mortality]{\label{fig:hkd_cdrf}\includegraphics[width=0.32\textwidth]{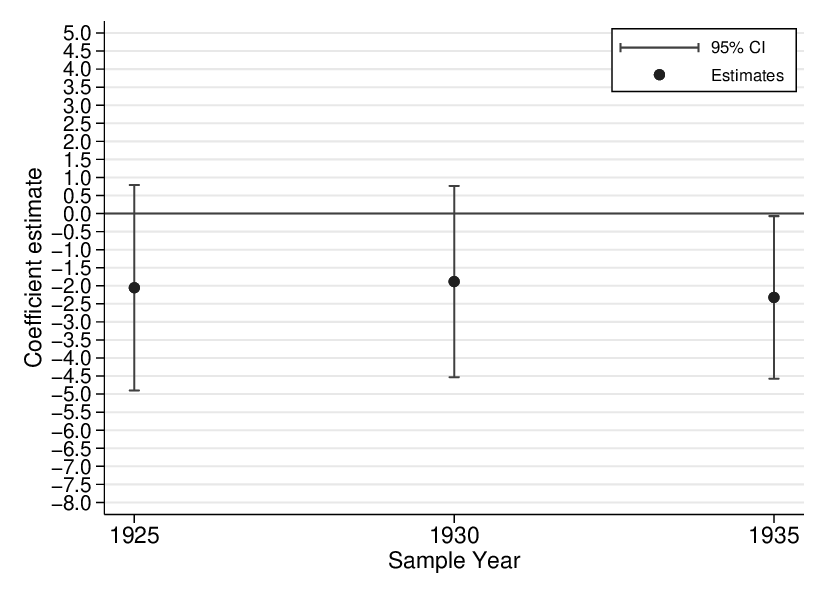}}\\
\subfloat[Sex Ratio]{\label{fig:hkd_sr}\includegraphics[width=0.32\textwidth]{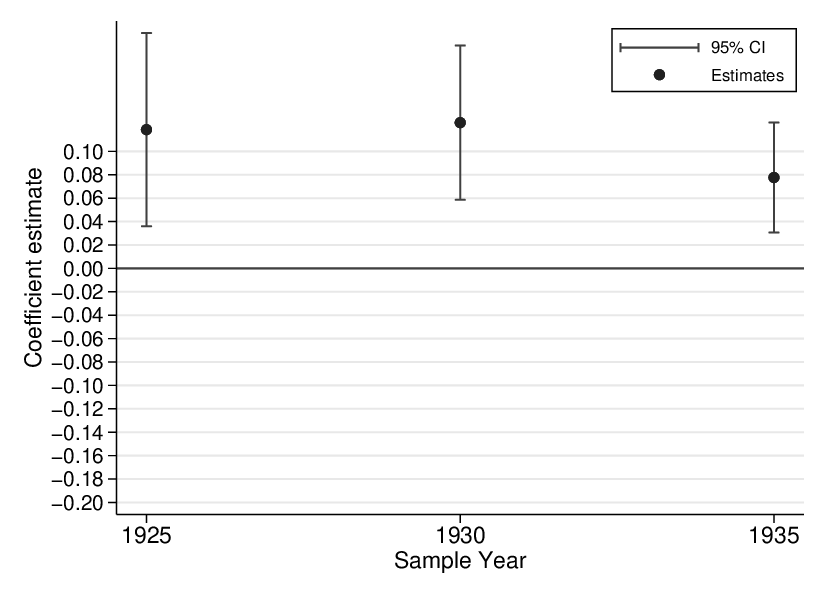}}
\subfloat[Size]{\label{fig:hkd_size}\includegraphics[width=0.32\textwidth]{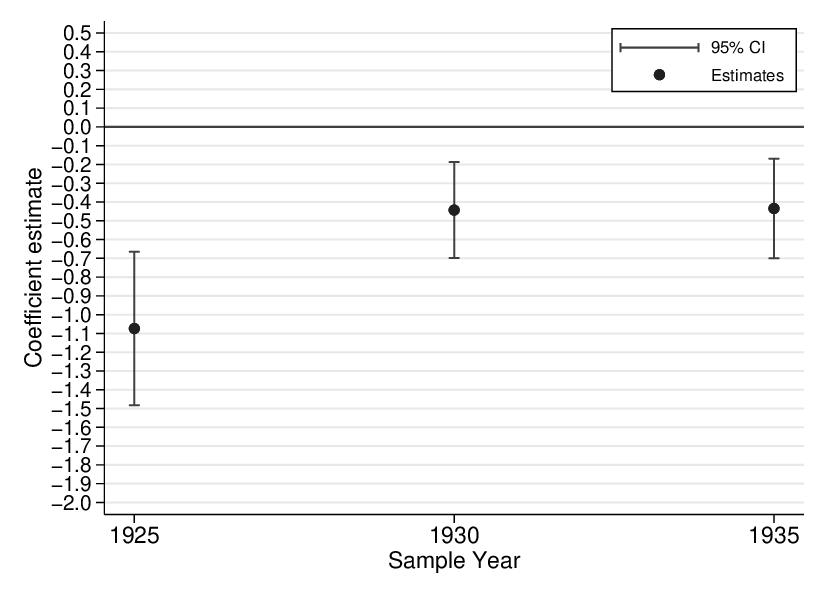}}
\subfloat[IMR]{\label{fig:hkd_imr}\includegraphics[width=0.33\textwidth]{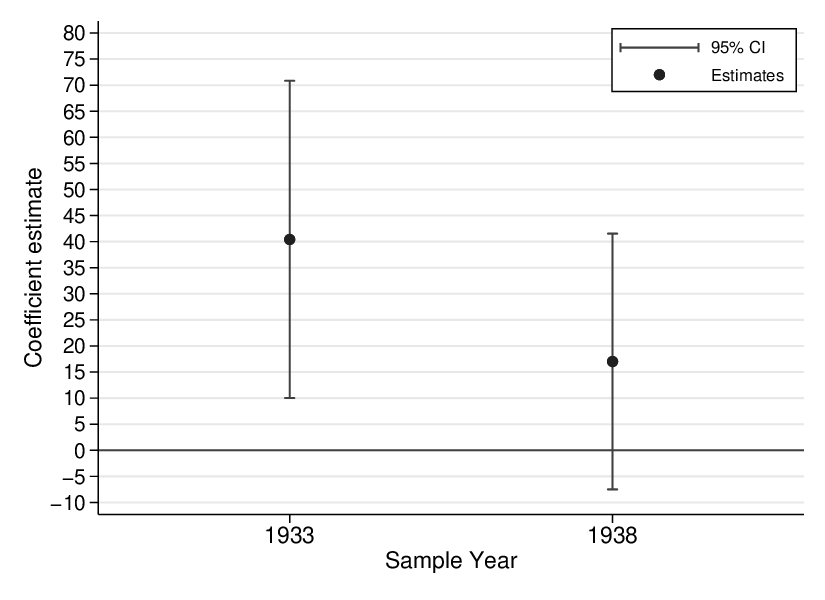}}
\caption{Additional Cross-Sectional Results: Including a Hokkaido Dummy}
\label{fig:r_hkd}
\scriptsize{\begin{minipage}{450pt}
\setstretch{0.9}
The lines with caps show the estimates and their 95 percent confidence intervals, respectively.
Standard errors are based on the heteroskedasticity-robust covariance matrix estimator.\\
Notes:
Figures~\ref{fig:hkd_lnpop},~\ref{fig:hkd_mfr},~\ref{fig:hkd_cdrf},~\ref{fig:hkd_sr},~\ref{fig:hkd_size}, and~\ref{fig:hkd_size} show the cross-sectional results for the local population (log-transformed), marital fertility rate, female mortality rate, sex ratio, average household size, and infant mortality rate, respectively.
All the estimates are based on the IV estimator.
All the regressions include city and town indicators, railway accessibility, port accessibility, river accessibility, elevation, and a Hokkaido dummy.
Source: Created by the author.
\end{minipage}}
\end{figure}
%----------

I assess the influence of regional heterogeneity in female labor patterns.
Nishinarita (1985) argues that, due to its geographical characteristics, coal mines in Hokkaido (a northernmost island) tended to employ men from rural villages in the T\=ohoku (northeastern) region of the main island as their main miners (see also Ogino 1993, p.~32).
This forced the coal mines in Hokkaido to use a sloping-face haulage method instead of using female miners (\textit{ato-yama}).
As a result, female miners in the Hokkaido mines were primarily engaged in coal selection rather than working in the pits.
It is unlikely that this regional heterogeneity could affect the location of mines because the location of coal mines depends on the presence or absence of coal-bearing strata (Section~\ref{sec:sec_ea_es}).
In addition, the number of mines in Hokkaido is relatively small, so such regional heterogeneity in labor supply is less likely to affect the main estimates.
Despite this, to test whether such features lead to omitted variable bias, an indicator variable for municipalities in Hokkaido was added to the regressions.
Figure~\ref{fig:r_hkd} shows the results from the expanded regressions.
The results are materially similar to those of the main results presented in the main text.

\subsection{Alternative Identification Strategy for Estimating the Impacts of Coal Mines}\label{sec:app_did}

There was little time variation in the coal mines during the sample period (Section~\ref{sec:sec_data_mine}).
This makes it difficult to apply the DID approach for the entire panel dataset to estimate the impacts of coal mines on the local outcomes.
Nevertheless, I consider the DID approach for a few available mining areas where coal mines opened during the sample period.
The Iwate coal mine in Iwate Prefecture and the Miyoshimuen coal mine in Akita Prefecture are not listed in the 1922 and 1931 editions of the NDFM but are listed in the 1936 edition.
It may provide relatively clean experimental conditions for the DID estimation, although the available sample size is very limited.

For municipal $i$ at time $t \in \{1925, 1930, 1935\}$, the two-way fixed effect (TWFE) model for the DID estimation is specified as follows:
\begin{eqnarray}\label{eq_did}
\footnotesize{
\begin{split}
\textit{y}_{i,t} = \psi + \delta \textit{D}_{i,t} + \mathbf{x}'_{i,t}\bm{\phi} + \eta_{i} + \lambda_{t} + \upsilon_{i,t},
\end{split}
}
\end{eqnarray}
where $\eta$, $\lambda$, and $\upsilon$ are the municipal fixed effect, year fixed effect, and random error term, respectively.
$\textit{D}_{i,t}$ is a treatment indicator variable that takes the value one for the treatment group in the post-treatment period and zero otherwise.\footnote{Because all treated municipalities are treated in the same period (1935), the TWFE estimator ($\hat{\delta}$) is equivalent to a two-group, two-period difference-in-differences estimator that compares the post-treatment period with the average of the two pre-treatment periods. It identifies the average treatment effect on the treated municipalities in 1935 under the assumptions of parallel trends and no anticipation. The problems with TWFE estimators discussed in the recent literature arise mainly when the timing of treatment varies across units, which is not the case here (de Chaisemartin and D'Haultf{\oe}uille 2023; Roth et al. 2023).}
$\mathbf{x}$ includes the accessibility measures on the stations and seaports.
Accessibility measures on the rivers are not included as they are time-constant variables.
However, the municipal fixed effect effectively controls for all these time-constant unobservable factors, including geological strata and terrain that may be correlated with mine locations.
The cluster-robust variance-covariance matrix estimator is used to address heteroskedasticity and serial dependence (Arellano 1987).

\begin{figure}[htb]
\centering
\captionsetup{justification=centering}
\subfloat[Population]{\label{fig:pta_pop}\includegraphics[width=0.33\textwidth]{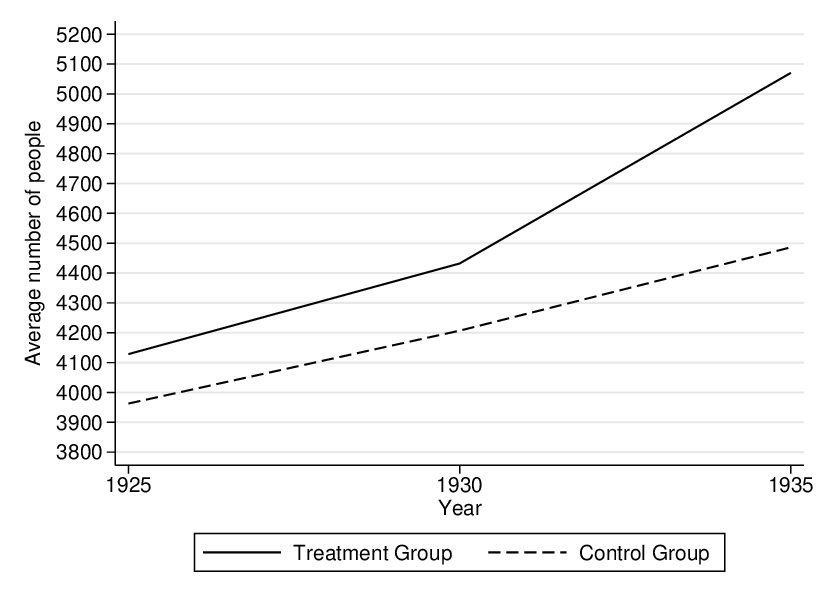}}
\subfloat[Marriage]{\label{fig:pta_cmarr}\includegraphics[width=0.33\textwidth]{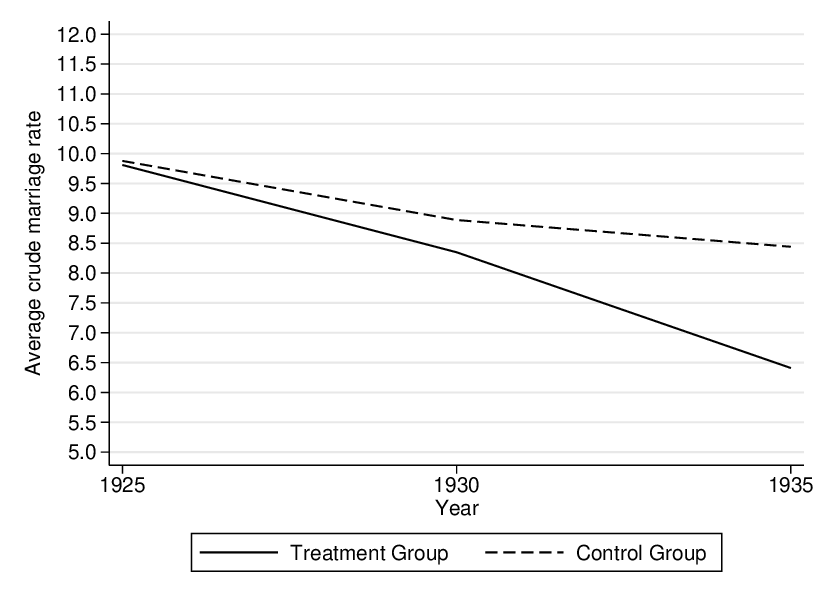}}
\subfloat[Fertility]{\label{fig:pta_mfr}\includegraphics[width=0.33\textwidth]{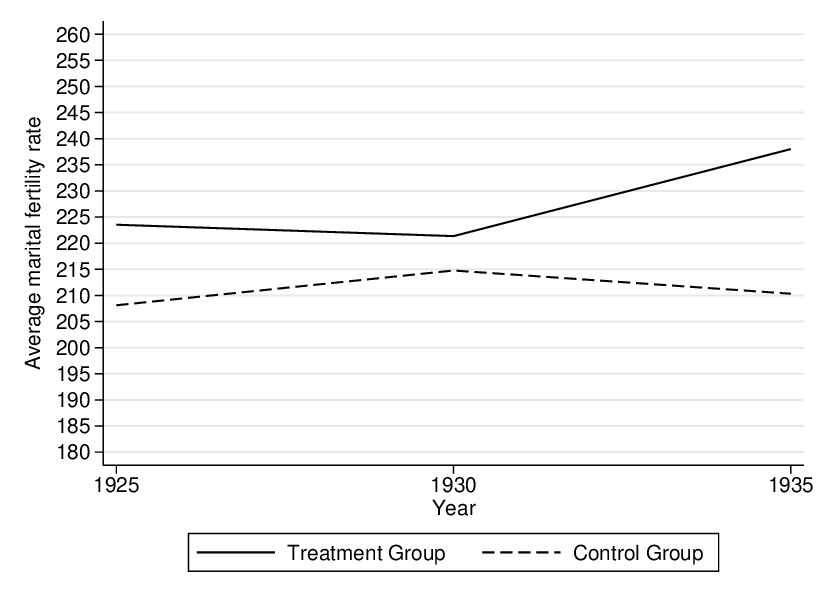}}\\
\subfloat[Mortality]{\label{fig:pta_cdrf}\includegraphics[width=0.33\textwidth]{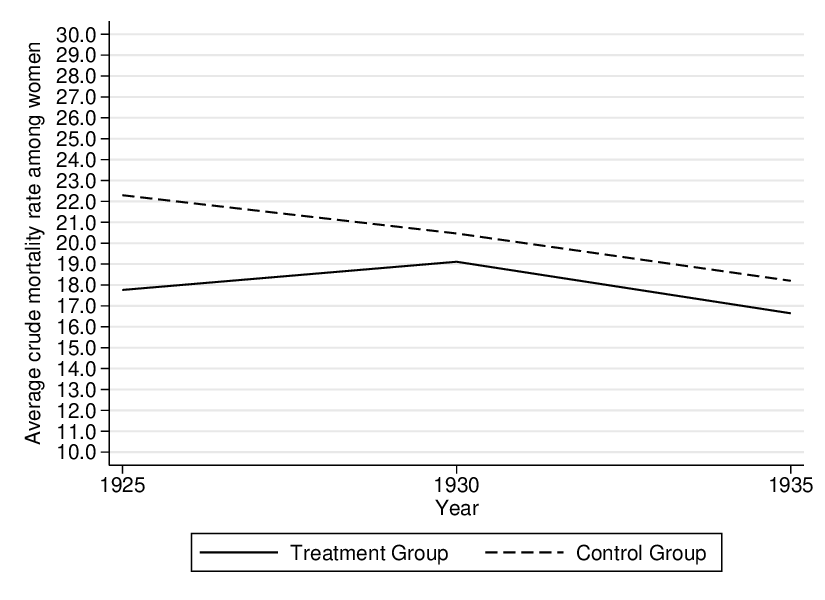}}
\subfloat[Sex Ratio]{\label{fig:pta_sr}\includegraphics[width=0.33\textwidth]{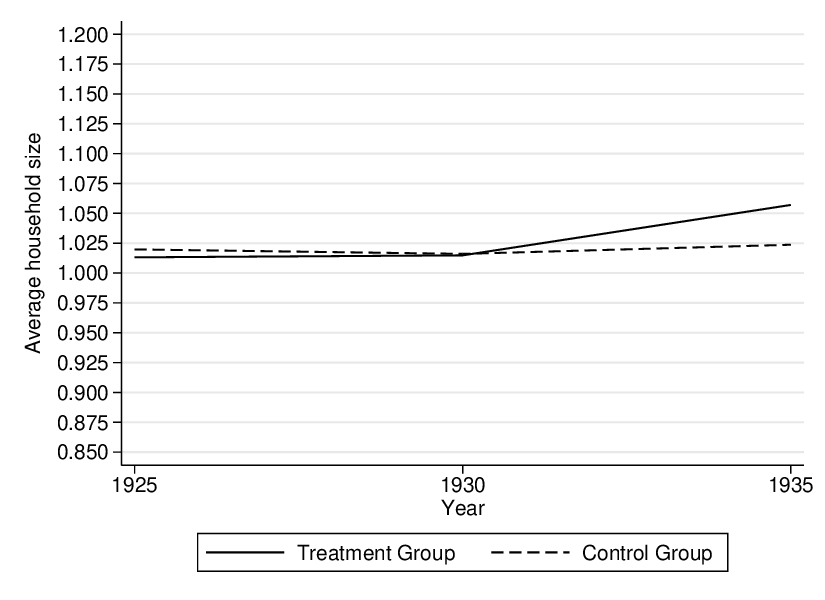}}
\subfloat[Household Size]{\label{fig:pta_size}\includegraphics[width=0.33\textwidth]{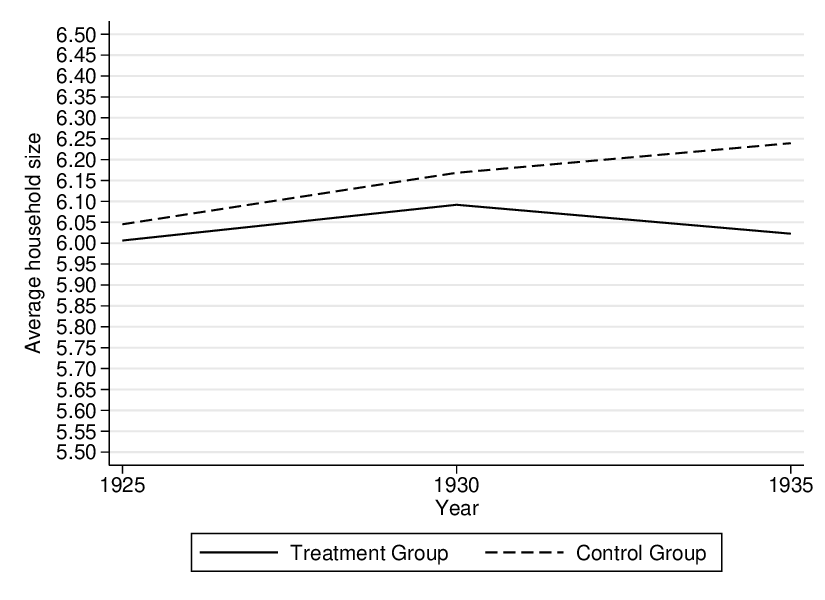}}
\caption{Trends in the Two Mining Areas in Iwate and Akita Prefectures}
\label{fig:pta}
\scriptsize{\begin{minipage}{450pt}
\setstretch{0.9}
Notes:
This figure shows the time-series plots of the outcomes of the treatment and control groups in Iwate and Akita prefectures used in the regressions shown in Table~\ref{tab:r_did_tohoku}.
Figures~\ref{fig:pta_pop},~\ref{fig:pta_cmarr},~\ref{fig:pta_mfr},~\ref{fig:pta_cdrf},~\ref{fig:pta_sr}, and~\ref{fig:pta_size} show the sample mean of the number of people, crude marriage rate, marital fertility rate, crude death rate among women, sex ratio, and household size, respectively.\\
Source: Created by the author.
\end{minipage}}
\end{figure}

The identifying assumption is that, in the absence of the mine openings, outcomes in the treatment and control groups would have followed parallel trends.
Figure~\ref{fig:pta} shows the time-series plots of the outcome variables used in the DID analysis.
Although there are moderate fluctuations due to the small sample size in each group, these outcomes do not show peculiar trends; they are materially similar in the pretreatment years.
The outcomes for the treatment group in the post-treatment period (i.e., 1935) show clear kinks in most cases, whereas those for the control group show only marginal changes.
A similar check of the early-life mortality rates is unavailable because data on the early-life deaths in 1925 are unavailable.

%----------
%Figure D4
%----------

Table~\ref{tab:r_did_tohoku} provides the results.
Overall, the DID results align with the baseline cross-sectional findings.
The estimate in Column (1) indicates that, after the opening of these coal mines, the population of the treatment group increased by approximately $9.2$\% ($\exp(0.088)-1$) relative to that of the control group.
Although the estimated magnitude is much smaller than the baseline magnitude listed in Table~\ref{tab:r_pop}\footnote{Notice that the estimates are compared with those for the results in 1935 or 1938 because the DID estimator captures the average treatment effect in the treated municipalities at the post-treatment period.}, this is consistent with the fact that both Iwate and Miyoshimuen coal mines are small-scale mines with around $100$ miners.
The estimated magnitudes for all the other dependent variables are also smaller than the baseline cross-sectional estimates.
This aligns with the little influence of the small-scale mines (Online Appendix~\ref{sec:r_inf}).
The estimated coefficients for the sex ratio and average household size are statistically significant (Columns 7 and 8), supporting the main finding that the influx of male workers increased the local population.

%------------------
%Table D2
\begin{landscape}
\begin{table}[htbp]
\def\arraystretch{1.0}
\begin{center}
\captionsetup{justification=centering}
\caption{Alternative Estimation Strategy: Difference-in-Differences Setting \\using Two-Mining Areas in Iwate and Akita Prefectures}
\label{tab:r_did_tohoku}
\scriptsize
\scalebox{0.95}[1]{
{\setlength\doublerulesep{2pt}
\begin{tabular}{lD{.}{.}{-2}D{.}{.}{-2}D{.}{.}{-2}D{.}{.}{-2}D{.}{.}{-2}D{.}{.}{-2}D{.}{.}{-2}D{.}{.}{-2}}
\toprule[1pt]\midrule[0.3pt]
&\multicolumn{8}{c}{Dependent Variable}\\
\cmidrule(rrrrrrrr){2-9}
&\multicolumn{1}{c}{}
&\multicolumn{3}{c}{Fertility Channel}
&\multicolumn{2}{c}{Mortality Channel}
&\multicolumn{2}{c}{Migration Channel}\\
\cmidrule(rrr){3-5}\cmidrule(rr){6-7}\cmidrule(rr){8-9}
&\multicolumn{1}{c}{(1) Population}
&\multicolumn{1}{c}{(2) Marriage}
&\multicolumn{1}{c}{(3) Crude}
&\multicolumn{1}{c}{(4) Marital}
&\multicolumn{1}{c}{(5) Male}
&\multicolumn{1}{c}{(6) Female}
&\multicolumn{1}{c}{(7) Sex Ratio}
&\multicolumn{1}{c}{(8) Size}\\
\hline
$\textit{MineDeposit} \times \textit{Post}$
&0.088$***$	&-1.574	&-1.793	&17.042	&1.885	&1.188	&0.042$***$	&-0.155$***$\\
&(0.024)		&(2.091)	&(1.568)	&(12.782)	&(1.973)	&(2.992)	&(0.011)		&(0.033) \\\hline
Municipal and Year FEs
&\multicolumn{1}{c}{Yes}&\multicolumn{1}{c}{Yes}&\multicolumn{1}{c}{Yes}&\multicolumn{1}{c}{Yes}&\multicolumn{1}{c}{Yes}&\multicolumn{1}{c}{Yes}&\multicolumn{1}{c}{Yes}&\multicolumn{1}{c}{Yes}\\
Railway accessibility
&\multicolumn{1}{c}{Yes}&\multicolumn{1}{c}{Yes}&\multicolumn{1}{c}{Yes}&\multicolumn{1}{c}{Yes}&\multicolumn{1}{c}{Yes}&\multicolumn{1}{c}{Yes}&\multicolumn{1}{c}{Yes}&\multicolumn{1}{c}{Yes}\\
Port accessibility
&\multicolumn{1}{c}{Yes}&\multicolumn{1}{c}{Yes}&\multicolumn{1}{c}{Yes}&\multicolumn{1}{c}{Yes}&\multicolumn{1}{c}{Yes}&\multicolumn{1}{c}{Yes}&\multicolumn{1}{c}{Yes}&\multicolumn{1}{c}{Yes}\\
Observations
&\multicolumn{1}{c}{207}&\multicolumn{1}{c}{207}&\multicolumn{1}{c}{207}&\multicolumn{1}{c}{207}&\multicolumn{1}{c}{207}&\multicolumn{1}{c}{207}&\multicolumn{1}{c}{207}&\multicolumn{1}{c}{207}\\
Clusters
&\multicolumn{1}{c}{69}&\multicolumn{1}{c}{69}&\multicolumn{1}{c}{69}&\multicolumn{1}{c}{69}&\multicolumn{1}{c}{69}&\multicolumn{1}{c}{69}&\multicolumn{1}{c}{69}&\multicolumn{1}{c}{69}\\
Sample Mean of the DV (TG in 1925; 30)
&\multicolumn{1}{c}{8.531}&\multicolumn{1}{c}{6.409}&\multicolumn{1}{c}{40.134}&\multicolumn{1}{c}{238.015}
&\multicolumn{1}{c}{19.184}&\multicolumn{1}{c}{16.645}&\multicolumn{1}{c}{1.057}&\multicolumn{1}{c}{6.023}\\
Std. Dev of the DV (TG in 1925; 30)
&\multicolumn{1}{c}{0.002}&\multicolumn{1}{c}{0.124}&\multicolumn{1}{c}{0.235}&\multicolumn{1}{c}{40.028}
&\multicolumn{1}{c}{1.457}&\multicolumn{1}{c}{5.240}&\multicolumn{1}{c}{0.014}&\multicolumn{1}{c}{0.364}\\\midrule[0.3pt]\bottomrule[1pt]
\end{tabular}
}
}
{\scriptsize
\begin{minipage}{600pt}
\setstretch{0.85}
***, **, and * represent statistical significance at the 1\%, 5\%, and 10\% levels, respectively.
Standard errors based on the cluster-robust covariance matrix estimator are reported in parentheses.\\
Notes:
This table shows the estimate ($\hat{\delta}$) from Equation~\ref{eq_did}, which aims to capture the average treatment effect of coal mine opening on the municipalities included in the treatment group in Akita and Iwate Prefectures.
This means that the estimate reflects both the impacts of the coal mine opening and the enforcement of female labor regulation.
Columns (1)--(8) show the results for the log-transformed population, crude marriage rate, crude birth rate, marital fertility rate, crude death rates for men and women, sex ratio, and household size, respectively.
The treatment group includes only two of the $69$ cross-sectional units, making IV estimation impossible for each dependent variable.
Given the small sample size, the port accessibility variable is omitted from the regressions for the early-life mortality outcomes due to collinearity.
If I conduct these regressions without the port accessibility variable, the estimates ($\hat{\delta}$) of the early-life mortality rates are not statistically significant.
\end{minipage}
}
\end{center}
\end{table}
\end{landscape}
%------------------

Note again that this table shows the suggested impacts of small-scale coal mines on the local demographic outcomes in the coal-mining municipalities in Akita and Iwate Prefectures in 1935.
This means that the estimate reflects both the influence of coal mine opening and the enforcement of regulations.
Despite this, both coal mines employed few female miners: only $9$ and $44$, in Iwate and Miyoshimuen, respectively.
Therefore, the regulations would be unlikely to have a large influence in these mines.
To summarize, the local population growth in both mining areas is mainly driven by the influx of male workers.

\subsection{Alternative Minerals as a Placebo: Heavy Metal Mines}\label{sec:secd_hm}

%------------------
%Table D3
%\begin{landscape}
\begin{table}[htbp]
\def\arraystretch{1.0}
\begin{center}
\captionsetup{justification=centering}
\caption{Placebo Results: Heavy Metal Mine as an Alternative Treatment}
\label{tab:r_gsc}
\scriptsize
\scalebox{0.88}[1]{
{\setlength\doublerulesep{2pt}
\begin{tabular}{lD{.}{.}{-2}D{.}{.}{-2}D{.}{.}{-2}D{.}{.}{-2}D{.}{.}{-2}D{.}{.}{-2}}
\toprule[1pt]\midrule[0.3pt]
\multicolumn{7}{l}{\textbf{Panel A: 1925 sample}}\\
&\multicolumn{6}{c}{Dependent Variable}\\
\cmidrule(rrrrrr){2-7}
&\multicolumn{1}{c}{(1) Population}
&\multicolumn{1}{c}{(2) Fertility}
&\multicolumn{1}{c}{(3) MR (Male)}
&\multicolumn{1}{c}{(4) MR (Female)}
&\multicolumn{1}{c}{(5) Sex Ratio}
&\multicolumn{1}{c}{(6) Size}\\\hline
\textit{MineDeposit}
&0.224$***$    &-0.080    &-0.137        &0.234    &-0.027$**$        &-0.195$***$\\
&(0.045)        &(2.451)    &(0.583)    &(1.011)        &(0.012)    &(0.059)\\\hline
City and town indicators
&\multicolumn{1}{c}{Yes}&\multicolumn{1}{c}{Yes}&\multicolumn{1}{c}{Yes}&\multicolumn{1}{c}{Yes}&\multicolumn{1}{c}{Yes}&\multicolumn{1}{c}{Yes}\\
Control variables
&\multicolumn{1}{c}{Yes}&\multicolumn{1}{c}{Yes}&\multicolumn{1}{c}{Yes}&\multicolumn{1}{c}{Yes}&\multicolumn{1}{c}{Yes}&\multicolumn{1}{c}{Yes}\\
Observations
&\multicolumn{1}{c}{3246}&\multicolumn{1}{c}{3246}&\multicolumn{1}{c}{3246}&\multicolumn{1}{c}{3246}&\multicolumn{1}{c}{3246}&\multicolumn{1}{c}{3246}\\
Estimator
&\multicolumn{1}{c}{OLS}&\multicolumn{1}{c}{OLS}&\multicolumn{1}{c}{OLS}&\multicolumn{1}{c}{OLS}&\multicolumn{1}{c}{OLS}&\multicolumn{1}{c}{OLS}\\
Sample Mean of the DV
&\multicolumn{1}{c}{8.041}&\multicolumn{1}{c}{182.345}&\multicolumn{1}{c}{21.392}&\multicolumn{1}{c}{20.295}&\multicolumn{1}{c}{1.009}&\multicolumn{1}{c}{5.213}\\
Std. Dev. of the DV
&\multicolumn{1}{c}{0.671}&\multicolumn{1}{c}{33.665}&\multicolumn{1}{c}{5.342}&\multicolumn{1}{c}{6.486}&\multicolumn{1}{c}{0.092}&\multicolumn{1}{c}{0.754}\\
&&&&&&\\
\multicolumn{7}{l}{\textbf{Panel B: 1930 sample}}\\
&\multicolumn{6}{c}{Dependent Variable}\\
\cmidrule(rrrrrr){2-7}
&\multicolumn{1}{c}{(1) Population}
&\multicolumn{1}{c}{(2) Fertility}
&\multicolumn{1}{c}{(3) MR (Male)}
&\multicolumn{1}{c}{(4) MR (Female)}
&\multicolumn{1}{c}{(5) Sex Ratio}
&\multicolumn{1}{c}{(6) Size}\\\hline
\textit{MineDeposit}
&0.181$***$    &-0.455    &-0.328        &-0.937$**$    &-0.008        &-0.164$***$\\
&(0.054)        &(3.371)    &(0.394)    &(0.378)        &(0.007)    &(0.056)\\\hline
City and town indicators
&\multicolumn{1}{c}{Yes}&\multicolumn{1}{c}{Yes}&\multicolumn{1}{c}{Yes}&\multicolumn{1}{c}{Yes}&\multicolumn{1}{c}{Yes}&\multicolumn{1}{c}{Yes}\\
Control variables
&\multicolumn{1}{c}{Yes}&\multicolumn{1}{c}{Yes}&\multicolumn{1}{c}{Yes}&\multicolumn{1}{c}{Yes}&\multicolumn{1}{c}{Yes}&\multicolumn{1}{c}{Yes}\\
Observations
&\multicolumn{1}{c}{2270}&\multicolumn{1}{c}{2270}&\multicolumn{1}{c}{2270}&\multicolumn{1}{c}{2270}&\multicolumn{1}{c}{2270}&\multicolumn{1}{c}{2270}\\
Estimator
&\multicolumn{1}{c}{OLS}&\multicolumn{1}{c}{OLS}&\multicolumn{1}{c}{OLS}&\multicolumn{1}{c}{OLS}&\multicolumn{1}{c}{OLS}&\multicolumn{1}{c}{OLS}\\
Sample Mean of the DV
&\multicolumn{1}{c}{8.162}&\multicolumn{1}{c}{183.497}&\multicolumn{1}{c}{19.598}&\multicolumn{1}{c}{18.629}&\multicolumn{1}{c}{1.015}&\multicolumn{1}{c}{5.510}\\
Std. Dev. of the DV
&\multicolumn{1}{c}{0.629}&\multicolumn{1}{c}{32.368}&\multicolumn{1}{c}{4.630}&\multicolumn{1}{c}{4.475}&\multicolumn{1}{c}{0.202}&\multicolumn{1}{c}{0.947}\\
&&&&&&\\
\multicolumn{7}{l}{\textbf{Panel C: 1935 sample}}\\
&\multicolumn{6}{c}{Dependent Variable}\\
\cmidrule(rrrrrr){2-7}
&\multicolumn{1}{c}{(1) Population}
&\multicolumn{1}{c}{(2) Fertility}
&\multicolumn{1}{c}{(3) MR (Male)}
&\multicolumn{1}{c}{(4) MR (Female)}
&\multicolumn{1}{c}{(5) Sex Ratio}
&\multicolumn{1}{c}{(6) Size}\\\hline
\textit{MineDeposit}
&0.180$***$    &2.085    &-0.530    &-0.411    &-0.029$***$    &-0.016\\
&(0.042)        &(2.000)    &(0.335)    &(0.335)    &(0.005)        &(0.043)\\\hline
City and town indicators
&\multicolumn{1}{c}{Yes}&\multicolumn{1}{c}{Yes}&\multicolumn{1}{c}{Yes}&\multicolumn{1}{c}{Yes}&\multicolumn{1}{c}{Yes}&\multicolumn{1}{c}{Yes}\\
Control variables
&\multicolumn{1}{c}{Yes}&\multicolumn{1}{c}{Yes}&\multicolumn{1}{c}{Yes}&\multicolumn{1}{c}{Yes}&\multicolumn{1}{c}{Yes}&\multicolumn{1}{c}{Yes}\\
Observations
&\multicolumn{1}{c}{3,936}&\multicolumn{1}{c}{3,936}&\multicolumn{1}{c}{3,936}&\multicolumn{1}{c}{3,936}&\multicolumn{1}{c}{3,936}&\multicolumn{1}{c}{3,936}\\
Estimator
&\multicolumn{1}{c}{OLS}&\multicolumn{1}{c}{OLS}&\multicolumn{1}{c}{OLS}&\multicolumn{1}{c}{OLS}&\multicolumn{1}{c}{OLS}&\multicolumn{1}{c}{OLS}\\
Sample Mean of the DV
&\multicolumn{1}{c}{8.086}&\multicolumn{1}{c}{179.269}&\multicolumn{1}{c}{18.774}&\multicolumn{1}{c}{17.843}&\multicolumn{1}{c}{1.013}&\multicolumn{1}{c}{5.454}\\
Std. Dev. of the DV
&\multicolumn{1}{c}{0.683}&\multicolumn{1}{c}{32.863}&\multicolumn{1}{c}{5.072}&\multicolumn{1}{c}{4.917}&\multicolumn{1}{c}{0.085}&\multicolumn{1}{c}{0.740}\\
&&&&&&\\
\multicolumn{7}{l}{\textbf{Panel D: 1925-1935 Panel Data Sample}}\\
&\multicolumn{6}{c}{Dependent Variable}\\
\cmidrule(rrrrrr){2-7}
&\multicolumn{1}{c}{(1) Population}
&\multicolumn{1}{c}{(2) Fertility}
&\multicolumn{1}{c}{(3) MR (Male)}
&\multicolumn{1}{c}{(4) MR (Female)}
&\multicolumn{1}{c}{(5) Sex Ratio}
&\multicolumn{1}{c}{(6) Size}\\\hline
\textit{MineDeposit}
&0.019	&1.570	&0.821	&-0.189	&0.006	&0.156$***$	\\
&(0.020)	&(2.908)	&(0.639)	&(0.529)	&(0.013)	&(0.043)		\\\hline
Municipal and Year FEs
&\multicolumn{1}{c}{Yes}&\multicolumn{1}{c}{Yes}&\multicolumn{1}{c}{Yes}&\multicolumn{1}{c}{Yes}&\multicolumn{1}{c}{Yes}&\multicolumn{1}{c}{Yes}\\
Control variables
&\multicolumn{1}{c}{Yes}&\multicolumn{1}{c}{Yes}&\multicolumn{1}{c}{Yes}&\multicolumn{1}{c}{Yes}&\multicolumn{1}{c}{Yes}&\multicolumn{1}{c}{Yes}\\
Observations
&\multicolumn{1}{c}{4,614}&\multicolumn{1}{c}{4,614}&\multicolumn{1}{c}{4,614}&\multicolumn{1}{c}{4,614}&\multicolumn{1}{c}{4,614}&\multicolumn{1}{c}{4,614}\\
Clusters
&\multicolumn{1}{c}{1,538}&\multicolumn{1}{c}{1,538}&\multicolumn{1}{c}{1,538}&\multicolumn{1}{c}{1,538}&\multicolumn{1}{c}{1,538}&\multicolumn{1}{c}{1,538}\\
Estimator
&\multicolumn{1}{c}{FE}&\multicolumn{1}{c}{FE}&\multicolumn{1}{c}{FE}&\multicolumn{1}{c}{FE}&\multicolumn{1}{c}{FE}&\multicolumn{1}{c}{FE}\\
Sample Mean of the DV
&\multicolumn{1}{c}{8.419}&\multicolumn{1}{c}{174.826}&\multicolumn{1}{c}{19.430}&\multicolumn{1}{c}{18.973}&\multicolumn{1}{c}{0.967}&\multicolumn{1}{c}{4.953}\\
 (TG in 1925 and 1930)&&&&&&\\
Std. Dev. of the DV
&\multicolumn{1}{c}{0.675}&\multicolumn{1}{c}{31.943}&\multicolumn{1}{c}{5.028}&\multicolumn{1}{c}{4.003}&\multicolumn{1}{c}{0.124}&\multicolumn{1}{c}{0.752}\\\midrule[0.3pt]\bottomrule[1pt]
\end{tabular}
}
}
{\scriptsize
\begin{minipage}{440pt}
\setstretch{0.85}
***, **, and * represent statistical significance at the 1\%, 5\%, and 10\% levels, respectively.
Standard errors based on the heteroskedasticity-robust covariance matrix estimator are reported in parentheses in Panels A-C.
Standard errors based on the cluster-robust covariance matrix estimator are reported in parentheses in Panel D.\\
Notes:
This table shows the results for the GSC mines sample.
Panels A, B, and C show the results for the 1925, 1930, and 1935 samples, respectively.
Panel D presents the results for the 1925-1935 ever- and never-treated panel data sample.
Column 1 shows the results for the log-transformed population.
Column 2 shows the results for the marital fertility rate.
Columns 3 and 4 show the results for the male and female mortality rates, respectively.
Columns 5 and 6 show the results for the sex ratio and average household size, respectively.
\end{minipage}
}
\end{center}
\end{table}
%\end{landscape}
%---------------

I use data on heavy metal mines as a placebo sample to assess the validity of my main results for the coal mines.
Specifically, I focus on three heavy metals: gold, silver, and copper (GSC).
This is because GSC mines account for more than $80$\% of all metal mines and generally extract these metals from a single deposit (Figure~\ref{fig:hist_mines}).
Although the NDFM documented other types of metal mines, they were much smaller in number: for example, the numbers of lead, zinc, pig iron, and tin mines recorded were only 1, 5, 4, and 4 in 1931, respectively.

Evidence indicates that coal mining requires many more workers than heavy metal mining because coal is a bulky mineral (Section~\ref{sec:sec2}).
Table~\ref{tab:sum_key} indicates that the scale of GSC mines was smaller than that of coal mines.
For example, the number of male (female) miners in 1922 is $1,641$ ($598$) for coal mines and $1,269$ ($168$) for GSC mines.
In addition, unlike coal mines, the GSC mines were scattered across the archipelago and less likely to agglomerate (Figure~\ref{fig:map_mines}).
The ``enclave effect'' suggests that some GSC mines are less likely to interact with local economies (Arag\'on and Rud 2013, 2016).\footnote{A potentially relevant study is Stijns (2005); using a country-level dataset, the author provided evidence suggesting that the impact of resource abundance on socioeconomic outcomes may vary among resource types.}
This means that the impact of GSC mines on the regional population growth could be smaller than that of coal mines.
To assess this, the OLS estimates of equation~\ref{eq1} for the coal and GSC mines are compared with those for the coal mines, unlike coal and fuel minerals, no specific stratum can predict the spatial distribution of GSC mines.

Trimming steps for the GSC analytical sample are summarized in Table~\ref{tab:dup}.
The ``enclave'' location of these mines increases the percentage of the control group, thereby increasing the number of observations relative to those for the coal mine samples.
The number of observations for the 1925, 1930, and 1935 samples are $3,246$, $2,270$, and $3,936$, respectively.
Similar to the coal mine data, the reduction in the number of heavy metal mines in 1931 NDFM might be due to a temporary decline in the workforce in small-scale GSC mines (i.e., falling below the NDFM threshold).
Again, the GSC mines are not concentrated in one place; a reduction in the number of mines reduced the number of controlled municipalities by a much greater margin than in the coal mine sample.

Panels A-C of Table~\ref{tab:r_gsc} list the cross-sectional results.
Panels A, B, and C summarize the results for the 1925, 1930, and 1935 samples, respectively.
As expected, the estimates for the local population listed in Column 1 range from $0.180$ to $0.224$, which are less than half of the estimates for the coal mine sample (Column 3 in Table~\ref{tab:r_pop}).
The results for the other demographic variables differ greatly from those for the coal mines.
The estimates for the fertility and mortality rates are close to zero and statistically insignificant in most cases (Columns (2)--(4)).
The estimate for female mortality in 1930 is statistically significantly negative.
While this might capture the wealth effects of the GSC mines, the estimate becomes less clear in 1935.
Similarly, Columns (5) and (6) do not provide systematic results on the sex ratio and average household size in the GSC mining area.
These results imply that regional growth in the GSC mining area was driven by mechanisms different from those in the coal mines.
This reflects the fundamental difference between coal and heavy metal mines as materials.

Panel D of Table~\ref{tab:r_gsc} lists the results for the ever-and never-treated panel data sample.
The estimates are close to zero in all the dependent variables and statistically insignificant in most cases.
This means that the enforcement of labor regulations had no meaningful impact on the GSC mining areas.
This is consistent with the fact that the GSC mines had not employed many female miners in the pits and thus were not influenced by the regulation.
This result supports the evidence that my main panel results for the coal mines capture the impacts of labor regulation enforcement rather than the trends in the local demographic outcomes.

\subsection{Time-Series Plots of the IMR and FDR}\label{sec:app_ts}

\begin{figure}[h]
\centering
\captionsetup{justification=centering,margin=1.5cm}
\includegraphics[width=0.5\textwidth]{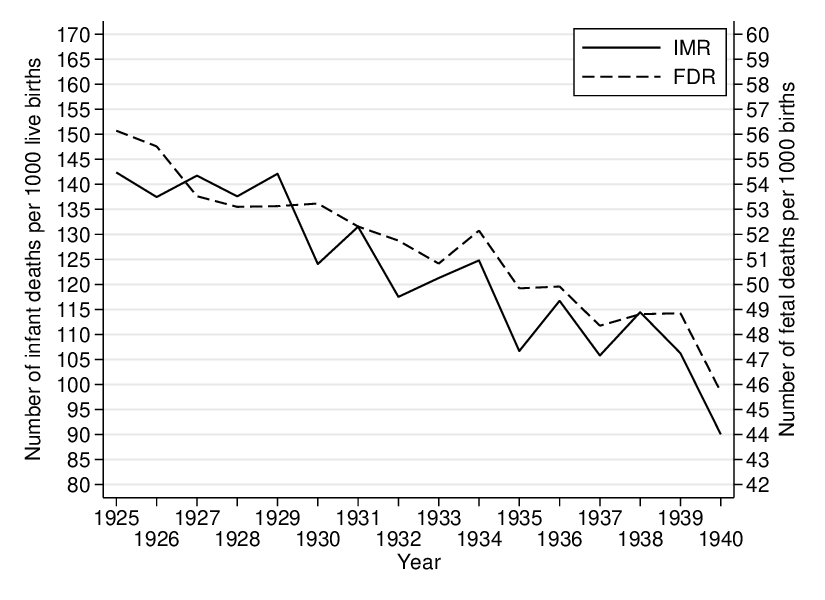}
\caption{The Infant Mortality and Fetal Death Rates\\ between 1925 and 1940}
\label{fig:ts_imr_fdr}
\scriptsize{\begin{minipage}{400pt}
\setstretch{0.85}
Notes:
This figure shows the national average infant mortality (IMR; solid line) and fetal death rates (FDR; dashed line) in Japan between 1925 and 1940.
The IMR is the number of infant deaths per 1,000 live births.
The FDR is defined as the number of fetal deaths per 1,000 births.\\
Sources:
Created by the author.
Data on the number of infant and fetal deaths are from the Statistics Bureau of the Cabinet (1926--1942).
\end{minipage}}
\end{figure}

Figure~\ref{fig:ts_imr_fdr} shows the national average IMR (solid line) and FDRs (dashed line).
Both series show secular decreasing trends between 1925 and 1940.

%---------------
%Figure D5
%---------------

\subsection{Sensitivity to Influential Observations}\label{sec:r_inf}

I assess the concern that municipalities with large-scale coal mines could have a critical influence, and thus my baseline results are simply based on these small numbers of large-scale mines.
Generally, estimating varying effects presents several practical issues.
For example, stratifying the treatment group disturbs the inference due to the small sample size in each subsample.
In fact, I confirmed that stratification by the median of the number of workers renders instrumental variable estimation impossible due to the small number of coal mines in each subsample.
Relatedly, determining a specific threshold for mine scale is an essentially difficult empirical issue.
Therefore, estimating the heterogeneous treatment effects in detail is beyond the scope of this study.
Despite this, it would be necessary to assess whether my baseline results are driven by very large-scale mines in a small subset of municipalities within the treatment group.

Panel A of Table~\ref{tab:r_did_pe_inf} presents the results of the panel data analysis using a subsample excluding large-scale mines (above the 95th percentile of the 1925 workers).
The results are materially similar to my baseline results presented in the main text.
This confirms that the baseline results are not driven by some influential large-scale mines.
Panel B of Table~\ref{tab:r_did_pe_inf} shows the results for a subsample excluding small-scale mines (fewer than the 5th percentile of the 1925 workers).
This also confirms that my baseline findings are not derived by the inclusion of small-scale mines.

%------------------
%Table D4
\begin{landscape}
\begin{table}[htbp]
\def\arraystretch{0.95}
\begin{center}
\captionsetup{justification=centering}
\caption{Additional Results: Excluding Large-Scale or Small-Scale Coal Mines\\ from the Ever-Treated and Ever-Controlled Panel Dataset}
\label{tab:r_did_pe_inf}
\scriptsize
\scalebox{1.0}[1]{
{\setlength\doublerulesep{2pt}
\begin{tabular}{lD{.}{.}{-2}D{.}{.}{-2}D{.}{.}{-2}D{.}{.}{-2}D{.}{.}{-2}D{.}{.}{-2}D{.}{.}{-2}D{.}{.}{-2}}
\toprule[1pt]\midrule[0.3pt]
\multicolumn{9}{l}{\textbf{Panel A: Large-Scale Coal Mines are Excluded}}\\
&\multicolumn{8}{c}{Dependent Variable}\\
\cmidrule(rrrrrrrr){2-9}
&\multicolumn{1}{c}{}
&\multicolumn{3}{c}{Fertility Channel}
&\multicolumn{2}{c}{Mortality Channel}
&\multicolumn{2}{c}{Migration Channel}\\
\cmidrule(rrr){3-5}\cmidrule(rr){6-7}\cmidrule(rr){8-9}
\textbf{Panel A-1: Equation~\ref{pe_did1}}
&\multicolumn{1}{c}{(1) Population}
&\multicolumn{1}{c}{(2) Marriage}
&\multicolumn{1}{c}{(3) Crude}
&\multicolumn{1}{c}{(4) Marital}
&\multicolumn{1}{c}{(5) Male}
&\multicolumn{1}{c}{(6) Female}
&\multicolumn{1}{c}{(7) Sex Ratio}
&\multicolumn{1}{c}{(8) Size}\\\hline
$\textit{MineDeposit} \times \textit{Post}$
&0.076		&1.602$**$	&4.038$***$	&21.174$***$    &0.320	&1.251   &-0.038     &0.004\\
&(0.048)		&(0.802)		&(1.564)		&(8.141)        	&(1.420)	&(1.436)  &(0.029)&(0.078) \\\hdashline[0.5pt/3pt]
First-stage $F$-statistic
&\multicolumn{1}{c}{35.842}&\multicolumn{1}{c}{35.842}
&\multicolumn{1}{c}{35.842}&\multicolumn{1}{c}{35.842}
&\multicolumn{1}{c}{35.842}&\multicolumn{1}{c}{35.842}
&\multicolumn{1}{c}{35.842}&\multicolumn{1}{c}{35.842}\\
&&&&&&&&\\
\textbf{Panel A-2: Pre-trend}
&\multicolumn{1}{c}{(1) Population}
&\multicolumn{1}{c}{(2) Marriage}
&\multicolumn{1}{c}{(3) Crude}
&\multicolumn{1}{c}{(4) Marital}
&\multicolumn{1}{c}{(5) Male}
&\multicolumn{1}{c}{(6) Female}
&\multicolumn{1}{c}{(7) Sex Ratio}
&\multicolumn{1}{c}{(8) Size}\\
\hline
$\textit{MineDeposit} \times I\text{(Year = 1925)}$
&-0.037		&0.618		&2.571		&14.742		&-3.506$*$	&-2.582	&0.024  &-0.028   \\
&(0.040)		&(0.948)		&(1.935)		&(9.973)		&(1.936)		&(1.810)	&(0.016) &(0.050) \\
$\textit{MineDeposit} \times I\text{(Year = 1935)}$
&0.057		&1.910$*$		&5.320$***$	&28.526$***$	&-1.428		&-0.036	&-0.026     &-0.010\\
&(0.040)		&(0.984)		&(1.987)		&(10.372)		&(1.681)		&(1.635)	&(0.026)  &(0.081)\\\hdashline[0.5pt/3pt]
First-stage $F$-statistic (1925)
&\multicolumn{1}{c}{18.847}&\multicolumn{1}{c}{18.847}&\multicolumn{1}{c}{18.847}&\multicolumn{1}{c}{18.847}
&\multicolumn{1}{c}{18.847}&\multicolumn{1}{c}{18.847}&\multicolumn{1}{c}{18.847}&\multicolumn{1}{c}{18.847}\\
First-stage $F$-statistic (1935)
&\multicolumn{1}{c}{20.744}&\multicolumn{1}{c}{20.744}&\multicolumn{1}{c}{20.744}&\multicolumn{1}{c}{20.744}
&\multicolumn{1}{c}{20.744}&\multicolumn{1}{c}{20.744}&\multicolumn{1}{c}{20.744}&\multicolumn{1}{c}{20.744}\\\hline
Municipal and year FEs
&\multicolumn{1}{c}{Yes}&\multicolumn{1}{c}{Yes}&\multicolumn{1}{c}{Yes}&\multicolumn{1}{c}{Yes}&\multicolumn{1}{c}{Yes}&\multicolumn{1}{c}{Yes}&\multicolumn{1}{c}{Yes}&\multicolumn{1}{c}{Yes}\\
Control variables
&\multicolumn{1}{c}{Yes}&\multicolumn{1}{c}{Yes}&\multicolumn{1}{c}{Yes}&\multicolumn{1}{c}{Yes}&\multicolumn{1}{c}{Yes}&\multicolumn{1}{c}{Yes}&\multicolumn{1}{c}{Yes}&\multicolumn{1}{c}{Yes}\\
Observations
&\multicolumn{1}{c}{2,745}&\multicolumn{1}{c}{2,745}&\multicolumn{1}{c}{2,745}&\multicolumn{1}{c}{2,745}
&\multicolumn{1}{c}{2,745}&\multicolumn{1}{c}{2,745}&\multicolumn{1}{c}{2,745}&\multicolumn{1}{c}{2,745}\\
Clusters
&\multicolumn{1}{c}{915}&\multicolumn{1}{c}{915}&\multicolumn{1}{c}{915}&\multicolumn{1}{c}{915}&\multicolumn{1}{c}{915}&\multicolumn{1}{c}{915}&\multicolumn{1}{c}{915}&\multicolumn{1}{c}{915}\\
&&&&&&\\
\multicolumn{9}{l}{\textbf{Panel B: Small-Scale Coal Mines are Excluded}}\\
&\multicolumn{8}{c}{Dependent Variable}\\
\cmidrule(rrrrrrrr){2-9}
&\multicolumn{1}{c}{}
&\multicolumn{3}{c}{Fertility Channel}
&\multicolumn{2}{c}{Mortality Channel}
&\multicolumn{2}{c}{Migration Channel}\\
\cmidrule(rrr){3-5}\cmidrule(rr){6-7}\cmidrule(rr){8-9}
\textbf{Panel B-1: Equation~\ref{pe_did1}}
&\multicolumn{1}{c}{(1) Population}
&\multicolumn{1}{c}{(2) Marriage}
&\multicolumn{1}{c}{(3) Crude}
&\multicolumn{1}{c}{(4) Marital}
&\multicolumn{1}{c}{(5) Male}
&\multicolumn{1}{c}{(6) Female}
&\multicolumn{1}{c}{(7) Sex Ratio}
&\multicolumn{1}{c}{(8) Size}\\\hline
$\textit{MineDeposit} \times \textit{Post}$
&0.086$*$		&0.855	&3.322$**$	&15.789$**$	&-0.985	&1.296   &-0.039     &0.028\\
&(0.050)		&(0.804)	&(1.532)		&(7.841)		&(1.438)	&(1.469)  &(0.030)&(0.080) \\\hdashline[0.5pt/3pt]
First-stage $F$-statistic
&\multicolumn{1}{c}{33.674}&\multicolumn{1}{c}{33.674}
&\multicolumn{1}{c}{33.674}&\multicolumn{1}{c}{33.674}
&\multicolumn{1}{c}{33.674}&\multicolumn{1}{c}{33.674}
&\multicolumn{1}{c}{33.674}&\multicolumn{1}{c}{33.674}\\
&&&&&&&&\\
\textbf{Panel B-2: Pre-trend}
&\multicolumn{1}{c}{(1) Population}
&\multicolumn{1}{c}{(2) Marriage}
&\multicolumn{1}{c}{(3) Crude}
&\multicolumn{1}{c}{(4) Marital}
&\multicolumn{1}{c}{(5) Male}
&\multicolumn{1}{c}{(6) Female}
&\multicolumn{1}{c}{(7) Sex Ratio}
&\multicolumn{1}{c}{(8) Size}\\
\hline
$\textit{MineDeposit} \times I\text{(Year = 1925)}$
&-0.050		&0.772		&1.748		&10.560		&-1.478		&-0.499	&0.017  &-0.037   \\
&(0.041)		&(0.984)		&(1.928)		&(9.924)		&(1.916)		&(1.831)	&(0.016) &(0.049) \\
$\textit{MineDeposit} \times I\text{(Year = 1935)}$
&0.061		&1.237		&4.188$**$	&21.023$**$	&-1.717		&1.049	&-0.030     &-0.010\\
&(0.041)		&(0.976)		&(1.919)		&(9.906)		&(1.716)		&(1.684)	&(0.027)  &(0.081)\\\hdashline[0.5pt/3pt]
First-stage $F$-statistic (1925)
&\multicolumn{1}{c}{17.565}&\multicolumn{1}{c}{17.565}&\multicolumn{1}{c}{17.565}&\multicolumn{1}{c}{17.565}
&\multicolumn{1}{c}{17.565}&\multicolumn{1}{c}{17.565}&\multicolumn{1}{c}{17.565}&\multicolumn{1}{c}{17.565}\\
First-stage $F$-statistic (1935)
&\multicolumn{1}{c}{18.463}&\multicolumn{1}{c}{18.463}&\multicolumn{1}{c}{18.463}&\multicolumn{1}{c}{18.463}
&\multicolumn{1}{c}{18.463}&\multicolumn{1}{c}{18.463}&\multicolumn{1}{c}{18.463}&\multicolumn{1}{c}{18.463}\\\hline
Municipal and year FEs
&\multicolumn{1}{c}{Yes}&\multicolumn{1}{c}{Yes}&\multicolumn{1}{c}{Yes}&\multicolumn{1}{c}{Yes}&\multicolumn{1}{c}{Yes}&\multicolumn{1}{c}{Yes}&\multicolumn{1}{c}{Yes}&\multicolumn{1}{c}{Yes}\\
Control variables
&\multicolumn{1}{c}{Yes}&\multicolumn{1}{c}{Yes}&\multicolumn{1}{c}{Yes}&\multicolumn{1}{c}{Yes}&\multicolumn{1}{c}{Yes}&\multicolumn{1}{c}{Yes}&\multicolumn{1}{c}{Yes}&\multicolumn{1}{c}{Yes}\\
Observations
&\multicolumn{1}{c}{2,388}&\multicolumn{1}{c}{2,388}&\multicolumn{1}{c}{2,388}&\multicolumn{1}{c}{2,388}
&\multicolumn{1}{c}{2,388}&\multicolumn{1}{c}{2,388}&\multicolumn{1}{c}{2,388}&\multicolumn{1}{c}{2,388}\\
Clusters
&\multicolumn{1}{c}{796}&\multicolumn{1}{c}{796}&\multicolumn{1}{c}{796}&\multicolumn{1}{c}{796}&\multicolumn{1}{c}{796}&\multicolumn{1}{c}{796}&\multicolumn{1}{c}{796}&\multicolumn{1}{c}{796}\\
\midrule[0.3pt]\bottomrule[1pt]
\end{tabular}
}
}
{\scriptsize
\begin{minipage}{590pt}
\setstretch{0.85}
***, **, and * represent statistical significance at the 1\%, 5\%, and 10\% levels, respectively.
Standard errors based on the cluster-robust covariance matrix estimator are reported in parentheses.\\
Notes:
Panel A (B) shows the results for the ever-treated and never-treated panel data sample, which excludes coal mines with workers above (below) the 95th (5th) percentile among the 1925 workers.
For example, the panel data sample in Panel A includes coal mines that were consistently observed from 1925 to 1935 and had fewer than the 95th percentile of workers among the 1925 mines, and defines ever-treated municipalities within a 5 km radius of those coal mines as the treatment group.
Municipalities within 5km to 30km of the coal mines are defined as never-treated municipalities.
To exclude duplicates, municipalities within a 5km radius of the GSC mines and coal mines that opened in 1930 or 1935 are removed from the panels.\\
Panels A-1 and B-1 present the estimate ($\hat{\varpi}$) from the second-stage structural equation~\ref{pe_did1}.
Panels A-2 and B-2 summarize the results from the expanded regression of equation~\ref{pe_did1}, which includes interaction terms between the ever-treated indicator variable and year dummies (with 1930 as the reference year).
Columns (1)--(8) show the results for the log-transformed population, crude marriage rate, crude birth rate, marital fertility rate, crude death rates for men and women, sex ratio, and average household size, respectively.
\end{minipage}
}
\end{center}
\end{table}
\end{landscape}
%------------------

%-------------------------------------------------------------------------------
% References used in the Appendices
%-------------------------------------------------------------------------------
\newpage
\renewcommand{\refname}{References used in the Appendices}

%-------------------------------------------------------------------------------
% Documents and statistical reports used in the Appendices
%-------------------------------------------------------------------------------
\renewcommand{\refname}{Documents used in the Appendices}

%-------------------------------------------------------------------------------
%-------------------------------------------------------------------------------
\end{document}